\documentclass[aps, two column, prb, superscriptaddress]{revtex4-2}
\usepackage{amsfonts}
\usepackage{amsmath}
\usepackage{subfigure}
\usepackage{graphicx, mathtools}
\usepackage[normalem]{ulem}
\usepackage{epsfig}
\usepackage{bm}
\usepackage{braket}
\usepackage{natbib}
\usepackage{enumitem}
\usepackage{longtable}
\usepackage{color}
\usepackage{xfrac}
\usepackage[UKenglish]{babel}
\usepackage[colorlinks=true,linkcolor=blue,urlcolor=blue,citecolor=blue,anchorcolor=blue]{hyperref}
\makeatletter
\usepackage{tikz}
\usetikzlibrary{decorations.markings}
\tikzset{->-/.style={decoration={markings, mark=at position #1 with {\arrow{>}}}, postaction={decorate}}}
\graphicspath{ {/home/soumya/figures/} }

\makeatother

\begin{document}

\title{Stripping the planar quantum compass model to its basics}

\author{Soumya Sur}\email{soumyasur@imsc.res.in}

\author{M. S. Laad}\email{mslaad@imsc.res.in}

\author{S. R. Hassan}\email{shassan@imsc.res.in}

\affiliation{Institute of Mathematical Sciences, Taramani, Chennai 600113, India}
\affiliation{Homi Bhabha National Institute Training School Complex,
Anushakti Nagar, Mumbai 400085, India}

\date{\rm\today}

\begin{abstract}
We introduce a novel mean-field theory (MFT) around the exactly soluble two-leg ladder limit for the planar quantum compass model (QCM). In contrast to usual MFT, our construction respects the stringent constraints imposed by emergent, lower (here $d=1$) dimensional symmetries of the QCM. Specializing our construction to the QCM on a periodic four-leg ladder, we find that a first-order transition separates two mutually dual Ising nematic phases, in good accord with state-of-the-art numerics for the planar QCM. One pseudo-spin-flip excitation in the ordered phase turns out to be two (Jordan-Wigner) fermion bound states, reminiscent of spin waves in spin-$1/2$ Heisenberg chains. We discuss the novel implications of our work on (1) the emergence of coupled orbital and magnetic ordered and liquidlike disordered phases, and (2) a rare instance of orbital-spin separation in $d>1$, in the context of a Kugel-Khomskii view of multi-orbital Mott insulators.
\end{abstract}
\maketitle

\section{Introduction}

\indent Frustration in quantum many body systems continues to evince interest for over four decades now. The wide range of exotic physical responses exhibited by transition metal oxides (TMO) are believed to originate from a complex interplay between strongly coupled charge, orbital, spin, and lattice degrees of freedom \cite{Imada-Fujimori}. Adding frustration results in exponential degeneracy of classical ground states. Quantum fluctuations are believed to select novel ordered states out of this manifold by ``order by disorder" mechanism \cite{Moessner-Sondhi,Lacorix}. The upshot is the possible emergence of truly novel, unconventional ordered phases of matter, characterized by fractionalized excitations in spatial dimension, $d>1$. These issues have been addressed extensively using a variety of analytic and numerical techniques each having their own strengths and limitations \cite{Lacorix}. While exact solutions in $d>1$ are rare, an exception is the celebrated Kitaev honeycomb model (KHM) \cite{Kitaev}. In absence of exact solutions in $d>1$, controlled approximations that capture the main essence of a given problem remain an attractive option. \\
	\indent The quantum compass model (QCM) on a square lattice, defined as 
\begin{align}
H_{QC}=\frac{1}{4}\sum_{\vec{r}}[J_{x}\sigma^{x}_{\vec{r}}\sigma^{x}_{\vec{r}+\hat{x}}+J_{z}\sigma^{z}_{\vec{r}}\sigma^{z}_{\vec{r}+\hat{z}}]
\end{align}
is a particularly illustrative case in point. The ``orbital only" QCM has received extensive attention as a model for orbital order in TMO based Mott insulators \cite{Kugel-Khomskii, Nussinov-review}, and in the context of qubit implementation in Josephson junction arrays \cite{Doucot}. Moreover, QCM on a square lattice is dual to the Xu-Moore model for $p+ip$ superconductor arrays \cite{Nussinov} and Kitaev's toric code (TC) model in transverse field \cite{Vidal}. These dualities are useful to unveil the hidden connections between classical and topological orders \cite{Chen1} in higher dimensions. On the analytic front, QCM has been extensively studied in $1d$ chains \cite{Andrzej1} and on two-leg ladders \cite{Andrzej2, Mattis}; the latter one has been very recently studied \cite{Hart} in the context of many-body localization.\\
\indent Inspired by the ``chain-mean-field" approach \cite{Chen2}, one may wonder whether the exactly soluble two-leg ladder QCM may serve as an appropriate template to investigate the QCM on a square lattice as a collection of coupled two-leg ladders. Reference~\cite{Chen2} finds a first-order quantum phase transition (QPT) between two states with $|\langle \sigma_{\vec{r}}^{x}\rangle|=m^{x}>0\ (|J_{x}|>\alpha |J_{z}|)$ and $|\langle \sigma^{z}_{\vec{r}}\rangle|=m^{z}>0\ (|J_{x}|<\alpha |J_{z}|)$ with $\alpha\neq 1$; thus, the chain-mean-field theory (MFT) violates the rigorous self-duality of square lattice QCM, which requires $\alpha=1$.\\
	\indent The (expected) inability of chain-MFT to respect self-duality can be traced back to its inherent inability to treat the lower ($d=1$) dimensional symmetries (LDS) \cite{Nussinov} rigorously. In the QCM, various $1d$ LDS constrain the finite temperature responses. These LDS are : (i) $P_{j}=\prod_{i}\sigma^{z}_{ij}$, acting on the row $j$ and (ii) $Q_{i}=\prod_{j}\sigma^{x}_{ij}$ acting along the column $i$ of the square lattice. Both $P_{j}$ and $Q_{i}$ commute with $H_{QC}$~, but $[P_{j},Q_{i}]\neq 0$ $\forall\ i,j$ while $[P_{j},P_{j'}]=[Q_{i},Q_{i'}]=0\ \forall\ i,i',j,j'$. Also, $[P_{i}P_{j},Q_{k}]=[Q_{i}Q_{j},P_{k}]=0$. Hence, all the eigenstates are two fold degenerate~\cite{Doucot} and there are exactly $2^{L}$ low energy states of $H_{QC}$ (here $L$ denotes linear size of the system).~At $J_{x}=J_{z}$, $H_{QC}$ is also invariant under a global ($d=2$) reflection symmetry $\sigma^{x}\leftrightarrow\sigma^{z}$ (implemented by an operator $R=\prod_{\vec{r}}\exp{\big[\frac{i\pi}{2\sqrt{2}}(\sigma^{x}_{\vec{r}}+\sigma^{z}_{\vec{r}})\big]}$). In the thermodynamic limit, all the $2^{L}$ low lying states collapse into each other for $J_{x}\neq J_{z}$ (and $2^{L+1}$ states for $J_{x}=J_{z}$) \cite{Dorier}, leading to infinite but sub-extensive degeneracy (scales with linear size) of the compass ground state.\\
	\indent Proliferation of non-local defects (like domain walls in $1d$ Ising model), generated by these $1d$ LDS, completely obliterates any conventional magnetic order at any temperature, $T>0$ due to Elitzur's theorem \cite{Batista}. Remarkably, the directional spin nematic order \cite{Nussinov}, described by an Ising like variable, $\langle\mathcal{D}\rangle=\langle \sigma^{x}_{\vec{r}}\sigma^{x}_{\vec{r}+\hat{x}}-\sigma^{z}_{\vec{r}}\sigma^{z}_{\vec{r}+\hat{z}}\rangle$ survives the strong fluctuations implied by these $1d$ LDS, even at finite $T$. Thus, any approach must obey (i) these LDS and (ii) self-duality. These are stringent constraints for any analytical approximation.\\
\section{Formulation}
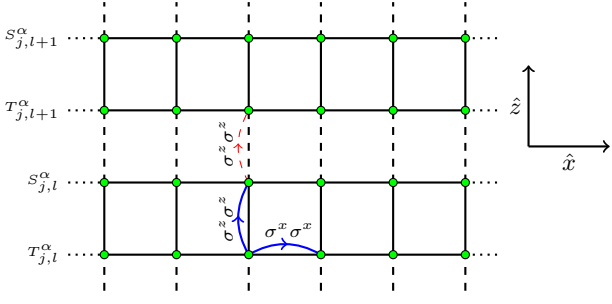
\begin{figure}
\begin{tikzpicture}[scale=1.2]
\draw [thick] (0,0) --(4.0,0); \draw [thick] (0,0.8) --(4.0,0.8);
\draw [thick] (0,1.6) --(4.0,1.6); \draw [thick] (0,2.4) --(4.0,2.4);
\foreach \i in {0,0.8,1.6,2.4,3.2,4.0}
     {\draw [thick] (\i,0) --(\i,0.8); \draw [thick] (\i,1.6) --(\i,2.4); 
      \draw [dashed, thick] (\i,0.8) --(\i,1.6);
       \draw [dashed, thick] (\i,2.4) --(\i,2.8); 
       \draw [dashed, thick] (\i,-0.4) --(\i,0);}
\foreach \i in {0,0.8,1.6,2.4}  
      {\draw [dotted, thick] (-0.4,\i) --(0.0,\i); 
       \draw [dotted, thick] (4.0,\i) --(4.4,\i);}
\draw [->-=.54, thick, blue] (1.6,0.0) to [bend left] (2.4,0.0);
\draw [->-=.54, thick, blue] (1.6,0.0) to [bend left] (1.6,0.8);
\draw [->-=.54, dashed, red] (1.6,0.8) to [bend left] (1.6,1.6);
\foreach \i in {0,0.8,1.6,2.4,3.2,4.0}
 \foreach \j in {0,0.8,1.6,2.4}
      \draw [fill=green] (\i,\j) circle [radius=0.045];
\node [left] at (-0.4,0.0) {\tiny $T^{\alpha}_{j,l}$};
\node [left] at (-0.4,0.8) {\tiny $S^{\alpha}_{j,l}$};
\node [left] at (-0.4,1.6) {\tiny $T^{\alpha}_{j,l+1}$};
\node [left] at (-0.4,2.4) {\tiny $S^{\alpha}_{j,l+1}$};
\node [above] at (2.05,0.1) {\scriptsize $\sigma^{x}\sigma^{x}$};
\node [left,rotate=90] at (1.35,0.77) {\scriptsize $\sigma^{z}\sigma^{z}$};
\node [left,rotate=90] at (1.35,1.6) {\scriptsize $\sigma^{z}\sigma^{z}$};
\draw [->=0.5, thick, black] (4.7,1.2) --(5.6,1.2);
\draw [->=0.5, thick, black] (4.7,1.2) --(4.7,2.1);
\node [below] at (5.15,1.22) {$\hat{x}$}; \node [left] at (4.72,1.65) {$\hat{z}$};
\end{tikzpicture}
\caption{QCM on a periodic square lattice is represented as collection of coupled two-leg QC ladders; Intra-(inter) two-leg ladder exchange interactions are shown in thick blue (dashed red) arrows.}\label{fig1}
\end{figure}

\indent In this communication, we construct a novel ``mean-field like" approach as a first step toward the full $2d$ QCM. Specifically, we exploit Mattis's exact solution for the $H_{QC}$ on a two-leg ladder \cite{Mattis} by coupling such two-leg ladders as shown in Fig.\ref{fig1}. Now, the Hamiltonian is $H_{QC}=\sum_{l=1}^{N/2} H^{(0)}_{l}+H_{int}$ , where $H^{(0)}_{l},H_{int}$ denote intra-ladder and inter-ladder interaction terms \cite{footnote1}.
\begin{align}
H^{(0)}_{l}=\sum_{j=1}^{N}\bigg[J_{x}\big(S^{x}_{jl}S^{x}_{j+1,l}+T^{x}_{jl}T^{x}_{j+1,l}\big)+J_{z}S^{z}_{jl}T^{z}_{jl}\bigg]\label{eq1}
\end{align}
\begin{align}
H_{int}=J_{z}\sum_{j=1}^{N}\sum_{l=1}^{N/2}S^{z}_{jl}T^{z}_{j,l+1}\label{eq2}
\end{align}
We now use the following canonical transformations \cite{Mattis},
\begin{align}
&\big(S^{x}_{jl}\ ,\ S^{y}_{jl}\ ,\ S^{z}_{jl}\big)=\big(P^{x}_{jl}\ ,\ 2P^{y}_{jl}Q^{x}_{jl}\ ,\ 2P^{z}_{jl}Q^{x}_{jl}\big)\nonumber\\
&\big(T^{x}_{jl}\ ,\ T^{y}_{jl}\ ,\ T^{z}_{jl}\big)=\big(-2Q^{z}_{jl}P^{x}_{jl}\ ,\ 2Q^{y}_{jl}P^{x}_{jl}\ ,\ Q^{x}_{jl}\big)\label{eq3}
\end{align}
These transformations are nothing but the two site version of well known Kramer-Wannier (KW) duality with an additional rotation of spin basis. To see this, we rotate only the $P^{\alpha}$ basis about $y$-axis by an angle $\pi/2$, ($P^{z}_{jl}\rightarrow \mathcal{P}_{jl}^{x},\ P^{x}_{jl}\rightarrow -\mathcal{P}_{jl}^{z}$) , then the Eq.~\eqref{eq3} implies (i): $2S^{z}_{jl}T^{z}_{jl}=\mathcal{P}^{x}_{jl},\ S^{x}_{jl}=-\mathcal{P}^{z}_{jl},\ T^{x}_{jl}=2(-\mathcal{P}^{z}_{jl})(-Q^{z}_{jl})$ and (ii): $2S^{x}_{jl}T^{x}_{jl}=-Q^{z}_{jl},\ T^{z}_{jl}=Q^{x}_{jl},\ S^{z}_{jl}=2Q^{x}_{jl}\mathcal{P}^{x}_{jl}$ . Both of these are conventional expressions of two site KW duality (apart from the minus signs which could be absorbed by a further rotation, $R_{x}(\pi)$ of $\mathcal{P},\ Q$ spins).\\
\indent Now using Eq.~\eqref{eq3}, $H^{(0)}_{l}$ and $H_{int}$ read following,
\begin{subequations}
\begin{align}
H^{(0)}_{l}=J_{x}\sum_{j=1}^{N}P^{x}_{jl}P^{x}_{j+1,l}(1+4Q^{z}_{jl}Q^{z}_{j+1,l})+\frac{J_{z}}{2}\sum_{j=1}^{N}P^{z}_{jl}\label{eq4a}
\end{align}
\begin{align}
H_{int}=2J_{z}\sum_{j=1}^{N}\sum_{l=1}^{N/2}P^{z}_{jl}Q^{x}_{jl}Q^{x}_{j,l+1}\label{eq4b}
\end{align}
\end{subequations}
When $H_{int}$ is absent, we get a collection of $1d$ transverse field Ising models (TFIM) with spins ($P^{x}_{jl}$) coupled to static $\mathbb{Z}_{2}$ variables $(Q^{z}_{jl})$. With $H_{int}$ , the $Q^{z}$s become fully dynamical, pre-empting exact solubility. At this stage, MF decoupling of $P^{\mu}$ and $Q^{\nu}$s in both Eq.~\eqref{eq4a} and \eqref{eq4b} gives,
\begin{subequations}
\begin{align}
H_{1}=\sum_{l=1}^{N/2}\bigg[J_{x}\sum_{j=1}^{N}&(1+4\langle Q^{z}_{jl}Q^{z}_{j+1,l}\rangle)P^{x}_{jl}P^{x}_{j+1,l}\nonumber\\
&+\frac{J_{z}}{2}\sum_{j=1}^{N}(1+4\langle Q^{x}_{jl}Q^{x}_{j,l+1}\rangle)P^{z}_{jl}\bigg]\label{eq5a}\\
&\hspace{-2.6cm}H_{2}=J_{x}\sum_{j=1}^{N}\sum_{l=1}^{N/2}(4\langle P^{x}_{jl}P^{x}_{j+1,l}\rangle)Q^{z}_{jl}Q^{z}_{j+1,l}\nonumber\\
&+J_{z}\sum_{j=1}^{N}\sum_{l=1}^{N/2}(2\langle P^{z}_{jl}\rangle)Q^{x}_{jl}Q^{x}_{j,l+1}\label{eq5b}
\end{align}
\end{subequations}
Now $H_{1}$ is a collection of $\sfrac{N}{2}$ $1d$ TFIMs with couplings determined by two-spin correlations of the $Q^{\nu}$, $(\nu=x,z)$, while $H_{2}$ is another $2d$ QCM (but on a $N\times\sfrac{N}{2}$ rectangular lattice) whose coefficients are the correlators of the $1d$ TFIM. Thus, it may look as if we have complicated the problem. However, this is not so, as we explain now.\\
\indent First we notice that (i) $\Delta^{x}_{jl}=4\langle P^{x}_{jl}P^{x}_{j+1,l}\rangle$ corresponds to $\langle \sigma^{x}_{\vec{r}}\sigma^{x}_{\vec{r}+\hat{x}}\rangle$ in the original spin language; similarly, (ii) $\Delta^{z}_{jl}=2\langle P^{z}_{jl}\rangle\sim \langle \sigma^{z}_{\vec{r}}\sigma^{z}_{\vec{r}+\hat{z}}\rangle$, (iii) $\Theta^{x}_{jl}=4\langle Q^{z}_{jl}Q^{z}_{j+1,l}\rangle\sim \langle\sigma^{x}_{\vec{r}}\sigma^{x}_{\vec{r}+\hat{z}}\sigma^{x}_{\vec{r}+\hat{x}}\sigma^{x}_{\vec{r}+\hat{x}+\hat{z}}\rangle$, (iv) $\Omega^{z}_{jl}=4\langle Q^{x}_{jl}Q^{x}_{j,l+1}\rangle\sim \langle \sigma^{z}_{\vec{r}}\sigma^{z}_{\vec{r}+2\hat{z}}\rangle$. Remarkably, {\it all} these MF averages thus respect the rigorous $1d$ LDS of the QCM. This feature, counter-intuitive for {\it any} MFT, will play a central role in our analysis, as we show below.\\
\indent We now notice that if we restrict ourselves to just {\it two coupled QC ladders}, Eq.~(\ref{eq5b}) then reads 
\begin{align}
H_{2}=\sum_{j=1}^{N}\sum_{l=1,2}\bigg[J_{x}\Delta^{x}_{jl}Q^{z}_{jl}Q^{z}_{j+1,l}+J_{z}\Delta^{z}_{jl}Q^{x}_{jl}Q^{x}_{j,l+1}\bigg]\label{eq6}
\end{align} 
which is {\it precisely} another two-leg QC ladder for $Q^{\nu}$. Here we assume periodic boundary conditions both along leg and rung directions of the ladder.\\
\indent Thus we can use the Mattis's transformations for the $Q^{\nu}$s in Eq.~\eqref{eq6}. Writing $Q^{x}_{j,1}=-2W^{z}_{j}V^{x}_{j},\ Q^{z}_{j,1}=W^{x}_{j},$ and $Q^{x}_{j,2}=-V^{x}_{j},\ Q^{z}_{j,2}=-2V^{z}_{j}W^{x}_{j},$ Eq.~\eqref{eq6} reads
\begin{align}
H_{2}=J_{x}\sum_{j=1}^{N}\bigg[\Delta^{x}_{j,1}+&(4V^{z}_{j}V^{z}_{j+1})\Delta^{x}_{j,2}\bigg]W^{x}_{j}W^{x}_{j+1}\nonumber\\
&+\frac{J_{z}}{2}\sum_{j=1}^{N}(\Delta^{z}_{j,1}+\Delta^{z}_{j,2})W^{z}_{j} \label{eq7}
\end{align}
So $H_{2}$ is another $1d$ TFIM of $W^{\mu}$ spins coupled to static $\mathbb{Z}_{2}$ fields $V^{z}_{j}$. These local $\mathbb{Z}_{2}$ variables are $V^{z}_{j}=8S^{x}_{j,2}T^{x}_{j,2}S^{x}_{j,1}T^{x}_{j,1}$ . Interestingly, these are just one of the $1d$ LDS of $H_{QC}$ for the {\it two coupled QC ladders}. We assume $V^{z}_{j}=\pm\sfrac{1}{2}$, $\forall\ j$, which is strictly valid only for $T=0^{+}$. This restores translational invariance, giving $\Delta^{\mu}_{jl}\equiv\Delta_{\mu}\ (\mu=x,z)$, $\Theta^{x}_{jl}\equiv\Theta_{x}$ and $\Omega^{z}_{jl}\equiv\Omega_{z}$. So finally,
\begin{subequations}
\begin{align}
H^{(l)}_{1}&=J_{x}(1+\Theta_{x})\sum_{j=1}^{N}P^{x}_{jl}P^{x}_{j+1,l}+\frac{J_{z}}{2}(1+\Omega_{z})\sum_{j=1}^{N}P^{z}_{jl}\label{eq8a}\\
& H_{2}=2J_{x}\Delta_{x}\sum_{j=1}^{N}W^{x}_{j}W^{x}_{j+1}+J_{z}\Delta_{z}\sum_{j=1}^{N}W^{z}_{j}\label{eq8b}
\end{align}
\end{subequations} 
We solve the Hamiltonians Eqs.~\eqref{eq8a} and \eqref{eq8b} using the exact solution of $1d$ TFIM (see Appendix \ref{App A}). The four self-consistency equations are written in the following compact manner. Define two vectors $\mathbf{M}^{x},\ \mathbf{M}^{z}$, with two components $M^{x}_{a}=\Delta_{x}$, $M^{x}_{b}=\Theta_{x}$, $M^{z}_{a}=\Delta_{z}$, and $M^{z}_{b}=\Omega_{z}$.
\begin{align}
M^{x}_{\sigma}=\int_{0}^{\pi}\frac{dk}{\pi}\frac{(h_{\sigma}\cos{k}-1)\tanh{\big(\beta E^{\sigma}_k/2\big)}}{\sqrt{1+h^{2}_{\sigma}-2h_{\sigma}\cos{k}}} \label{eq9}
\end{align} 
\begin{align}
\ \ \ M^{z}_{\sigma}=\int_{0}^{\pi}\frac{dk}{\pi}\frac{(h^{-1}_{\sigma}\cos{k}-1)\tanh{\big(\beta E^{\sigma}_k/2\big)}}{\sqrt{1+h^{-2}_{\sigma}-2h^{-1}_{\sigma}\cos{k}}}\label{eq10}
\end{align} 
where $E^{a}_{k}=\frac{|J_{x}|}{2}(1+\Theta_{x})\sqrt{1+h^{2}_{a}-2h_{a}\cos{k}}$ , $E^{b}_{k}=|J_{x}||\Delta_{x}|\sqrt{1+h^{2}_{b}-2h_{b}\cos{k}}$ , $h_{a}=J_{z}(1+\Omega_{z})/J_{x}(1+\Theta_{x})$ , $h_{b}=J_{z}\Delta_{z}/J_{x}\Delta_{x}$ , and $\beta=1/T$ .\\
\indent We could equivalently have applied the Mattis's relations (for $Q^{\nu}$) before Eq.~\eqref{eq7} to Eqs.~\eqref{eq4a} and \eqref{eq4b} and then done a MF decoupling, again leading to Eqs.~\eqref{eq8a} and \eqref{eq8b}. Notice that the MF self-consistency Eqs.~\eqref{eq9} and \eqref{eq10} faithfully reflect the exact self-duality of the square lattice QCM at $J_{x}=J_{z}$ ($M^{x}_{\sigma}\leftrightarrow M^{z}_{\sigma}$ when $J_{x},\ \Delta_{x},\ \Theta_{x}\leftrightarrow J_{z},\ \Delta_{z},\ \Omega_{z}$). This is a very positive feature of the present approach, in contrast to that of Ref. \cite{Chen2}, which violates this constraint. This important difference can be traced back to the fact that our MF decoupling is performed at a specific two-(particle) spin channel that preserves the LDS. We also find that this particular decoupling channel is the {\it only} one (in the present approach, based on Mattis's relations) which preserves LDS and self-duality. Other channels (e.g., conventional single site MF decoupling) will violate these symmetries and this may produce qualitatively incorrect results at finite $T$. Keeping the LDS in mind, one could in principle use the above MF construction for a general $N$-leg compass ladder (for example, see the eight-leg case in Appendix \ref{App E}). For $N=2^{p}$, one has to apply Mattis's relations (Eq.~\eqref{eq3}) $(p-1)$ times and similarly $(p-1)$ MF decouplings [like Eqs.~\eqref{eq5a} and \eqref{eq5b}] to get a collection of analytically soluble TFIMs [like Eqs.~\eqref{eq8a} and \eqref{eq8b}]. In this way, total $2p$ different MF order parameters are generated. We find nearest neighbour nematics ($\Delta_{x},\ \Delta_{z}$) and various non-local static spin-correlations like $\Omega_{m}^{z}=\langle\sigma^{z}_{j}\sigma^{z}_{j+2^{m}\hat{z}}\rangle$ and $\Theta^{x}_{m}=\langle \theta^{x}_{j}\theta^{x}_{j+2\hat{z}}...\theta^{x}_{j+(2^{m}-4)\hat{z}}\theta^{x}_{j+(2^{m}-2)\hat{z}}\rangle$, $\ m=1,$ $2,...,p-1$ with $\theta^{x}_{j}=\sigma_{i}^{x}\sigma^{x}_{i+\hat{x}}\sigma^{x}_{i+\hat{z}}\sigma^{x}_{i+\hat{x}+\hat{z}}$ to emerge from the MF decoupling procedure. The number of such MF order parameters (to be determined self-consistently) diverges as $O(\ln(N))$ as $N$ becomes large, which make the above method impractical for doing calculations when the number of ladder legs become large. Hence, a MFT which can be directly used for the full 2d compass model (and preserves the LDS) requires further ingenious way(s) which are beyond the scope of present work.\\

\section{Results}
\begin{figure}[htb]
\includegraphics[height=50mm,width=70mm]{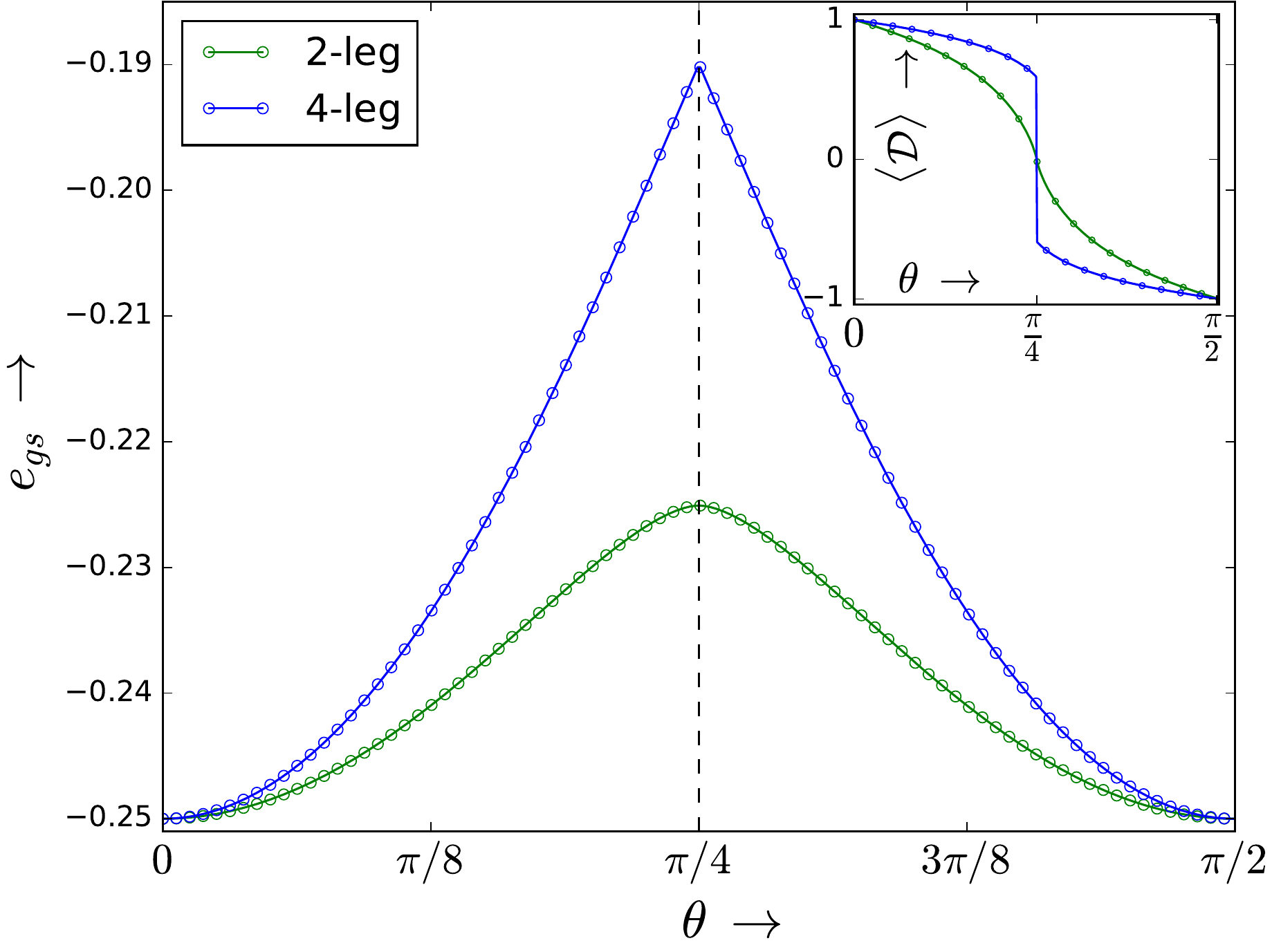}
\caption{(a) Ground state energy per site ($e_{gs}$) vs. anisotropy parameter ($\theta$) for (i) two-leg ladder [~$\theta=\tan^{-1}(|J_{z}|/2|J_{x}|)$~], (ii) four-leg ladder [~$\theta=\tan^{-1}(|J_{z}|/|J_{x}|)$~]; this MF $e_{gs}$ agrees closely with Refs.~\cite{Orus, Vidal, Dorier}. (b) Nematic order parameter ($\langle\mathcal{D}\rangle$) at $T=0$ for (i) and (ii) is shown in the inset.}
\label{Fig.2}
\end{figure}

\indent We now present our results. The two-leg QC ladder exhibits a quantum critical point (see Ref.~\cite{Mattis}) separating a ``magnetically ordered" and ``quantum disordered" phase at $T=0$. In original spin variables, this is also a continuous transition between $xx$-ordered phase to $zz$-ordered phase, with Ising nematic order parameter $\langle \mathcal{D}\rangle=+(-)D_{0}$ for $2|J_{x}|>(<)|J_{z}|$, clearly shown in Fig.~\ref{Fig.2}.~This QPT belongs to the well known $2d$ classical Ising universality class.
On the other side, four-leg QC ladder reveals a clear first-order transition between the two Ising nematic phases above, (see Fig.~\ref{Fig.2}) precisely at the self-dual point ($J_{x}=J_{z}$). This agrees fully with both exact arguments and numerical results \cite{Rynbach, Orus}. Interestingly, the MF ground state energy per lattice site, $e_{gs}(\theta)$, as a function of $\theta=\tan^{-1}(|J_{z}|/|J_{x}|)$ (see Eq.~\eqref{A.8}) exhibits a clear cusp (see Fig.~\ref{Fig.2}) at $\theta=\pi/4\ (J_{x}=J_{z})$, and agrees closely, both in its functional form and magnitude, with the PCUT results of Vidal~{\it et~al.} \cite{Vidal}, iPEPS results of Or\'us {\it et al.} \cite{Orus}, and Green function Monte-Carlo results of Dorier {\it et al.} \cite{Dorier}. In fact, at $\theta=\pi/4$, the point of maximum frustration, our $e_{gs}=-0.19$ is very close to the $e_{gs}\approx -0.2$ found from the above numerical techniques. Considering the approximations made here, this is remarkable accord. Thus, most of the ground state correlation energy for the full $2d$ QCM already seems to be captured by a four-leg ladder. Equivalences between the QCM and Xu-Moore as well as the transverse field-TC model also imply first-order QPTs at self-dual points in these models \cite{Vidal}. Even more interestingly, we also uncover a hitherto unnoticed (to our best knowledge) duality between a plaquette order, $\Theta_{x}$ and a next-near neighbour $zz$-correlation, $\Omega_{z}$ (this ``hidden" duality could also be proven analytically, see Appendix \ref{App B}) in Fig.~\ref{Fig.3}. Both of these also exhibit a clear jump at $J_{x}=J_{z}$~. While possible ``hidden" dimer order has been studied earlier \cite{Brzezicki} in the $2d$ QCM, the duality between $\Theta_{x}$ and $\Omega_{z}$ is a new finding.\\
\begin{figure}
\includegraphics[height=55mm,width=75mm]{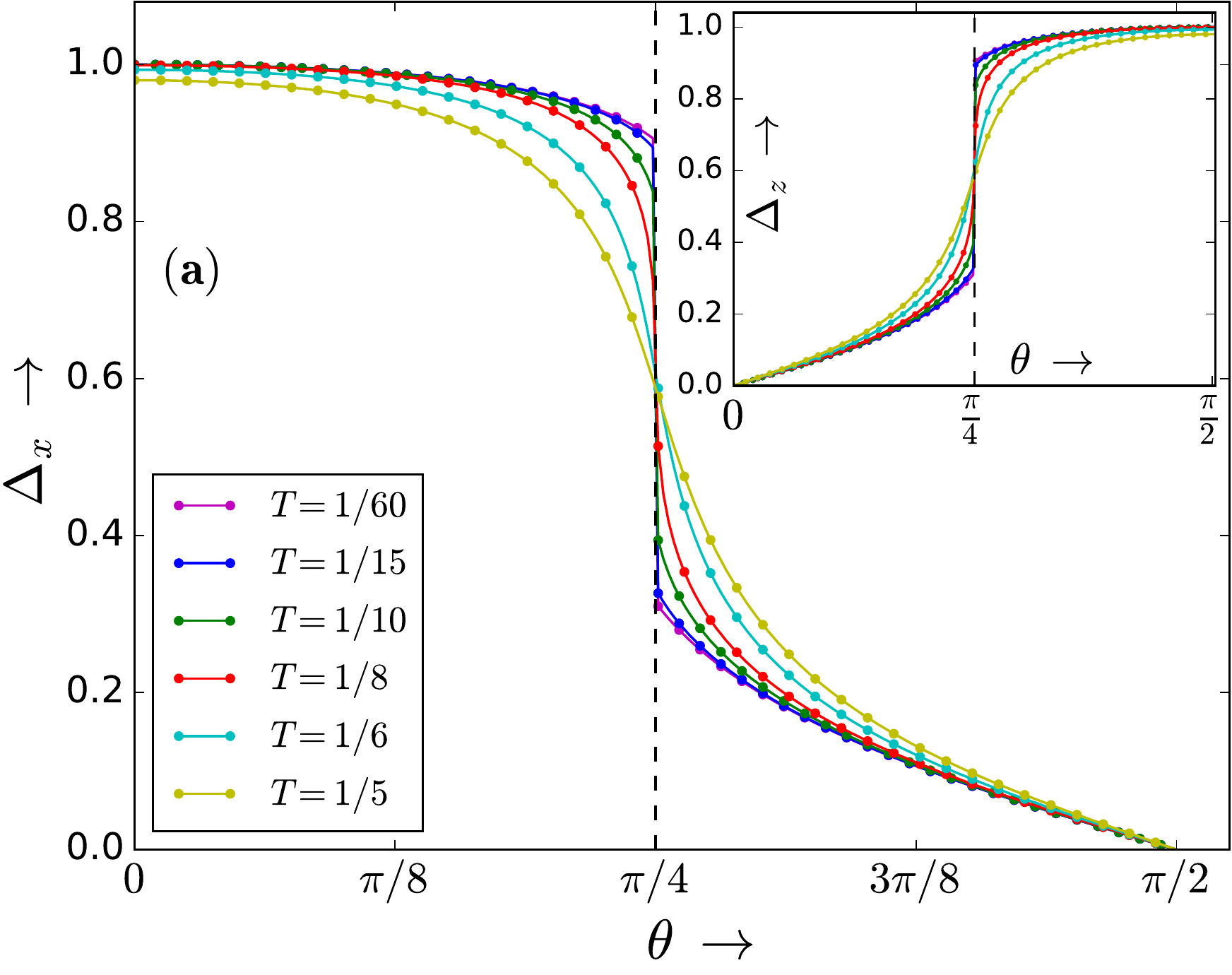}
\includegraphics[height=55mm,width=75mm]{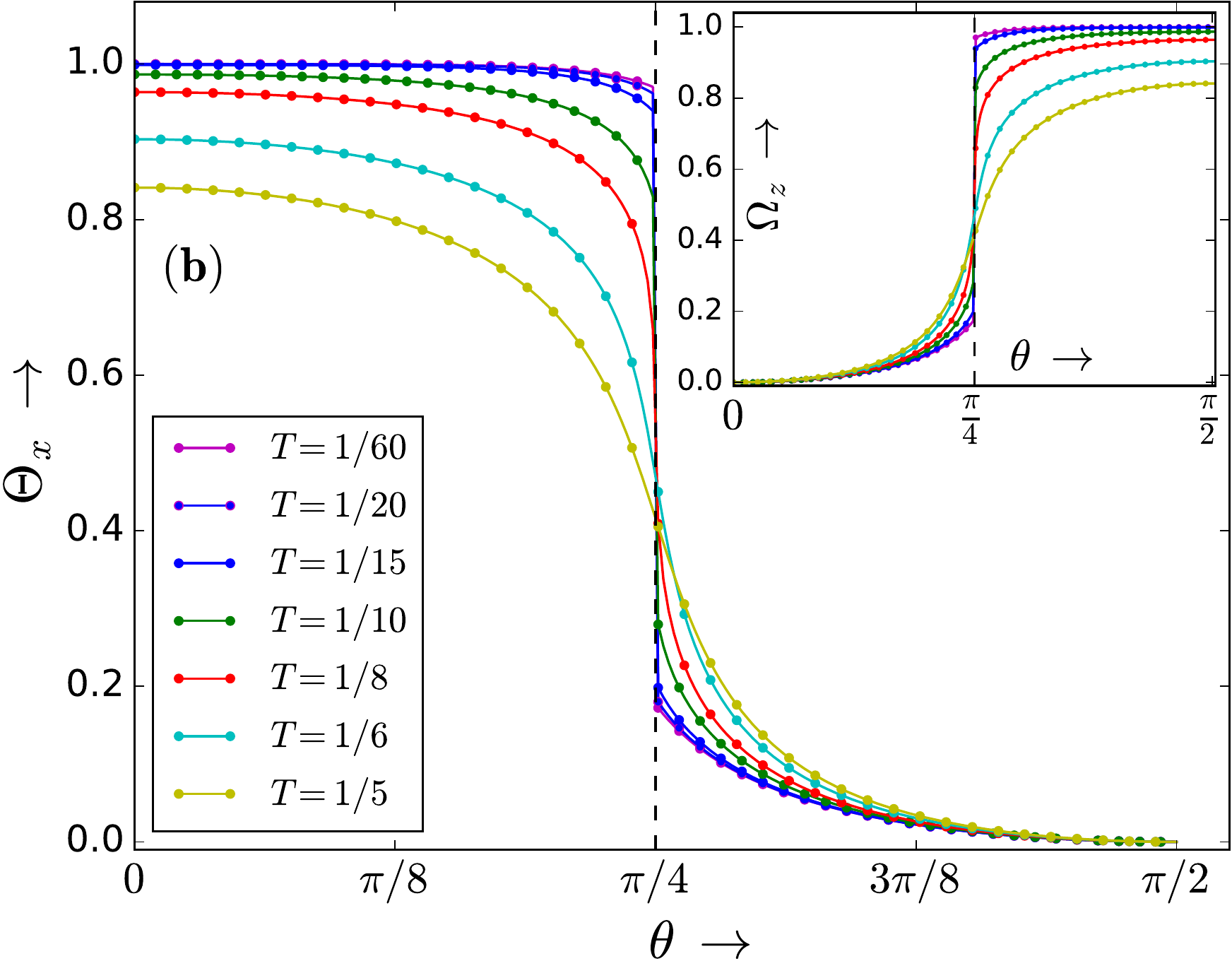}
\caption{Various MF averages for the four-leg QC ladder at finite $T$,~(a) $\Delta_{x}$ and dual $\Delta_{z}$ (shown in the inset), (b) the other dual pair $\Theta_{x}$ and $\Omega_{z}$ (shown in the inset) are plotted as function of $\theta$, for different $T$ values. All these averages continue showing a sudden jump at $\theta=\pi/4$ ($J_{x}=J_{z}$) up to $T=T_{c}\approx 0.125$, where the discontinuity just vanishes.}
\label{Fig.3}
\end{figure}
What about the excitations?~In the $xx$-ordered state, the pseudo-spin fluctuation spectrum, given by $\mathrm{Im}\chi_{(P^{x}P^{x})} (q,\omega)$, is the two Jordan-Wigner (JW) fermion continuum. Thus, a pseudo-spin-flip in the QCM is a two fermion bound state, reminiscent of the des Cloizeaux-Pearson spin wave spectrum of the $S=1/2$ AF Heisenberg chain \cite{Faddeev}. Thus, excitations of the nematic order are (i) the dispersive orbital pseudospin excitations, which are two-fermion composites, and (ii) ``individual spin-flips" which, in our case, corresepond to flipping a $V^{z}_{j}$ on local ``sites", $j$: for coupled ladders, these are precisely ``column flips". These two distinct types of excitations have also been observed by F. Trousselet \emph{et al} \cite{Trousselet}, in finite size compass clusters perturbed by Heisenberg interactions. The $\langle \sigma^{z}_{\vec{r}}\sigma^{z}_{\vec{r}+n}\rangle$ correlations along  the legs ($x$-direction) rigorously vanish, both at zero and finite $T$, and $\langle\sigma^{x}_{\vec{r}}\sigma^{x}_{\vec{r}+n}\rangle$ correlations along $x$, $z$-directions are consistent with a $T=0$, $\mathbb{Z}_{2}$ symmetry ($P_{j}=\prod_{i}\sigma^{z}_{ij}$) broken ordered state \cite{Orus} when $|J_{x}|>|J_{z}|$, as well as with short-range correlation at any finite $T$ (see Appendix \ref{App D}). The nematic ($\Delta_{x,z}\ ,\ \Omega_{z}$) and plaquette ($\Theta_{x}$) correlators nevertheless dominate at finite $T$; this is a consequence of the present MF decoupling, which is carried out by decoupling the ``four-spin interaction" at a ``two-spin" level, in a way that preserves the LDS. By construction, our MFT treats spatial correlations along $x$-direction almost exactly (cf. two coupled $1d$ TFIMs), but approximates correlations along $z$-direction to just three lattice spacings. Thus, by construction, the correlation length along $x$ ($\xi_{x}$) is much larger than that along $z$ ($\xi_{z}$) : this assumption is in qualitative accord with Ref.~\cite{Czarnik-Supplementary}, which indeed finds $\xi_{x}>\xi_{z}$. Thus, even though we cannot treat long-range correlations along $z$ accurately, the emergent LDS of the QCM come to the rescue.\\
\indent At $J_{x}=J_{z}$, $T=T_{c}\approx 0.1225$ (for periodic four-leg ladder), the second order endpoint has a Landau mean-field character (see Appendix \ref{App C}). As expected from any MFT, $T_{c}$ is overestimated, in comparison with the quantum Monte Carlo \cite{Wenzel}, high-$T$ expansion \cite{Oitmaa}, and more recent variational tensor network \cite{Czarnik-Supplementary} approaches. Beyond-MF fluctuations (see below) will certainly depress the MF $T_{c}$, but estimating this reduction and ``restoring" the classical $2d$ Ising critical exponents is beyond the scope of the present work. Finally, about fluctuations beyond MFT, these are mainly of two types: (1) interaction between the JW fermions, and (2) sudden local quenches due to flipping of one or more of $V^{z}_{j}$s. In Appendix \ref{App F}, we detail how beyond-MFT fluctuation effects converge almost everywhere in the parameter space, except possibly at $J_{x}=J_{z},\ T=T_{c}$. Thus we expect our main results to survive beyond-MF fluctuation effects except close to $T_{c}$.\\
\indent About entanglement properties, we follow the approach of O. Hart \emph{et al.} \cite{Hart} for our case and ask about the time dependent entanglement entropy $S(t)$ after starting from a translationally invariant initial state (when all $V^{z}_{j}=\pm 1/2$) and partitioning the full four-leg ladder leg-wise, into two equal length four-leg ladders. Signatures of strong entanglement (at long times) in the system (if any) are generally also reflected in certain long time dynamical spin correlations which could be mapped to Loschmidt-echo-like objects. In addition, these dynamical spin responses are also experimentally measurable. We consider the following correlator as an example, $\langle S^{z}_{j,1}(t)S^{z}_{j+n,1}(0)\rangle$ at long times. 
\begin{align}
\langle S^{z}_{j,1}(t)& S^{z}_{j+n,1}(0)\rangle\\
& \approx\frac{1}{4}\delta_{n,0}\big\langle p^{z}_{j,1}(t)p^{z}_{j,1}(0)\big\rangle\big\langle w^{z}_{j}(t)v^{x}_{j}(t)w^{z}_{j}(0)v^{x}_{j}(0) \big\rangle\nonumber
\end{align}
Here $p^{\alpha}_{j}=2P^{\alpha}_{j}$ (similarly for $w^{\beta}, v^{\gamma}$) are Pauli spins, $(p^{\alpha})^{2}=1$. The trace is performed over all $\lbrace v^{z}_{j}\rbrace$ configurations, so $v^{x}_{j}$ should be applied twice at the same position to return back to same $\lbrace v^{z}_{j}\rbrace$ sector. This makes the above correlator completely local. Now, 
\begin{align*}
\big\langle w^{z}_{j}(t)v^{x}_{j}(t)w^{z}_{j}(0)v^{x}_{j}(0) \big\rangle=\big\langle e^{itH(\lbrace v^{z}_{j}\rbrace)}e^{-itH_{j}^{'}(\lbrace v^{z}_{j}\rbrace; -v^{z}_{j})}\big\rangle
\end{align*}
Here $H_{j}^{'}=w^{z}_{j}H(\lbrace v^{z}_{j}\rbrace; -v^{z}_{j})w^{z}_{j}$, and $H(\lbrace v^{z}_{j}\rbrace; -v^{z}_{j})$ represents the Hamiltonian where one $v^{z}_{j}$ is flipped. we have used the identity $w^{z}_{j}e^{-iHt}w^{z}_{j}=e^{-it(w^{z}_{j}Hw^{z}_{j})}$, to get the above relation. The above expectation value is nothing but Loschmidt-echo-like quantity generally occurs in the ``sudden quench" context. As we discuss in Appendix \ref{App F}, (i) the presence of gap everywhere (below $T_{c}$) and (ii) the first-order transition at $J_{x}=J_{z}$, lead to suppression of infrared singular behaviour in these types of local dynamical correlators (away from $T_{c}$). Thus no power-law behaviour at long times obtains, and thus $S(t)$ will not grow logarithmically.\\
\indent Exactly at $J_{x}=J_{z}$, $T=T_{c}$, we are left with gapless JW-Bogoliubov fermions (see Fig.~\ref{figS4}), with a finite density of states (DOS) at low energies. Now, as detailed in Appendix \ref{App F}, a sudden quench as above leads to a non-perturbative shake-up of the JW Fermi sea via an Anderson orthogonality Catastrophe, leading to complete loss of the pole structure of the JW fermionic propagator in favour of a branch-cut structure. Interestingly, this novel feature is associated with ``strange" metallicity in a completely different context \cite{Anderson}, and leads to a vanishing (like a power law at long times) fidelity. This fully accords with numerics (F. Trousselet \emph{et al.} \cite{Trousselet}), which indeed finds vanishing fidelity at the compass point in a compass-Heisenberg model.\\
\indent There are also interesting connections to the celebrated KHM. KHM and it's variants are exactly soluble via Majorana or JW fermionization, due to gapless (gapped) Majoranas coupled to static $\mathbb{Z}_{2}$ gauge fields \cite{Kitaev, Chen-Nussinov}. Interestingly, the two-leg ladder QCM is exactly soluble, again because it is reduced to free (gapped or gapless) Dirac fermions coupled to a static $\mathbb{Z}_{2}$ fields on bonds \cite{Mattis}. However, it is of course not exactly soluble in $d>1$. Nevertheless, the nematic state with $\langle\mathcal{D}\rangle>0$ is an analytic continuation of the decoupled p-wave Kitaev chains, and pseudo-spin-flip excitations are two-(JW) fermion composites (bound states). Finally, QCM is also related to the (large-$J_{z}$ KHM) toric code model in a transverse field, which exhibits a first-order transition separating a topologically ordered phase from a conventional ordered one. A direct comparison between our results and Fig. 2 of Vidal \emph{et al.} \cite{Vidal} shows excellent accord in ground state energy, the magnetization, $m_{y}$, in their work is directly related to, and in excellent accord with, $\Delta_{z}$ in our work.\\
  
\section{M\lowercase{agnetism in} TMO \lowercase{with} $t_{2g}$ \lowercase{orbital degeneracy}}
\indent Buoyed by very good accord we find with both, iPEPS and PCUT results, as well as by comparison with an eight-leg ladder MFT (see Appendix~\ref{App E}), we {\it assume} that our MFT is a good approximation to the planar QCM. Consider a two dimensional TMO with active degenerate orbital degrees of freedom per lattice site. In cases, where the $MO_{6}$ octahedron is squashed, the $xy$ orbital is ``pushed above" the twofold degenerate $xz,\ yz$ orbitals, whence orbital degeneracy is relevant for the $d^{1}$ and $d^{4}$ configurations of the TM ion. When the local Hubbard interaction is large compared to the $d$ electron kinetic energy, the Mott insulator is effectively described by Kugel-Khomskii Hamiltonian \cite{Nevidomskyy, C.Chen},  
\begin{align}
H_{KK}=J_{1}&\sum_{i,\alpha=x,z}\bigg(\mathbf{S}_{i}\cdot\mathbf{S}_{i+\hat{\alpha}}+\frac{1}{4}\bigg)\tau^{\alpha}_{i}\tau^{\alpha}_{i+\hat{\alpha}}\nonumber\\
&+J_{2}\sum_{\langle\langle i,j\rangle\rangle}\mathbf{S}_{i}\cdot\mathbf{S}_{j}-J_{f}\sum_{\langle i,j\rangle}\mathbf{S}_{i}\cdot\mathbf{S}_{j} \label{eq11}
\end{align}
This model can also be motivated as a large-$U$ limit of a {\it spinful}, two-orbital Hubbard model with active $xz$, $yz$ degenerate orbitals per site. The spinless version at large-$U$ maps to the QCM \cite{Nasu}. Similarly, one can show that the spinful version maps onto the first term of $H_{KK}$.
For a $d^{1}$ TM ion $S=1/2$, while for the $d^{4}$ TM ion, $S=1$ (we assume that the crystal field splitting between $t_{2g}$ and $e_{g}$ states is larger than the Hund coupling, so can neglect $e_{g}$ states). Here $J_{2}$ is the diagonal anti-ferromagnetic
(AF) interaction while $J_{f}$ is the direct ferromagnetic (FM) term \cite{Rajiv R. Singh}.\\
\indent Suppose orbital order occurs before magnetic order ($T_{o}>T_{N}$). Starting from high temperature, the orbital ordering (OO) is captured by the ``orbital only" part of  Eq.~(\ref{eq11}); $H_{oo}=\frac{J_{1}}{4}\sum_{i}\big(\tau^{x}_{i}\tau^{x}_{i+\hat{x}}+\lambda \tau^{z}_{i}\tau^{z}_{i+\hat{z}}\big)$, where $\lambda$ can differ from unity, either via coupling to the spin fluctuations in Eq.~(\ref{eq11}) or a coupling to Jahn-Teller modes, or both \cite{Kugel-Khomskii}. The effective Heisenberg super-exchange now explicitly depends on the orbital correlations and the effective spin couplings read $J_{1x}=J_{1}\langle \tau^{x}_{i}\tau^{x}_{i+\hat{x}}\rangle-J_{f}$, $J_{1z}=J_{1}\langle \tau^{z}_{i}\tau^{z}_{i+\hat{z}}\rangle-J_{f}$. The resulting spin model,
\begin{align}
H_{s}=\sum_{i}\sum_{\alpha=x,z}J_{1\alpha}\mathbf{S}_{i}\cdot\mathbf{S}_{i+\hat{\alpha}}+J_{2}\sum_{\langle\langle i,j\rangle\rangle} \mathbf{S}_{i}\cdot\mathbf{S}_{j}
\end{align}
is {\it precisely} the $J_{1x}$-$J_{1z}$-$J_{2}$ Heisenberg model successfully used in the Fe-arsenide \cite{Applegate} context, but should obviously be more generally valid. Furthermore, the sign of the coupling strengths $J_{1\alpha}$ could be AF or FM type depending on the values of orbital correlations. If we deal with $S>1/2$, spin excitations below $T_{N}$ are qualitatively described by renormalized spin wave theory (RSWT) \cite{Takahashi} and are dressed propagating magnons of a stripe magnetic order with $\mathbf{q}=(\pi,0)$.\\
\indent In general, when magnetic and orbital ordering occur close together, it is not possible to decouple the spin-orbital coupling. Spin-orbital entanglement plays an important role in these cases. However, if we restrict ourselves to a subset of cases where $T_{oo}$ and $T_{N}$ are well separated, and orbital order happens at higher $T$, some interesting conclusions can be drawn without recourse to a full calculation.\\
\indent At high $T$, without any order, the orbital excitations are fractionalized orbitons, i.e. they are two JW fermion continuum, while the spin excitations are overdamped bosons. At $T_{oo}$, the JW fermions reconfine into boson like orbitons, while the spin excitations are still over damped bosons. Thus the part of the entropy is lost at $T_{oo}$. At $T_{N}$, stripe magnetic order sets in, and we are left with bosons like orbitons and usual magnons.\\
\indent Thus, at least in this subset of cases, the phenomenon of orbital-spin separation \cite{Schlappa} as well as reconfinement reveals itself. It depends crucially on (i) emergent $d=1$ symmetries \cite{Nussinov2009} in the QCM and (ii) well-separated orbital and magnetic ordering scales.
\section{Conclusion}
\indent We have constructed a novel MFT for the planar QCM and implemented it for the four-leg case. A distinguishing feature of our approach is that the MF decoupling is done at the level of two-spin averages, in contrast to usual MFT. For a four-leg ladder, our MFT preserves the LDS crucial to proper analysis of the QCM. In very good accord with exact arguments and numerics for the full $2d$ QCM, we find that a first-order transition separates two, mutually dual Ising nematic phases at $J_{x}=J_{z}$ . Our results reveal an exciting (pseudo-) spin fractionalization that may survive in the full planer QCM, and point a way to realization of (i) strongly coupled orbital and magnetic orders, and (ii) orbital-spin-charge separation in $2d$ multi-orbital Mott insulators. Considering fluctuations beyond ladder-MFT, as well as extension to $d=2$, are complex avenues for future work and subject of future study.\\
\section{Acknowledgements}
\indent We thank G. Baskaran for suggesting that we make a brief comparative remarks on the Kitaev model. We thank the Institute of Mathematical Sciences for financial support.
\\

\appendix

\section{Derivation of mean-field self-consistency equations and ground state energy}\label{App A}
Exact solution of $1d$ TFIM using Jordan-Wigner (JW) fermionization is well known (See \cite{Subir Sachdev-Supplementary} and references therein), we provide here only few steps to derive the mean-field self-consistency conditions. The mapping between spins ($S=\sfrac{1}{2}$) and spin-less JW fermions are following,
\begin{gather}
P^{+}_{jl}=a^{\dag}_{jl}\prod_{m<j}\big(2a^{\dag}_{ml}a_{ml}-1\big)\ ,\ \ (l=1,2)\nonumber\\
W^{+}_{j}=b^{\dag}_{j}\prod_{m<j}\big(2b^{\dag}_{m}b_{m}-1\big) \label{A.1}
\end{gather}
Using this, Eqs.~\eqref{eq8a} and \eqref{eq8b} become
\begin{align}
H=\frac{t_{\alpha}}{2}\sum_{j=1}^{N}(\psi_{j\alpha}-\psi^{\dag}_{j\alpha})&(\psi_{j+1,\alpha}+\psi_{j+1,\alpha}^{\dag})\nonumber\\
&+\frac{\mu_{\alpha}}{2}\sum_{j=1}^{N}\big(\psi^{\dag}_{j\alpha}\psi_{j\alpha}-\sfrac{1}{2}\big) 
\end{align}
Here $\psi_\alpha=(a,\ b)$ represent two JW fermionic modes, $t_{a}=\frac{J_{x}}{2}(1+\Theta_{x}),\ \mu_{a}=J_{z}(1+\Omega_{z})$, $t_{b}=J_{x}\Delta_{x},\ \mu_{b}=2J_{z}\Delta_{z}$ are hopping, pairing, and chemical potentials of the JW fermions. To obtain the bulk excitation spectrum, we simply go to momentum space (using translational invariance), and perform Bogoliubov transformation to diagonalize the momentum space ($-\pi\leq k<\pi$) Hamiltonian. The transformation is given by 
\begin{gather}
\psi_{k,\alpha}= u_{k}^{\alpha}\gamma_{k,\alpha}-iv_{k}^{\alpha}\gamma^{\dag}_{-k,\alpha}\nonumber\\
u_{k}^{\alpha}=\frac{1}{\sqrt{2}}\bigg(1+\frac{\xi^{\alpha}_{k}}{E^{\alpha}_{k}}\bigg)^{1/2},v_{k}^{\alpha}=\frac{\Delta_{k}^{\alpha}}{\sqrt{2}|\Delta_{k}^{\alpha}|}\bigg(1-\frac{\xi^{\alpha}_{k}}{E^{\alpha}_{k}}\bigg)^{1/2}\label{A.3}
\end{gather}
with $\xi^{\alpha}_{k}=-t_{\alpha}\cos{k}+(\mu_{\alpha}/2)$, $\Delta_{k}^{\alpha}=t_{\alpha}\sin{k}$, and  $E_{k}^{\alpha}=\sqrt{(\xi^{\alpha}_{k})^{2}+(\Delta^{\alpha}_{k})^{2}}$ .\\
\indent The diagonalized Hamiltonian is following, 
\begin{align}
H=\sum_{k=-\pi}^{\pi}E_{k}^{\alpha}\big(\gamma^{\dag}_{k,\alpha}\gamma_{k,\alpha}-\sfrac{1}{2}\big)
\end{align}
Next we compute various expectation values using the above exact solution. We have defined $\Delta_{x}=4\langle P^{x}_{jl}P^{x}_{j+1,l}\rangle\equiv 4\langle P^{x}_{j}P^{x}_{j+1}\rangle=\langle (a_{j}-a_{j}^{\dag})(a_{j+1}+a_{j+1}^{\dag})\rangle$. Similarly, we have $\Theta_{x}=4\langle Q^{z}_{jl}Q^{z}_{j+1,l}\rangle$. Applying Mattis's duality, $4\langle Q^{z}_{j,1}Q^{z}_{j+1,1}\rangle=4\langle W^{x}_{j}W^{x}_{j+1}\rangle$, $\langle Q^{z}_{j,1}Q^{z}_{j+1,1}\rangle=16\langle V^{z}_{j}V^{z}_{j+1}W^{x}_{j}W^{x}_{j+1}\rangle=4\langle W^{x}_{j}W^{x}_{j+1}\rangle$, so $\Theta_{x}=4\langle W^{x}_{j}W^{x}_{j+1}\rangle=\langle (b_{j}-b_{j}^{\dag})(b_{j+1}+b_{j+1}^{\dag})\rangle$ . So,
\begin{align}
\langle \psi_{j\alpha}\psi_{j+1,\alpha}\rangle+&h.c.=-\frac{1}{N}\sum_{k}(2u_{k}^{\alpha}v_{k}^{\alpha})\sin{k}\tanh{\bigg(\frac{\beta E_{k}^{\alpha}}{2}\bigg)}\nonumber\\
&=-\frac{1}{N}\sum_{k}\frac{t_{\alpha}}{E^{\alpha}_{k}}\sin^{2}{k}\tanh{\bigg(\frac{\beta E_{k}^{\alpha}}{2}\bigg)} \label{A.5}
\end{align}
\begin{align}
&\langle \psi^{\dag}_{j\alpha}\psi_{j+1,\alpha}\rangle+h.c\nonumber\\
&=\frac{1}{N}\sum_{k}(2\cos{k})[(u^{\alpha}_{k})^{2}n(E_{k}^{\alpha})+(v^{\alpha}_{k})^{2}(1-n(E^{\alpha}_{k}))]\nonumber\\
&=\frac{1}{N}\sum_{k}\cos{k}\bigg[1-\frac{\xi^{\alpha}_{k}}{E_{k}^{\alpha}}\bigg]\tanh{\bigg(\frac{\beta E_{k}^{\alpha}}{2}\bigg)} \label{A.6}
\end{align}
Subtracting \eqref{A.6} from \eqref{A.5}, we arrive at Eq.~\eqref{eq9}. In a similar way, Eq.~\eqref{eq10} could be found from (1): $\Delta_{z}=2\langle P^{z}_{j}\rangle=2\langle a^{\dag}_{j}a_{j}\rangle-1$ and (2): $\Omega_{z}=4\langle Q^{x}_{j,1}Q^{x}_{j,2}\rangle= 2\langle W^{z}_{j}\rangle=2\langle b^{\dag}_{j}b_{j}\rangle-1$ .\\
\indent The mean field ground state (GS) energy of the four-leg compass ladder is following,
\begin{align}
&E_{gs}=-4J_{x}\sum_{j=1}^{N}\sum_{l=1}^{2}\langle P^{x}_{jl}P^{x}_{j+1,l}\rangle\langle Q^{z}_{jl}Q^{z}_{j+1,l}\rangle\nonumber\\
&-2J_{z}\sum_{j=1}^{N}[\langle P^{z}_{j,1}\rangle+\langle P^{z}_{j,2}\rangle]\langle Q^{x}_{j,1}Q^{x}_{j,2}\rangle-\frac{1}{2}\sum_{k=-\pi}^{\pi}\big(2E^{a}_{k}+E^{b}_{k}\big) \label{A.7}
\end{align}
The first two position space summations come from the mean field decoupling of \eqref{eq4a} and \eqref{eq4b}. The last one involving momentum space sum, is the condensation energies of $p-$wave superconductor chains, there are two identical fermionic chains of ``$a$" (for $l=1,2$) and one of ``$b$" JW fermions. Simplifying \eqref{A.7}, the GS energy per lattice site ($e_{gs}$) could be written as 
\begin{align}
e_{gs}&=E_{gs}/4N=-\frac{1}{8}(J_{x}\Delta_{x}\Theta_{x}+J_{z}\Delta_{z}\Omega_{z})\nonumber\\
&\ \ -\int_{0}^{\pi}\frac{dk}{8\pi}\bigg[\ |J_{x}|(1+\Theta_{x})\sqrt{1+h_{a}^{2}-2h_{a}\cos{k}}\nonumber\\
&\ \ \ \ \ \ \ \ +|J_{x}||\Delta_{x}|\sqrt{1+h_{b}^{2}-2h_{b}\cos{k}}\ \bigg] \label{A.8}
\end{align} 
We have chosen $J_{x}=J\cos{\theta},\ J_{z}=J\sin{\theta},$ and $J=1$ for doing calculations.
\section{Exact derivation of the duality between $\Theta_{x}$ and $\Omega_{z}$}\label{App B}
First we establish the exact duality between $\Theta_{x}$ and $\Omega_{z}$ (in addition to the duality between $\Delta_{x}\ , \ \Delta_{z} $) for four-leg compass ladder in the restricted subspace where all local $\mathbb{Z}_{2}$ symmetry operators ($V^{z}_{j}$) are frozen to $\pm\sfrac{1}{2}$. To do that, we start from the Hamiltonian,
\begin{align}
&H_{QC}=J_{x}\sum_{j=1}^{N}\sum_{l=1}^{2}\big(S^{x}_{jl}S^{x}_{j+1,l}+T^{x}_{jl}T^{x}_{j+1,l}\big)\nonumber\\
&\ \ \ \ \ \ \ +J_{z}\sum_{j=1}^{N}\sum_{l=1}^{2}S^{z}_{jl}T^{z}_{jl}+J_{z}\sum_{j=1}^{N}\sum_{l=1}^{2}S^{z}_{jl}T^{z}_{j,l+1} \label{B.1}
\end{align}
We successively apply two Mattis's transformations, first apply Eq.~\eqref{eq3} and then
\begin{gather}
(Q^{x}_{j,1}\ ,\ Q^{y}_{j,1}\ ,\ Q^{z}_{j,1})=(-2W^{z}_{j}V^{x}_{j}\ ,\ 2W^{y}_{j}V^{x}_{j}\ ,\ W^{x}_{j})\nonumber\\ 
(Q^{x}_{j,2}\ ,\ Q^{y}_{j,2}\ ,\ Q^{z}_{j,2})=(-V^{x}_{j}\ ,\ 2V^{y}_{j}W^{x}_{j}\ ,\ -2V^{z}_{j}W^{x}_{j}) \label{B.2}
\end{gather} 
As explained before, both Eqs.~\eqref{eq3} and \eqref{B.2} are just two-site Kramers-Wannier (KW) dualities with additional rotations of spin basis (the two sites are positioned along rungs of the ladder). Then, Eq.~\eqref{B.1} reads
\begin{align}
H_{QC}=J_{x}\sum_{j=1}^{N}&\bigg[P^{x}_{j,1}P^{x}_{j+1,1}(1+4W^{x}_{j}W^{x}_{j+1})\nonumber\\
&+P^{x}_{j,2}P^{x}_{j+1,2}(1+4(4V^{z}_{j}V^{z}_{j+1})W^{x}_{j}W^{x}_{j+1})\bigg]\nonumber\\
&+\frac{J_{z}}{2}\sum_{j=1}^{N}\sum_{l=1,2}(1+2W^{z}_{j})P^{z}_{jl} 
\end{align}
Here $V^{z}_{j}$s are conserved operators. We define a projector ($\mathcal{P}$) onto the subspace where all $V^{z}_{j}=\pm\sfrac{1}{2}$ , i.e. $\mathcal{P}=\sum_{k}\ket{\phi_{k},\lbrace V^{z}_{j}=\pm\sfrac{1}{2}\rbrace}\bra{\phi_{k},\lbrace V^{z}_{j}=\pm\sfrac{1}{2}\rbrace}$ . The projected Hamiltonian looks following,
\begin{align}
H^{P}_{QC}=\mathcal{P}H_{QC}\mathcal{P}\nonumber=&J_{x}\sum_{j=1}^{N}\sum_{l=1,2}P^{x}_{jl}P^{x}_{j+1,l}(1+4W^{x}_{j}W^{x}_{j+1})\nonumber\\
&\ \ \ \ \ \ \ \ +\frac{J_{z}}{2}\sum_{j=1}^{N}\sum_{l=1,2}(1+2W^{z}_{j})P^{z}_{jl} \label{B.4}
\end{align}
We have the following correspondences between the averages of transformed \eqref{B.4} and original \eqref{B.1} model, it could be easily shown by inverting Eqs.~\eqref{eq3} and \eqref{B.2}. We see, $\Delta_{x}=4\langle P^{x}_{jl}P^{x}_{j+1,l}\rangle\equiv 4\langle S^{x}_{jl}S^{x}_{j+1,l}\rangle\sim \langle \sigma^{x}_{\vec{r}}\sigma^{x}_{\vec{r}+\hat{x}}\rangle$, $\Delta_{z}=2\langle P^{z}_{jl}\rangle\equiv 4\langle S^{z}_{jl}T^{z}_{jl}\rangle\sim \langle \sigma^{z}_{\vec{r}}\sigma^{z}_{\vec{r}+\hat{z}}\rangle$ , $\Theta_{x}=4\langle W^{x}_{j}W^{x}_{j+1}\rangle\equiv 16\langle S^{x}_{jl}T^{x}_{jl}S^{x}_{j+1,l}T^{x}_{j+1,l}\rangle\sim\langle\sigma^{x}_{\vec{r}}\sigma^{x}_{\vec{r}+\hat{z}}\sigma^{x}_{\vec{r}+\hat{x}}\sigma^{x}_{\vec{r}+\hat{x}+\hat{z}}\rangle$, and $\Omega_{z}=2\langle W^{z}_{j}\rangle\equiv 4\langle T^{z}_{j,1}T^{z}_{j,2}\rangle\sim\langle\sigma^{z}_{\vec{r}}\sigma^{z}_{\vec{r}+2\hat{z}}\rangle$ .\\
\indent We consider another four-leg ladder compass model, where spins ($\tilde{S}^{\mu},\ \tilde{T}^{\nu}$) reside on the ``dual'' lattice sites ($\tilde{l}=l,\ \tilde{j}=j+\sfrac{1}{2}$) with coupling strengths being swapped with the original model:  
\begin{align}
&H_{QC}^{dual}=J_{z}\sum_{\tilde{j}=1}^{N}\sum_{\tilde{l}=1}^{2}\big(\tilde{S}^{x}_{\tilde{j}\tilde{l}}\tilde{S}^{x}_{\tilde{j}+1,\tilde{l}}+\tilde{T}^{x}_{\tilde{j}\tilde{l}}\tilde{T}^{x}_{\tilde{j}+1,\tilde{l}}\big)\nonumber\\
&\ \ \ \ \ \ \ +J_{x}\sum_{\tilde{j}=1}^{N}\sum_{\tilde{l}=1}^{2}\tilde{S}^{z}_{\tilde{j}\tilde{l}}\tilde{T}^{z}_{\tilde{j}\tilde{l}}+J_{x}\sum_{\tilde{j}=1}^{N}\sum_{\tilde{l}=1}^{2}\tilde{S}^{z}_{\tilde{j}\tilde{l}}\tilde{T}^{z}_{\tilde{j},\tilde{l}+1} \label{B.5}
\end{align}
The dual spins and co-ordinates are denoted by tilde symbols. Applying transformations like Eqs~\eqref{eq3}, \eqref{B.2} on these `dual' spins, we get the following,
\begin{align}
H^{dual}_{QC}=J_{z}\sum_{\tilde{j}=1}^{N}&\bigg[\tilde{P}^{x}_{\tilde{j},1}\tilde{P}^{x}_{\tilde{j}+1,1}(1+4\tilde{W}^{x}_{\tilde{j}}\tilde{W}^{x}_{\tilde{j}+1})\nonumber\\
&+\tilde{P}^{x}_{\tilde{j},2}\tilde{P}^{x}_{\tilde{j}+1,2}(1+4(4\tilde{V}^{z}_{\tilde{j}}\tilde{V}^{z}_{\tilde{j}+1})\tilde{W}^{x}_{\tilde{j}}\tilde{W}^{x}_{\tilde{j}+1})\bigg]\nonumber\\
&+\frac{J_{x}}{2}\sum_{\tilde{j}=1}^{N}\sum_{\tilde{l}=1,2}(1+2\tilde{W}^{z}_{\tilde{j}})\tilde{P}^{z}_{\tilde{j}\tilde{l}} 
\end{align}
We project onto the subspace where all $\tilde{V}^{z}_{j}=\pm\sfrac{1}{2}$, so the projected Hamiltonian, $H^{P,dual}_{QC}=\tilde{\mathcal{P}}H^{dual}_{QC}\tilde{\mathcal{P}}$ is following,
\begin{align}
H^{P,dual}_{QC}&=J_{z}\sum_{\tilde{j}=1}^{N}\sum_{\tilde{l}=1}^{2}\tilde{P}^{x}_{\tilde{j}\tilde{l}}\tilde{P}^{x}_{\tilde{j}+1,\tilde{l}}(1+4\tilde{W}^{x}_{\tilde{j}}\tilde{W}^{x}_{\tilde{j}+1})\nonumber\\
&\ \ \ \ \ +\frac{J_{x}}{2}\sum_{\tilde{j}=1}^{N}\sum_{\tilde{l}=1}^{2}(1+2\tilde{W}^{z}_{\tilde{j}})\tilde{P}^{z}_{\tilde{j}\tilde{l}} \label{B.7}
\end{align}
Here we have $\Delta^{dual}_{x}=4\langle \tilde{P}^{x}_{\tilde{j}\tilde{l}}\tilde{P}^{x}_{\tilde{j}+1,\tilde{l}}\rangle\equiv 4\langle \tilde{S}^{x}_{\tilde{j}\tilde{l}}\tilde{S}^{x}_{\tilde{j}+1,\tilde{l}}\rangle\sim \langle \tilde{\sigma}^{x}_{\vec{r}^{*}}\tilde{\sigma}^{x}_{\vec{r}^{*}+\hat{x}}\rangle$, $\Delta^{dual}_{z}=2\langle \tilde{P}^{z}_{\tilde{j}\tilde{l}}\rangle\equiv 4\langle \tilde{S}^{z}_{\tilde{j}\tilde{l}}\tilde{T}^{z}_{\tilde{j}\tilde{l}}\rangle\sim \langle \tilde{\sigma}^{z}_{\vec{r}^{*}}\tilde{\sigma}^{z}_{\vec{r}^{*}+\hat{z}}\rangle$ , $\Omega^{dual}_{z}=2\langle \tilde{W}^{z}_{\tilde{j}}\rangle\equiv 4\langle \tilde{T}^{z}_{\tilde{j},1}\tilde{T}^{z}_{\tilde{j},2}\rangle\sim\langle\tilde{\sigma}^{z}_{\vec{r}^{*}}\tilde{\sigma}^{z}_{\vec{r}^{*}+2\hat{z}}\rangle$, and $\Theta^{dual}_{x}=4\langle \tilde{W}^{x}_{\tilde{j}}\tilde{W}^{x}_{\tilde{j}+1}\rangle\equiv 16\langle \tilde{S}^{x}_{\tilde{j}\tilde{l}}\tilde{T}^{x}_{\tilde{j}\tilde{l}}\tilde{S}^{x}_{\tilde{j}+1,\tilde{l}}\tilde{T}^{x}_{\tilde{j}+1,\tilde{l}}\rangle$ $\sim\langle\tilde{\sigma}^{x}_{\vec{r}^{*}}\tilde{\sigma}^{x}_{\vec{r}^{*}+\hat{z}}\tilde{\sigma}^{x}_{\vec{r}^{*}+\hat{x}}\tilde{\sigma}^{x}_{\vec{r}^{*}+\hat{x}+\hat{z}}\rangle$.\\
\indent We see that Eqs.~\eqref{B.4} and \eqref{B.7} could be connected by an ``infinite chain type" KW duality (like the one used in the context of $1d$ TFIM),
\begin{align}
&W^{z}_{j}\rightarrow 2\tilde{W}^{x}_{\tilde{j}}\tilde{W}^{x}_{\tilde{j}+1}\ \ \ ,\ \ \ W^{x}_{j}W^{x}_{j+1}\rightarrow \frac{1}{2}\tilde{W}^{z}_{\tilde{j}}\nonumber\\
&P^{z}_{jl}\rightarrow 2\tilde{P}^{x}_{\tilde{j}\tilde{l}}\tilde{P}^{x}_{\tilde{j}+1,\tilde{l}}\ \ \ \ ,\ \ \ P^{x}_{jl}P^{x}_{j+1,l}\rightarrow \frac{1}{2}\tilde{P}^{z}_{\tilde{j}\tilde{l}} \label{B.8}
\end{align}
Using Eq.~\eqref{B.8}, we see $\Delta_{x}$ maps to $\Delta^{dual}_{z}$. Similarly, we have $\Delta_{z}\rightarrow \Delta_{x}^{dual}$, $\Theta_{x}\rightarrow\Omega_{z}^{dual}$, $\Omega_{z}\rightarrow\Theta_{x}^{dual}$ . So we have proved the duality in a projected subspace of all $V^{z}_{j}=\pm\sfrac{1}{2}$ (and $\tilde{V}_{\tilde{j}}^{z}=\pm\sfrac{1}{2}$).\\

\indent We now extend the above proof to arbitrary subspaces (of $V^{z}_{j},\ \tilde{V}^{z}_{\tilde{j}}$) for planar compass model by taking a different route. We start from the Hamiltonian \eqref{B.1} (change the upper limit of $l$ to $N/2$) and apply Eq.~\eqref{eq3}, we get
\begin{align}
H_{QC}=&\frac{J_{x}}{4}\sum_{j=1}^{N}\sum_{l=1}^{N/2}p^{x}_{jl}p^{x}_{j+1,l}(1+q^{z}_{jl}q^{z}_{j+1,l})\nonumber\\
&\ \ \ +\frac{J_{z}}{4}\sum_{j=1}^{N}\sum_{l=1}^{N/2}p^{z}_{jl}(1+q^{x}_{jl}q^{x}_{j,l+1}) \label{B.9}
\end{align}
Here $p^{\mu}_{jl}=2P^{\mu}_{jl},\ q^{\nu}_{jl}=2Q^{\nu}_{jl}$. The self-duality is already visible from this expression as we have $1d$ TFIM of $p^{\mu}$ coupled to a compass model of $q^{\nu}$, both of them show self-dual property. To be rigorous mathematically, we prove it by applying the following set of duality relations successively on \eqref{B.9}, (we define $\tilde{l}=l,\ \tilde{j}=j+\sfrac{1}{2}$ )
\\

(1): We use the ``infinite-chain-like" KW duality for $p^{\mu}$ and $q^{\nu}$ spins,
\begin{align}
p^{x}_{jl}p^{x}_{j+1,l}\ \rightarrow\ \tilde{p}^{z}_{\tilde{j}\tilde{l}}\ \ ,\ \ p^{z}_{jl}\ \rightarrow\ \tilde{p}^{x}_{\tilde{j}\tilde{l}}\tilde{p}^{x}_{\tilde{j}+1,\tilde{l}}\nonumber\\
q^{z}_{jl}q^{z}_{j+1,l}\ \rightarrow\ \tau^{x}_{\tilde{j}\tilde{l}}\ \ ,\ \ q^{x}_{jl}\ \rightarrow\ \tau^{z}_{\tilde{j}\tilde{l}}\tau^{z}_{\tilde{j}+1,\tilde{l}} \tag{B10.a}\label{B.10a}
\end{align}
(2): Next we implement Xu-Moore duality relation \cite{Xu-Moore-Supplementary} for $\tau^{\nu}$ spins,
\begin{align}
\tilde{\tau}^{x}_{\tilde{j},\tilde{l}}=\tau^{z}_{\tilde{j}\tilde{l}}\tau^{z}_{\tilde{j}+1,\tilde{l}}\tau^{z}_{\tilde{j},\tilde{l}+1}\tau^{z}_{\tilde{j}+1,\tilde{l}+1}\ \ ,\ \ \tilde{\tau}^{z}_{\tilde{j},\tilde{l}}=\prod_{\vec{r}<(\tilde{j},\tilde{l})}\tau^{x}_{\vec{r}}\tag{B10.b}\label{B.10b}
\end{align}
Here $\vec{r}<(\tilde{j},\tilde{l})$ means the $x$ and $z$ co-ordinates of $\vec{r}$ is less than $\tilde{j}$ and $\tilde{l}$ respectively. 
\newline
(3): Finally, we use again ``infinite-chain-like" KW duality, for $\tilde{\tau}^{\mu}$ spins,
\begin{align}
\tilde{\tau}^{z}_{\tilde{j}\tilde{l}}\tilde{\tau}^{z}_{\tilde{j}+1,\tilde{l}}\ \rightarrow\ \tilde{q}^{x}_{\tilde{j}\tilde{l}}\ \ ,\ \ \tilde{\tau}^{x}_{\tilde{j}\tilde{l}}\rightarrow \tilde{q}^{z}_{\tilde{j}\tilde{l}}\tilde{q}^{z}_{\tilde{j}+1,\tilde{l}}\tag{B10.c}\label{B.10c}
\end{align}
Then Eq.~\eqref{B.9} reads following:
\begin{align}
H^{dual}_{QC}=&J_{z}\sum_{\tilde{j}=1}^{N}\sum_{\tilde{l}=1}^{N/2}\tilde{P}^{x}_{\tilde{j}\tilde{l}}\tilde{P}^{x}_{\tilde{j}+1,\tilde{l}}(1+4\tilde{Q}^{z}_{\tilde{j}\tilde{l}}\tilde{Q}^{z}_{\tilde{j}+1,\tilde{l}})\nonumber\\
&\ \ \ +\frac{J_{x}}{2}\sum_{\tilde{j}=1}^{N}\sum_{\tilde{l}=1}^{N/2}\tilde{P}^{z}_{\tilde{j}\tilde{l}}(1+4\tilde{Q}^{x}_{\tilde{j}\tilde{l}}\tilde{Q}^{x}_{\tilde{j},\tilde{l}+1}) \tag{B11} \label{B.11}
\end{align}
Here $\tilde{P}^{\mu}_{jl}=\tilde{p}^{\mu}_{jl}/2,\ \tilde{Q}^{\nu}_{jl}=\tilde{q}^{\nu}_{jl}/2$ . Using the inverse of Eq.~\eqref{eq3} for $\tilde{P}^{\mu},\ \tilde{Q}^{\nu}$, the Hamiltonian \eqref{B.11} reduces to the `dual' compass model (see \eqref{B.5}). We see $\Theta_{x}\ (\sim \langle q^{z}_{jl}q^{z}_{j+1,l}\rangle)$ maps to $\Omega_{z}^{dual}\ (\sim \langle \tilde{q}^{x}_{\tilde{j}\tilde{l}}\tilde{q}^{x}_{\tilde{j},\tilde{l}+1}\rangle)$ under the web of dualities \eqref{B.10a}-\eqref{B.10c}, similar mapping also holds true between $\Omega_{z}$ and $\Theta_{x}^{dual}$. So the duality between $\Theta_{x}$ and $\Omega_{z}$ holds generally true for compass like interactions, not just restricted to a particular subspace or four-leg ladder geometries.

\section{MFT results for finite temperature criticality (four-leg ladder case)}\label{App C}
\begin{figure}[h!]
\centering
\subfigure{\label{Fig.S1a}\includegraphics[height=50mm, width=70mm]{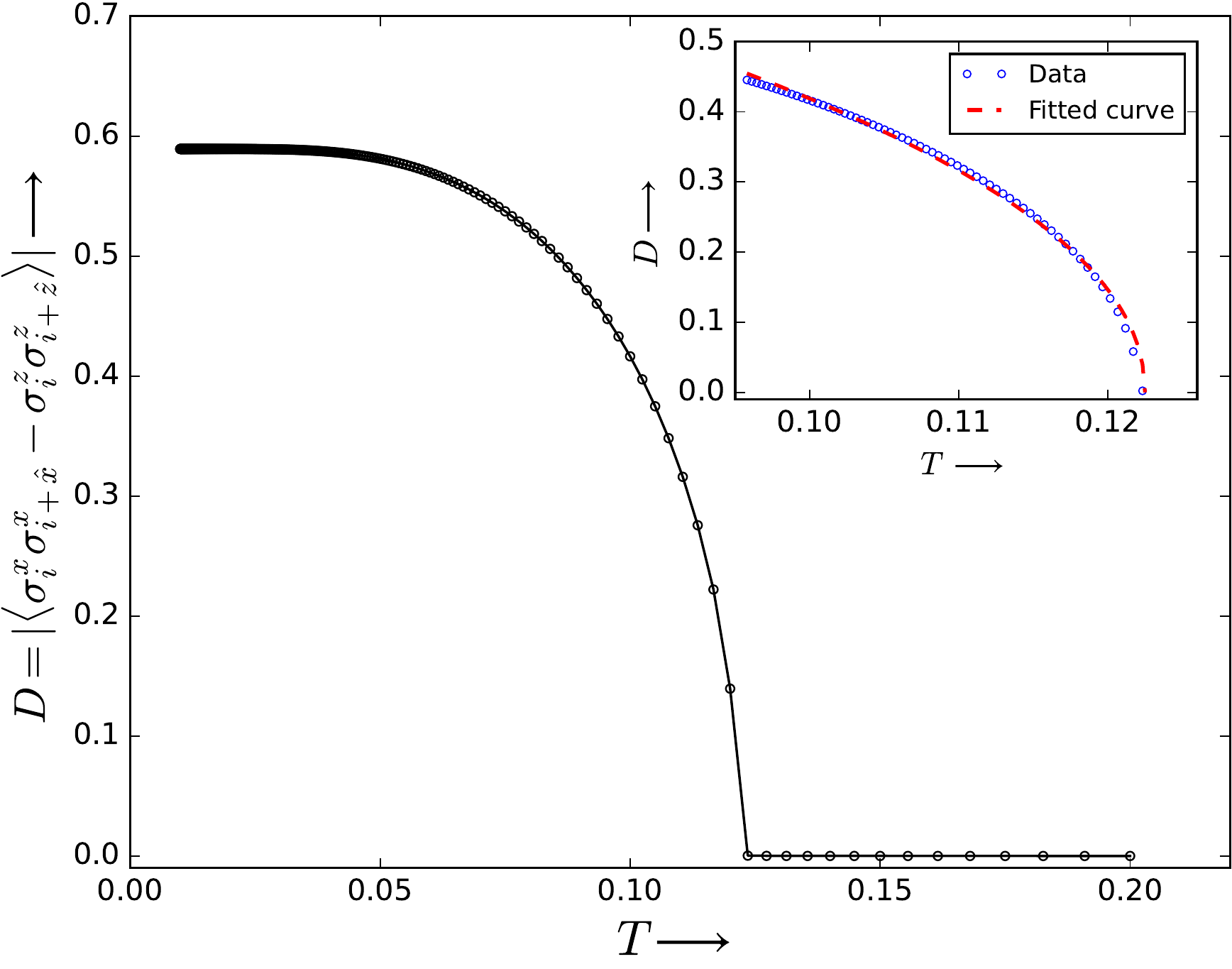}}
\subfigure{\label{Fig.S1b}\includegraphics[height=50mm, width=70mm]{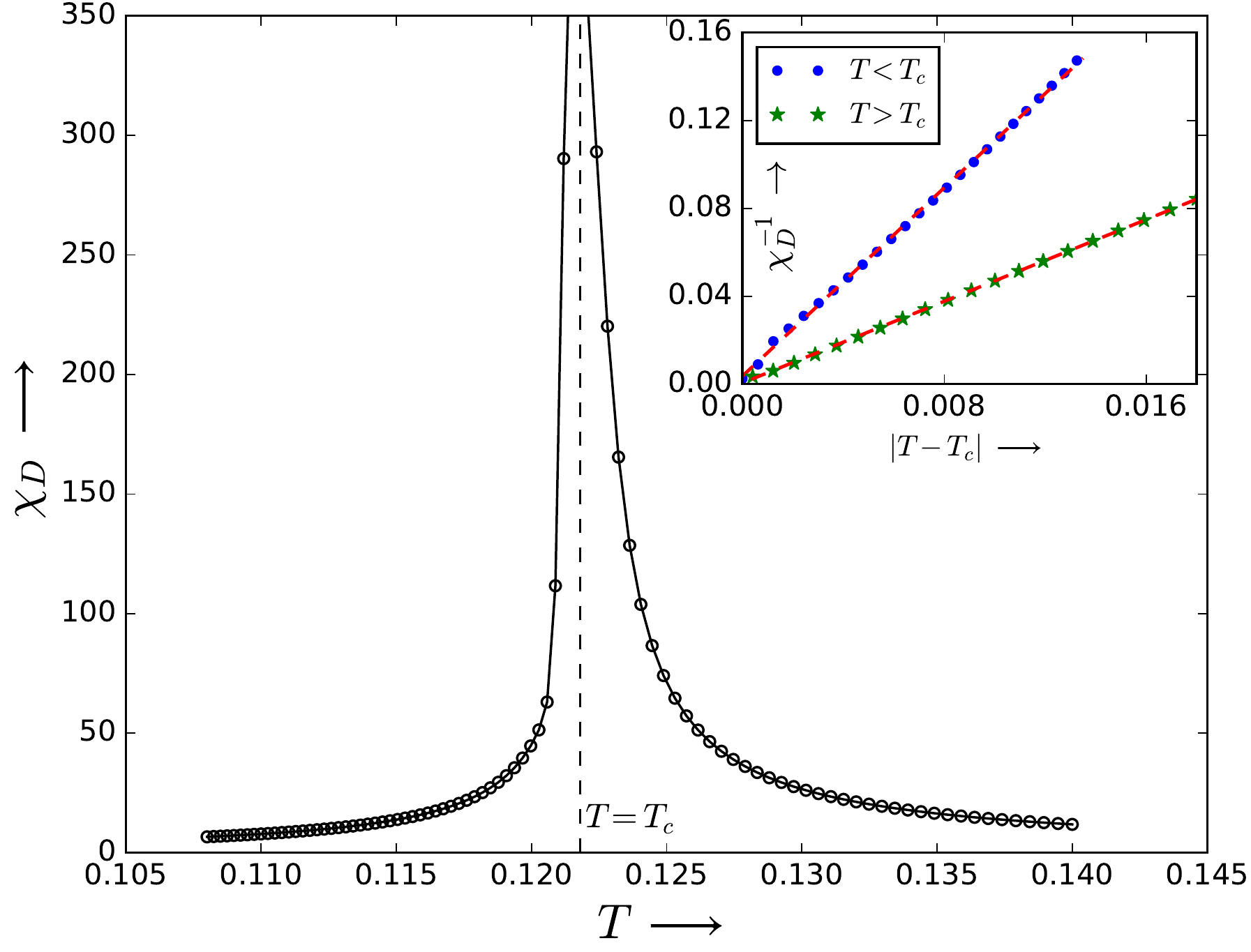}}
\subfigure{\label{Fig.S1c}\includegraphics[width=70mm]{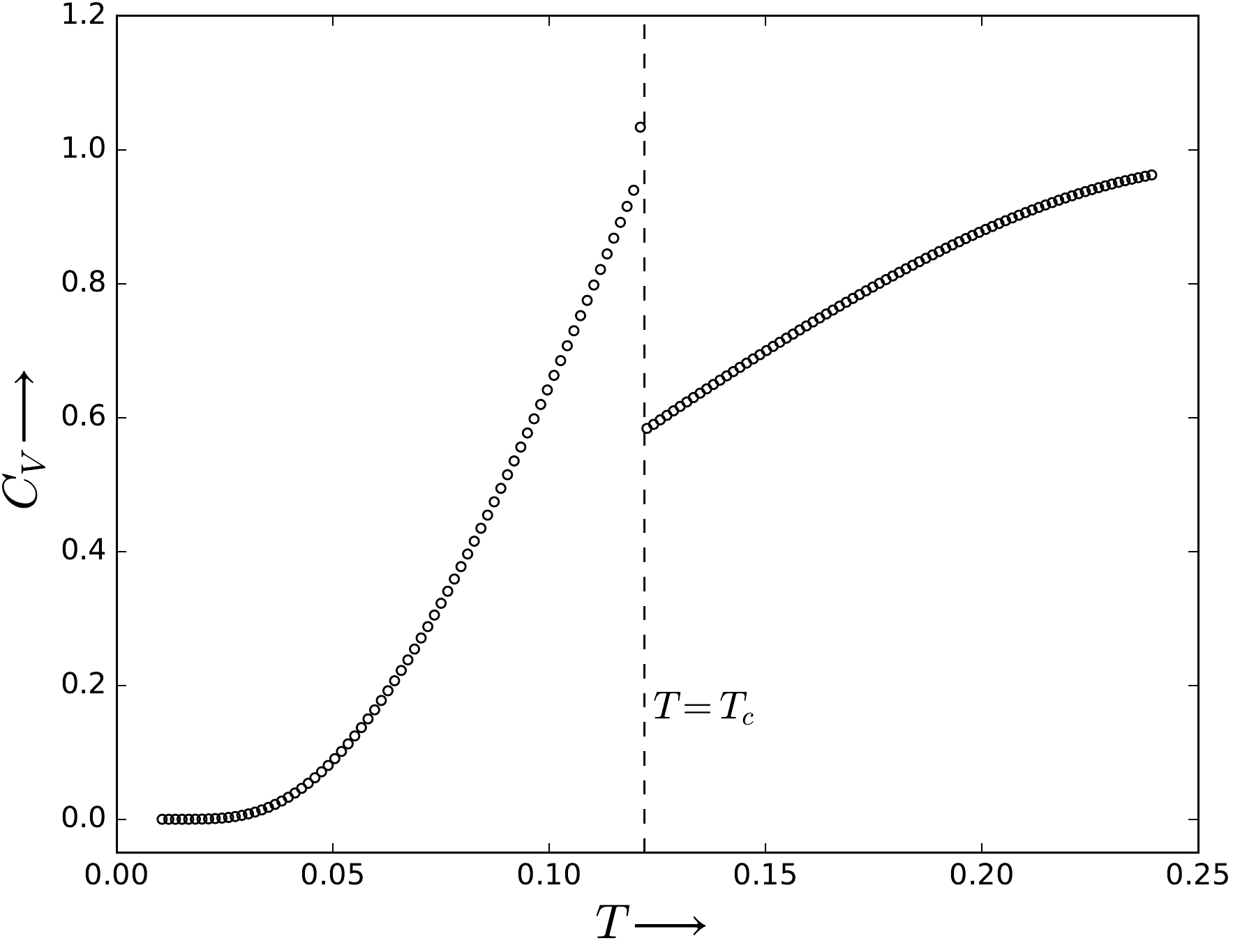}}
\caption{(a) Nematic order parameter ($D$) vs. temperature ($T$) is shown, the inset figure compares it's behaviour near $T_{c}=0.1225$ (blue dots) with the fitted curve ($\sim c(T_{c}-T)^{\beta}$, red dashed line) used for extracting the exponent $\beta$. (b) Nematic susceptibility ($\chi_{D}$) vs. $T$, the inset figure plots $\chi^{-1}_{D}$ vs. $|T-T_{c}|$, which are straight lines with almost zero intercept along $\chi^{-1}_{D}$ axis. Slope of $T<Tc$ line (blue dots) is almost twice as the slope of $T>T_{c}$ line (green stars). (c) MF specific heat ($C_{V}$) vs. $T$ plot showing a finite gap at $T_{c}$, similar to the Landau theory. In all these figures, we consider four-leg compass ladder.}
\end{figure}

In Fig.~\ref{Fig.3} we see that the jump discontinuity (at $J_{x}=J_{z}$) of various mean-field averages vanishes near $T\approx 0.125$ and the graphs for higher $T$ values show continuous behaviour. Here we provide some additional plots which show that the point $T\approx 0.125,\ J_{x}=J_{z}$ actually corresponds to a second order critical endpoint ($T_{c}$) to the first order transitions below $T_{c}$ .\\
\indent We find that the Ising like nematic order parameter, $|\langle \mathcal{D}\rangle|= |\langle \sigma^{x}_{\vec{r}}\sigma^{x}_{\vec{r}+\hat{x}}-\sigma^{z}_{\vec{r}}\sigma^{z}_{\vec{r}+\hat{z}}\rangle|$ continuously decreases with increasing $T$ and goes to zero at $T=T_{c}=0.1225$ ($J=1$) when $J_{x}$ and $J_{z}$ are equal [see Fig.~\ref{Fig.S1a}]. In our calculations, we have used $\theta=\tan^{-1}(J_{z}/J_{x})=\pi/4\pm 10^{-6}$. We also compute the ``nematic susceptibility", defined as $\chi_{D}=(\partial \langle \mathcal{D}\rangle/\partial h)\big|_{h\rightarrow 0}$ , where $h=1-(J_{z}/J_{x})$, is the anisotropy in coupling strengths which acts as a fictitious external field needed to compute susceptibility. We find that the susceptibility diverges near $T=T_{c}$ [see Fig.~\ref{Fig.S1b}]. The critical exponents, $\beta$ (defined as $|\langle \mathcal{D}\rangle|\sim (T_{c}-T)^{\beta}$) and $\gamma$ (defined as $\chi_{D}\sim A_{\pm}|T-T_{c}|^{-\gamma}$) are almost equal to the classical Landau theory exponents, we find $\beta=0.4776$ and $\gamma=1.0$. For the calculation of $\gamma$, we have fitted $\chi^{-1}_{D}$ with $a_{\pm}|T-T_{c}|+b_{\pm}$ ; we find $a_{-}=10.7$, $b_{-}=3.6\times 10^{-3}$ for $T<T_{c}$, and $a_{+}=4.648$, $b_{+}=5.02\times 10^{-4}$ for $T>T_{c}$. So even the ratio of $A_{\pm}$ ($\sim 1/a_{\pm}$) is almost like Landau theory. The specific heat ($C_{V}$) calculated from MFT shows a finite jump at $T_{c}$, meaning the corresponding exponent, $\alpha=0.0$ [see Fig.~\ref{Fig.S1c}]. This is also same as classical Landau theory. 
\\
\indent We find the critical exponents are same as classical Landau mean-field theory, whereas the numerical results \cite{Wenzel, Oitmaa, Czarnik-Supplementary} show signatures of $2d$ classical Ising universality class. Fluctuations around MFT play a significant role near this critical region (as we will explain below).
\section{Two-spin correlation functions}\label{App D}
One novel feature of our MFT is that the spatial correlations are partially retained, which are unusual in conventional MF descriptions. We evaluate here some important time independent two-spin correlation functions both at zero and finite temperatures. Explicit computations are not performed here as the MF Hamiltonians are just $1d$ TFIMs. We will use Mattis's dualities and some well-known results of $1d$ TFIM \cite{Subir Sachdev-Supplementary, Barouch-Supplementary} to derive the following correlation functions. \\

\textbf{(A)} $\langle \sigma^{x}_{\vec{r}}\sigma^{x}_{\vec{r}+n\hat{x}}\rangle$ for $n>>1$, $T=0$ :
\begin{align}
&\ 4|\langle S^{x}_{j,1}S^{x}_{j+n,1}\rangle|\equiv 4|\langle P^{x}_{j,1}P^{x}_{j+n,1}\rangle|\nonumber\\
&\text{for}\ |J_{x}|>|J_{z}|\ =\ (1-\lambda_{a}^{2})^{1/4}\bigg[1+\frac{1}{2\pi n^{2}}\frac{\lambda^{2n+2}_{a}}{1-\lambda_{a}^{2}}+\cdot\cdot\cdot\cdot\bigg] \tag{D1.a}\\
&\text{for}\ |J_{x}|<|J_{z}|\ =\ \frac{1}{\sqrt{\pi n}}\frac{\lambda^{1/2}_{a}}{(\lambda^{2}_{a}-1)^{1/4}}e^{-\sfrac{|n|}{\xi_{a}}} \tag{D1.b}
\end{align}
Here $\lambda_{a}=\frac{|J_{z}|(1+\Omega_{z})}{|J_{x}|(1+\Theta_{x})}$ , $\xi_{a}=1/|\ln{(\lambda_{a})}|$.
By applying Mattis's transformations Eqs.~\eqref{eq3} and \eqref{B.2}, we get $4T^{x}_{j,1}T^{x}_{j+n,1}=16Q^{z}_{j,1}P^{x}_{j,1}Q^{z}_{j+n,1}P^{x}_{j+n,1}=$\newline  $16W^{x}_{j}P^{x}_{j,1}W^{x}_{j+n}P^{x}_{j+n,1}$ . So at the MFT level,
\begin{align}
\ 4|\langle T^{x}_{j,1}T^{x}_{j+n,1}\rangle| &\approx 4|\langle P^{x}_{j,1}P^{x}_{j+n,1}\rangle|\times 4|\langle W^{x}_{j}W^{x}_{j+n}\rangle|\nonumber
\end{align}
\begin{align}
&4|\langle T^{x}_{j,1}T^{x}_{j+n,1}\rangle|=\nonumber\\
&\text{for}\ \ |J_{x}|>|J_{z}|\nonumber\\
&=\ \big\lbrace(1-\lambda_{a}^{2})(1-\lambda_{b}^{2})\big\rbrace^{1/4}\bigg[1+\frac{1}{2\pi n^{2}}\sum_{\alpha=a,b}\frac{\lambda^{2n+2}_{\alpha}}{1-\lambda_{\alpha}^{2}}+\cdot\cdot\cdot\bigg]\tag{D2.a}\\
&\text{for}\ \ |J_{z}|>|J_{x}|\nonumber\\
&=\ \frac{1}{\pi n}\frac{(\lambda_{a}\lambda_{b})^{1/2}e^{-|n|/\xi_{t}}}{(\lambda_{a}^{2}-1)^{1/4}(\lambda_{b}^{2}-1)^{1/4}}\ \ ,\ \ \xi_{t}^{-1}=\xi_{a}^{-1}+\xi_{b}^{-1} \tag{D2.b}
\end{align}
Here $\lambda_{b}=|J_{z}|\Delta_{z}/|J_{x}|\Delta_{x}$ and $\xi_{a,b}=1/|\ln{(\lambda_{a,b})}|$.
We have $S^{x}_{j,2}S^{x}_{j+n,2}=P^{x}_{j,2}P^{x}_{j+n,2}$ and $T^{x}_{j,2}T^{x}_{j+n,2}=64V^{z}_{j}V^{z}_{j+n}W^{x}_{j}P^{x}_{j,2}W^{x}_{j+n}P^{x}_{j+n,2}$ . We take all $V^{z}_{j}=\pm\sfrac{1}{2}$, this holds rigorously at $T=0$ . So we see $|\langle S^{x}_{jl}S^{x}_{j+n,l}\rangle|$ and $|\langle T^{x}_{jl}T^{x}_{j+n,l}\rangle|$ are same for $l=1,2$.\\

\textbf{(B)} $\langle \sigma^{x}_{\vec{r}}\sigma^{x}_{\vec{r}+m\hat{z}}\rangle$ for $m=1,2$, $T=0$ : \\

These operators do not commute with the $1d$ symmetries, $Z_{l}=\lim_{N\rightarrow\infty}\prod_{j=1}^{N}\sigma^{z}_{jl}$ ; so according to Elitzur's theorem \cite{Batista}, their expectation values should vanish rigorously above $T=0$. Although at $T=0$, this theorem permits spontaneous breaking of these $1d$ $\mathbb{Z}_{2}$ symmetries in the thermodynamic limit, resulting finite average values of these symmetry non-invariant operators. We now show that these non-zero averages result from the long range ordering of $P$ and $W$ Ising chains ($\langle P^{x}_{jl}\rangle,\ \langle W^{x}_{j}\rangle\neq 0$) in the $|J_{x}|>|J_{z}|$ region. At first, we consider the nearest neighbour (along ladder rungs) or $m=1$ case, 
\begin{align}
&|\langle\sigma^{x}_{j,1}\sigma^{x}_{j,2}\rangle|\sim\nonumber\\
&\ \ 4|\langle S^{x}_{j,1}T^{x}_{j,1}\rangle|\sim 2|\langle Q^{z}_{j,1}\rangle|\sim 2|\langle W_{j}^{x}\rangle|= (1-\lambda_{b}^{2})^{1/8} \tag{D3.a}
\end{align}
Similarly, it is easy to show $|\langle\sigma^{x}_{j,3}\sigma^{x}_{j,4}\rangle|=|\langle\sigma^{x}_{j,1}\sigma^{x}_{j,2}\rangle|\sim (1-\lambda_{b}^{2})^{1/8}$.
\begin{align}
\ &|\langle\sigma^{x}_{j,2}\sigma^{x}_{j,3}\rangle|\sim 4|\langle S^{x}_{j,1}T^{x}_{j,2}\rangle|\sim 2|\langle W^{x}_{j}\rangle|\times 4|\langle P^{x}_{j,1}P^{x}_{j,2}\rangle|\nonumber\\
&\sim 8|\langle W^{x}_{j}\rangle||\langle P^{x}_{j,1}\rangle|^{2}=(1-\lambda_{a}^{2})^{1/4}(1-\lambda^{2}_{b})^{1/8} \tag{D3.b}
\end{align}
Here the spins $P^{x}_{j,1},P^{x}_{j,2}$ are governed by identical Hamiltonians and they don't interact with each other, so we can write $|\langle P^{x}_{j,1}P^{x}_{j,2}\rangle|= |\langle P^{x}_{j,1}\rangle|^{2}$. Similarly, $|\langle\sigma^{x}_{j,1}\sigma^{x}_{j,4}\rangle|=|\langle\sigma^{x}_{j,2}\sigma^{x}_{j,3}\rangle|=(1-\lambda_{a}^{2})^{1/4}(1-\lambda^{2}_{b})^{1/8}$ . Due to periodic boundary condition, $(j,1)$ and $(j,4)$ are here nearest neighbour sites. Next we calculate the second neighbour or $m=2$ case,
\begin{align}
&|\langle\sigma^{x}_{j,1}\sigma^{x}_{j,3}\rangle|\sim 4|\langle T^{x}_{j,1}T^{x}_{j,2}\rangle|\sim 16|\langle Q^{z}_{j,1}P^{x}_{j,1}Q^{z}_{j,2}P^{x}_{j,2}\rangle|\nonumber\\
&\sim 4|\langle P^{x}_{j,1}P^{x}_{j,2}\rangle|\sim 4|\langle P^{x}_{j,1}\rangle|^{2}=(1-\lambda^{2}_{a})^{1/4} \tag{D3.c} \label{D.3c}
\end{align}
Finally, we see $|\langle\sigma^{x}_{j,2}\sigma^{x}_{j,4}\rangle|\sim 4|\langle S^{x}_{j,1}S^{x}_{j,2}\rangle|\sim 4|\langle P^{x}_{j,1}P^{x}_{j,2}\rangle|$ is identical to \eqref{D.3c}. As expected here, the second neighbour $(m=2)$ correlations are weaker than the nearest neighbour $(m=1)$ ones.\\

\textbf{(C)} $\langle \sigma^{\mu}_{\vec{r}}\sigma^{\mu}_{\vec{r}+n\hat{x}}\rangle$ for any $n$, $\mu=y,z$, and $T=0$ :\\

\indent These operators violate local ($d=0$) symmetries ($V^{z}_{j}\sim \prod_{l=1}^{4}\sigma^{x}_{jl}$) of the Hamiltonian. Elitzur's theorem tells that spontaneous breaking of local gauge symmetries is impossible even at $T=0$. So, these spatial correlations are ultra-local in nature, $\langle \sigma^{\mu}_{\vec{r}}\sigma^{\mu}_{\vec{r}+n\hat{x}}\rangle=\delta_{n,0}$. We now prove this statement using our MF construction. We show here two such examples,
\begin{align}
&\langle \sigma^{z}_{j,2}\sigma^{z}_{j+n,2}\rangle\sim 4\langle S^{z}_{j,1}S^{z}_{j+n,1}\rangle\sim 16\langle P^{z}_{j,1}Q^{x}_{j,1}P^{z}_{j+n,1}Q^{x}_{j+n,1}\rangle\nonumber\\
&\ \ \ \ \ \sim 64\langle P^{z}_{j,1}W^{z}_{j}V^{x}_{j}P^{z}_{j+n,1}W^{z}_{j+n}V^{x}_{j+n}\rangle=\delta_{n,0}\tag{D4.a}
\end{align}
Similarly,
\begin{align}
&\langle \sigma^{y}_{j,1}\sigma^{y}_{j+n,1}\rangle\sim 4\langle T^{y}_{j,1}T^{y}_{j+n,1}\rangle\sim 16\langle P^{x}_{j,1}Q^{y}_{j,1}P^{x}_{j+n,1}Q^{y}_{j+n,1}\rangle\nonumber\\
&\ \ \ \ \ \sim 64\langle P^{x}_{j,1}W^{y}_{j}V^{x}_{j}P^{x}_{j+n,1}W^{y}_{j+n}V^{x}_{j+n}\rangle=\delta_{n,0}\tag{D4.b}
\end{align}
The MF eigenstates are common eigenstates of $H_{QC}$ and $V^{z}_{j}$, so flipping of $V^{z}_{j}$ will map to different symmetry sector and thus makes the overlap zero. This is independent of any $|J_{z}|/|J_{x}|$ . \\

\textbf{Conclusion (1)}: When $T=0$ and $|J_{x}|>|J_{z}|$ , there is a long range magnetic order ($\langle\sigma^{x}_{\vec{r}}\rangle\neq 0$) in the system where spins are mostly ``aligned" in the $x$-direction. This ordering occurs due to spontaneous breaking of $d=1$ Ising symmetries. The magnetic order suddenly drops to zero at the self-dual point, $J_{x}=J_{z}$ and continues to be zero in $|J_{z}|>|J_{x}|$ region with a finite two-spin correlation length. \\

Now we will show that long-range magnetic order completely disappears as we go above $T=0$ and the two-spin correlation functions become short ranged. Elitzur's theorem forbids spontaneous breaking of d=1 symmetries, $Z_{l}$ at any $T>0$ (like in $1d$ TFIM), thus the long-range order vanishes. We define following parameters, $\Delta_{a}=|J_{x}|(1+\Theta_{x})(1-\lambda_{a})$, $\Delta_{b}=|J_{x}|\Delta_{x}(1-\lambda_{b})$, $v_{a}=2|J_{x}|(1+\Theta_{x})$, $v_{b}=2|J_{x}|\Delta_{x}$ . Using Mattis's relations and the rigorous finite $T$ results of 1d TFIM \cite{Subir Sachdev-Supplementary}, we find the following, \\

When $\Delta_{a},\Delta_{b}>>T>0\ (\text{or}\ |J_{x}|>|J_{z}|,\ T<<T_{c})$ :
\begin{align}
&4\langle S^{x}_{jl}S^{x}_{j+n,l}\rangle\sim \Delta^{1/4}_{a}e^{-|n|/\xi_{a}(T)} \tag{D5.a}\\
&4\langle T^{x}_{jl}T^{x}_{j+n,l}\rangle\sim \Delta_{a}^{1/4}\Delta_{b}^{1/4}e^{-|n|/\xi_{t}(T)} \tag{D5.b}
\end{align}
Here $\xi_{\alpha}^{-1}=\sqrt{2|\Delta_{\alpha}|T/\pi v_{\alpha}^{2}}e^{-|\Delta_{\alpha}|/T}$ with $\alpha=a,b$, and $\xi^{-1}_{t}=\xi^{-1}_{a}+\xi^{-1}_{b}$.
So the temperature induces a finite two-spin correlation length, destroying the $T=0$ magnetic order. \\

When $\Delta_{a},\Delta_{b}<0$, $|\Delta_{a}|,|\Delta_{b}|>>T>0\ $(or$\ |J_{z}|>|J_{x}|,\ T<<T_{c})$ :
\begin{gather}
4\langle S^{x}_{jl}S^{x}_{j+n,l}\rangle\sim \frac{T}{|\Delta_{a}|^{3/4}}e^{-|n|/\tilde{\xi}_{a}(T)}\tag{D6.a}\\
4\langle T^{x}_{jl}T^{x}_{j+n,l}\rangle\sim \frac{T^{2}}{(|\Delta_{a}||\Delta_{b}|)^{3/4}}e^{-|n|/\tilde{\xi}_{t}(T)} \tag{D6.b}
\end{gather}
Here $\tilde{\xi}^{-1}_{\alpha}=(|\Delta_{\alpha}|/v_{\alpha})+\sqrt{2|\Delta_{\alpha}|T/\pi v_{\alpha}^{2}}e^{-|\Delta_{\alpha}|/T}$ with $\alpha=a,b$, and $\tilde{\xi}^{-1}_{t}=\tilde{\xi}^{-1}_{a}+\tilde{\xi}^{-1}_{b}$.\\

When $|\Delta_{a}|,|\Delta_{b}|<<T\ (\text{or}\ J_{z}=J_{x},\ T\rightarrow T_{c}^{-})$ :
\\

This is the region near second order critical point where the nematic order parameter smoothly goes to zero (Fig.~\ref{Fig.S1a}) and the elementary JW fermionic excitations become gapless (Fig.~\ref{figS4}). Although fluctuations are important (as we show below) in this region, we continue using the MFT to see what minimal features could be extracted from it. We find
\begin{gather}
4\langle S^{x}_{jl}S^{x}_{j+n,l}\rangle\sim T^{1/4}e^{-|n|/\xi^{c}_{a}} \tag{D7.a}\\
4\langle T^{x}_{jl}T^{x}_{j+n,l}\rangle\sim  T^{1/2}e^{-|n|/\xi^{c}_{t}} \tag{D7.b}
\end{gather}
Here $(\xi^{c}_{\alpha})^{-1}\approx (\pi T_{c}/4v_{\alpha})$, $\alpha=a,b$, and $(\xi^{c}_{t})^{-1}=(\xi^{c}_{a})^{-1}+(\xi^{c}_{b})^{-1}$.
So the two-spin correlation length remains finite even in this critical region, reflecting that finite $T$ phase transition is non-magnetic in nature. In the finite $T$ calculations, we assumed that all the classical Ising variables, $V^{z}_{j}$ are frozen to $\pm\sfrac{1}{2}$, this is rigorously valid only at $T=0$. We should expect that at any finite $T$, $\langle V^{z}_{j}V^{z}_{j+n}\rangle\sim e^{-|n|/\xi_{v}}$, where $\xi^{-1}_{v}=\ln{\coth{(J_{s}/T)}}$. Here $J_{s}$ represents some unknown energy scale depending on $J_{x}\ ,\ J_{z}$ . We already have exponentially decaying finite $T$ correlations at the MF level. An Additional decay of $\langle V^{z}_{j}V^{z}_{j+n}\rangle$ will just bring quantitative changes in correlation lengths, the functional forms of these correlations remain same as MFT.
\section{Mean-field results of eight-leg compass ladder}\label{App E}
Construction of lower dimensional symmetry preserving MFT for eight-leg compass ladder is straightforward. We skip the lengthy algebra and provide only some crucial steps to derive the MF self-consistency relations. \\
\indent We start form the 8-leg compass ladder Hamiltonian ($H_{QC}$) and apply Mattis's transformations (labelling it M1). Like Eqs.~\eqref{eq5a},\eqref{eq5b}, we decouple the resulting four-spin interacting model in a two-spin channel which preserves all lower dimensional symmetries. This decoupling (calling it D1) results four identical $1d$ TFIMs ($H_{1}^{(l)},\ l=1,2,3,4$) and a four-leg compass ladder ($H'_{QC}$). Coupling constants in both of these Hamiltonians depend on bare strengths ($J_{x}, J_{z}$) plus various mean-field averages which have to be determined later using self-consistency conditions. Now a similar scheme would be repeated for $H'_{QC}$; we use M2 and D2 \cite{footnote1-Supplementary} which gives two more $1d$ TFIMs ($H^{(m)}_{2},\ m=1,2$) and a two-leg compass ladder ($H''_{C}$). Applying M3 on $H''_{C}$, we get the final $1d$ TFIM ($H_{3}$) coupled to static $\mathbb{Z}_{2}$ fields which are one of the gauge-like symmetries of $H_{QC}$.
\\
\indent We arrive at the following Hamiltonians,
\begin{gather}
H_{1}^{(l)}=J_{x}\sum_{j=1}^{N}(1+\Theta_{jl}^{x})P^{x}_{jl}P^{x}_{j+1,l}+\frac{J_{z}}{2}\sum_{j=1}^{N}(1+\Omega_{jl}^{z})P^{z}_{jl}\nonumber\\
\hspace{5.5cm}(l=1,2,3,4)\\
H^{(m)}_{2}=J_{x}\sum_{j=1}^{N}\alpha^{(m)}_{j}Q^{x}_{jm}Q^{x}_{j+1,m}+\frac{J_{z}}{2}\sum_{j=1}^{N}\mu^{(m)}_{j}Q^{z}_{jm}\nonumber\\
\hspace{6cm} (m=1,2)\\
H_{3}=J_{x}\sum_{j=1}^{N}\big[\beta_{j}^{(1)}+4V^{z}_{j}V^{z}_{j+1}\beta_{j}^{(2)}\big]W^{x}_{j}W^{x}_{j+1}\nonumber\\
\hspace{5cm}+\frac{J_{z}}{2}\sum_{j=1}^{N}\gamma_{j}W^{z}_{j}
\end{gather} 
Like in the four-leg case, here also we find the following MF order parameters, $\Delta_{jl}^{x}= 4\langle S^{x}_{jl}S^{x}_{j+1,l}\rangle$, $\Delta_{jl}^{z}= 4\langle S^{z}_{jl}T^{z}_{jl}\rangle$, $\Theta^{x}_{jl}=16\langle S^{x}_{jl}S^{x}_{j+1,l}T^{x}_{jl}T^{x}_{j+1,l}\rangle$, and $\Omega_{jl}^{z}=4\langle T^{z}_{jl}T^{z}_{j,l+1}\rangle$ (with $j=1\ \text{to}\ N$ and $l=1,2,3,4$). In addition, we find a rectangular loop-like object, $\Phi_{jl}^{x}=2^{8}\langle S^{x}_{jl}S^{x}_{j+1,l}T^{x}_{jl}T^{x}_{j+1,l}S^{x}_{j,l+1}S^{x}_{j+1,l+1}T^{x}_{j,l+1}T^{x}_{j+1,l+1}\rangle$ and a fourth neighbour $zz$-correlation, $\Pi^{z}_{j}=4\langle T^{z}_{j,3}T^{z}_{j,1}\rangle\ (\sim \langle \sigma^{z}_{\vec{r}}\sigma^{z}_{\vec{r}+4\hat{z}}\rangle)$, these two are also dual to each other as we will show below. \\
\indent The coupling constants in $H^{(m)}_{2}$ and $H_{3}$ are following, $\alpha^{(1)}_{j}=\Delta_{j,2}^{x}+\Delta^{x}_{j,1}\Phi^{x}_{j,1}$, $\alpha^{(2)}_{j}=\Delta_{j,4}^{x}+\Delta^{x}_{j,3}\Phi^{x}_{j,3}$, $\mu^{(1)}_{j}=\Delta^{z}_{j,1}+\Delta^{z}_{j,2}\Pi^{z}_{j}$, $\mu^{(2)}_{j}=\Delta^{z}_{j,3}+\Delta^{z}_{j,4}\Pi^{z}_{j}$, $\beta^{(1)}_{j}=\Delta_{j,1}^{z}\Theta^{x}_{j,2}$, $\beta^{(2)}_{j}=\Delta_{j,3}^{z}\Theta^{x}_{j,4}$, and $\gamma_{j}=\Delta^{z}_{j,2}\Omega^{z}_{j,1}+\Delta^{z}_{j,4}\Omega^{z}_{j,3}$ . The conserved $\mathbb{Z}_{2}$ fields are $V^{z}_{j}=\frac{1}{2}\prod_{l=1}^{4}(4S^{x}_{jl}T^{x}_{jl})$, which are one of the gauge-like symmetries of eight-leg compass ladder.\\

\begin{figure}
\centering
\includegraphics[height=55mm,width=75mm]{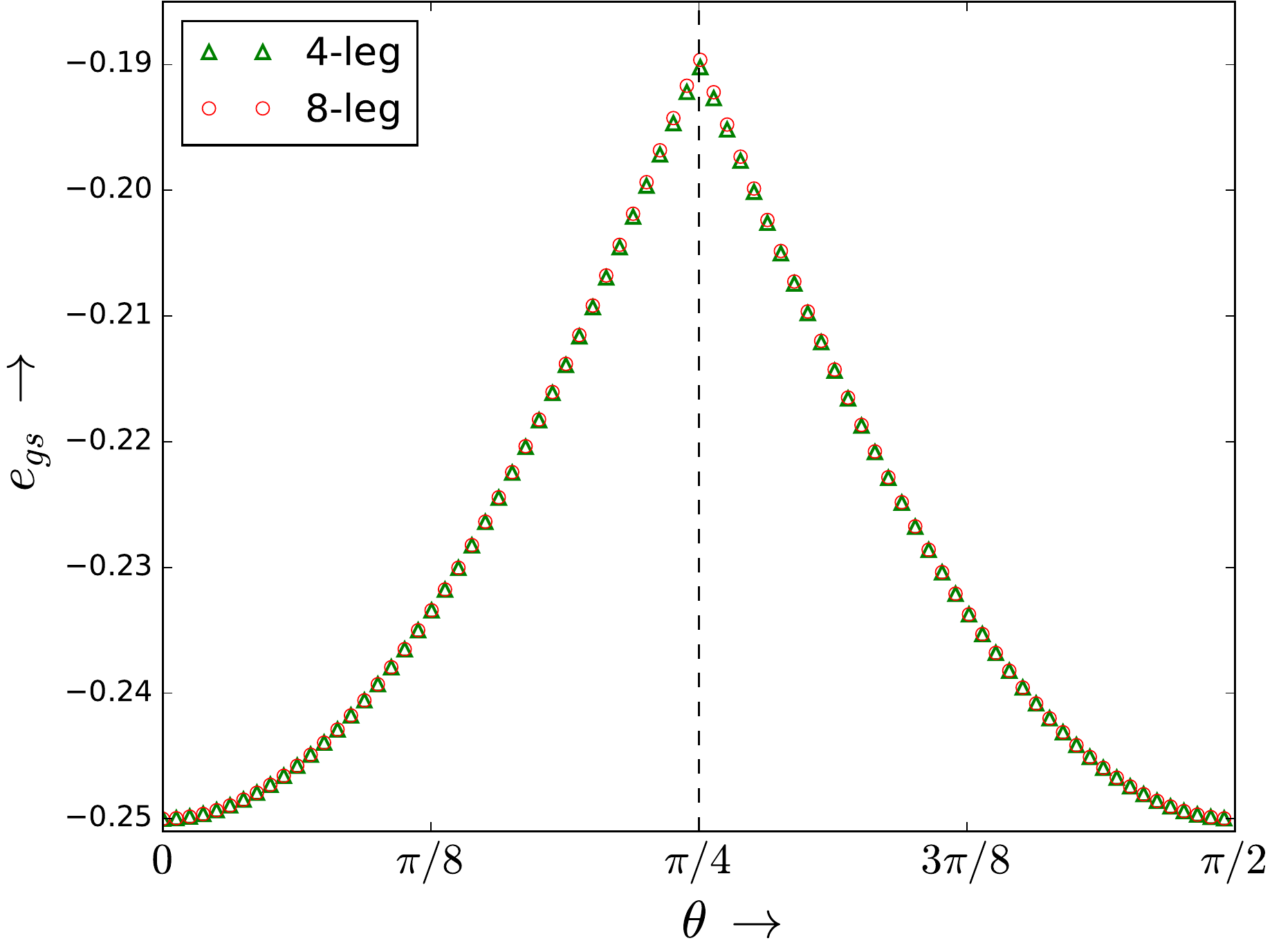}
\caption{Comparison between ground state energy density ($e_{gs}$) of four-leg (green triangles) and eight-leg (red circles) compass ladder as function of $\theta=\tan^{-1}(|J_{z}|/|J_{x}|)$, there is a negligibly small difference between two energies, only near $\theta=\pi/4\ (J_{x}=J_{z})$ .}
\label{figS2}
\end{figure}

\indent Next we assume translationally invariant ansatz for the MF order parameters, i.e. (a) take all $V^{z}_{j}=\pm \sfrac{1}{2}$, this helps us to get rid of $j$ dependence, and (b) $\Delta_{jl}^{x}\equiv \Delta_{x}$ and $\Delta_{jl}^{z}\equiv \Delta_{z}$ (independent of $j,\ l$). In the four-leg case, the condition (b) comes as a consequence of condition (a), but here we have to impose it separately. It is straightforward to verify that these conditions (a) and (b) will make all the order parameters (hence the coupling constants) to be completely independent of spatial co-ordinates ($j,\ l$).\\
\indent Finally, we arrive at the following self-consistently coupled, spatially uniform $1d$ TFIMs,
\begin{gather}
H^{(l)}_{1}=J_{x}(1+\Theta_{x})\sum_{j=1}^{N}P^{x}_{jl}P^{x}_{j+1,l}+\frac{J_{z}}{2}(1+\Omega_{z})\sum_{j=1}^{N}P^{z}_{jl}\nonumber\\
\hspace{5cm}(l=1,2,3,4) \label{E.4}\\
\hspace{-2cm}H_{2}^{(m)}=J_{x}\Delta_{x}(1+\Phi_{x})\sum_{j=1}^{N}Q_{jm}^{x}Q_{j+1,m}^{x}\nonumber\\
\hspace{2cm}+\frac{J_{z}}{2}\Delta_{z}(1+\Pi_{z})\sum_{j=1}^{N}Q^{z}_{jm}\ \ (m=1,2)\label{E.5}\\
H_{3}=2J_{1}\Delta_{x}\Theta_{x}\sum_{j=1}^{N}W_{j}W_{j+1}+J_{2}\Delta_{z}\Omega_{z}\sum_{j=1}^{N}W^{z}_{j}\label{E.6}
\end{gather}
Which could be solved easily as shown in the Appendix \ref{App A}, the final equations are same as in the four-leg case (see Eqs.~\eqref{eq9}, \eqref{eq10}), except $\textbf{M}^{x}=(\Delta_{x}, \Theta_{x},\Phi_{x})$ and $\textbf{M}^{z}=(\Delta_{z}, \Omega_{z},\Pi_{z})$ now have three components. 
\begin{align}
M^{x}_{\sigma}=\int_{0}^{\pi}\frac{dk}{\pi}\frac{(h_{\sigma}\cos{k}-1)\tanh{\big(\beta E^{\sigma}_k/2\big)}}{\sqrt{1+h^{2}_{\sigma}-2h_{\sigma}\cos{k}}}\\
M^{z}_{\sigma}=\int_{0}^{\pi}\frac{dk}{\pi}\frac{(h^{-1}_{\sigma}\cos{k}-1)\tanh{\big(\beta E^{\sigma}_k/2\big)}}{\sqrt{1+h^{-2}_{\sigma}-2h^{-1}_{\sigma}\cos{k}}}
\end{align} 
Here $E_{k}^{a}=\frac{|J_{x}|}{2}(1+\Theta_{x})\sqrt{1+h^{2}_{a}-2h_{a}\cos{k}}$ , $E_{k}^{b}=\frac{|J_{x}|}{2}|\Delta_{x}|(1+\Phi_{x})\sqrt{1+h^{2}_{b}-2h_{b}\cos{k}}$ , and $E_{k}^{c}=|J_{x}||\Delta_{x}|\Theta_{x}$ $\times\sqrt{1+h^{2}_{c}-2h_{c}\cos{k}}$ are elementary excitation spectrum of the above Ising chains \eqref{E.4}-\eqref{E.6} with $h_{a}=\frac{J_{z}(1+\Omega_{z})}{J_{x}(1+\Theta_{x})}$, $h_{b}=\frac{J_{z}\Delta_{z}(1+\Pi_{z})}{J_{x}\Delta_{x}(1+\Phi_{x})}$ , $h_{c}=\frac{J_{z}\Delta_{z}\Omega_{z}}{J_{x}\Delta_{x}\Theta_{x}}$ , and $\beta=1/T$ .\\
The ground state energy per site is given by following,
\begin{align}
e_{gs}=-\frac{1}{16}\big[J_{x}&\Delta_{x}\Theta_{x}(1+2\Phi_{x})+J_{z}\Delta_{z}\Omega_{z}(1+2\Pi_{z})\big]\nonumber\\
&-\frac{1}{2}\sum_{k=-\pi}^{\pi}\big[4E^{a}_{k}+2E^{b}_{k}+E^{c}_{k}\big]
\end{align}
\indent We now proceed to the results. We find that the ground state energy density is almost same as what is found in four-leg ladder case, except close to isotropic or self-dual point $J_{x}=J_{z}$ where very small changes are observed (see Fig.~\ref{figS2}). At the self-dual point, we find $e_{gs}=-0.189623$ for four-leg case and $e_{gs}=-0.190205$ for four-leg case $(J=1)$. This negligible difference strongly supports our previous argument (in the main text) that most of the ground state correlation energy of $2d$ compass model is captured in a four-leg ladder and increasing number of legs has only negligible effect. Although we have performed the mean-field decoupling twice (in eight-leg case), various loop/ plaquette correlators $(\Theta_{x},\Phi_{x}$) and beyond nearest neighbour $zz$-correlations ($\Omega_{z},\Pi_{z}$) (which don't violate gauge-like symmetries) actually capture the correlations along rungs in a self-consistent way. This should be a reason behind such nice convergence of $e_{gs}$.

\begin{figure}[htb]
\centering
\subfigure{\includegraphics[width=70mm]{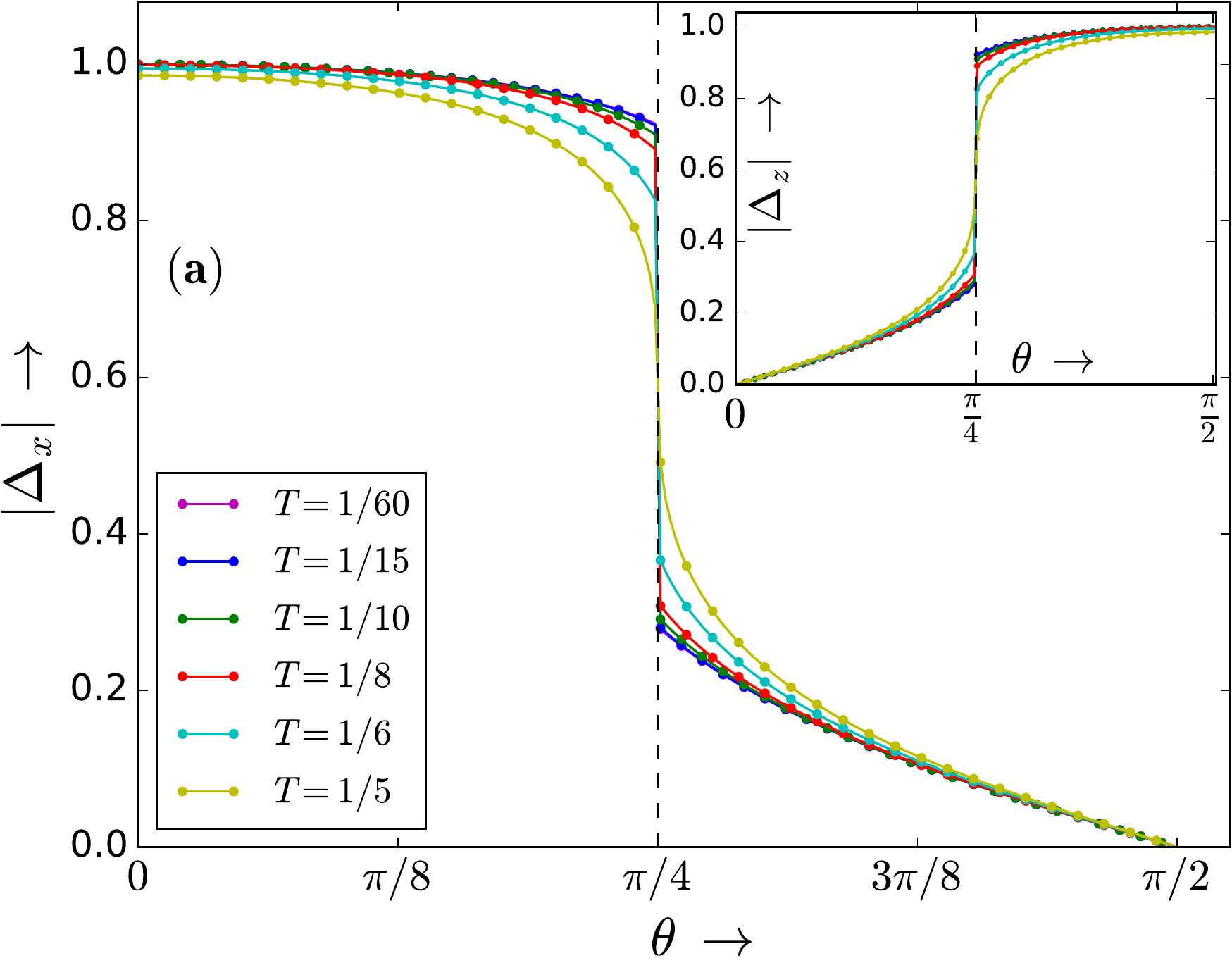}}
\subfigure{\includegraphics[width=70mm]{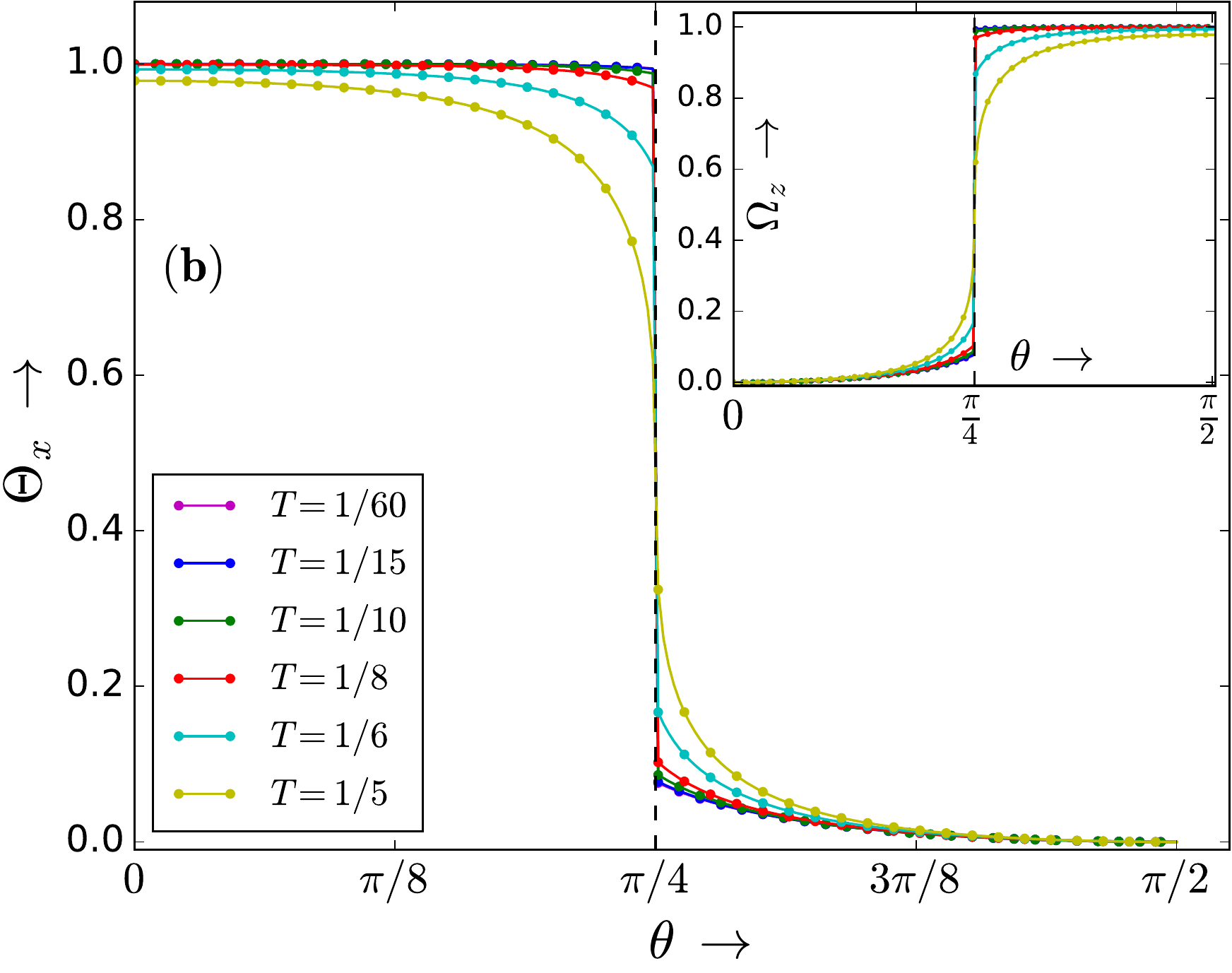}}
\subfigure{\includegraphics[width=70mm]{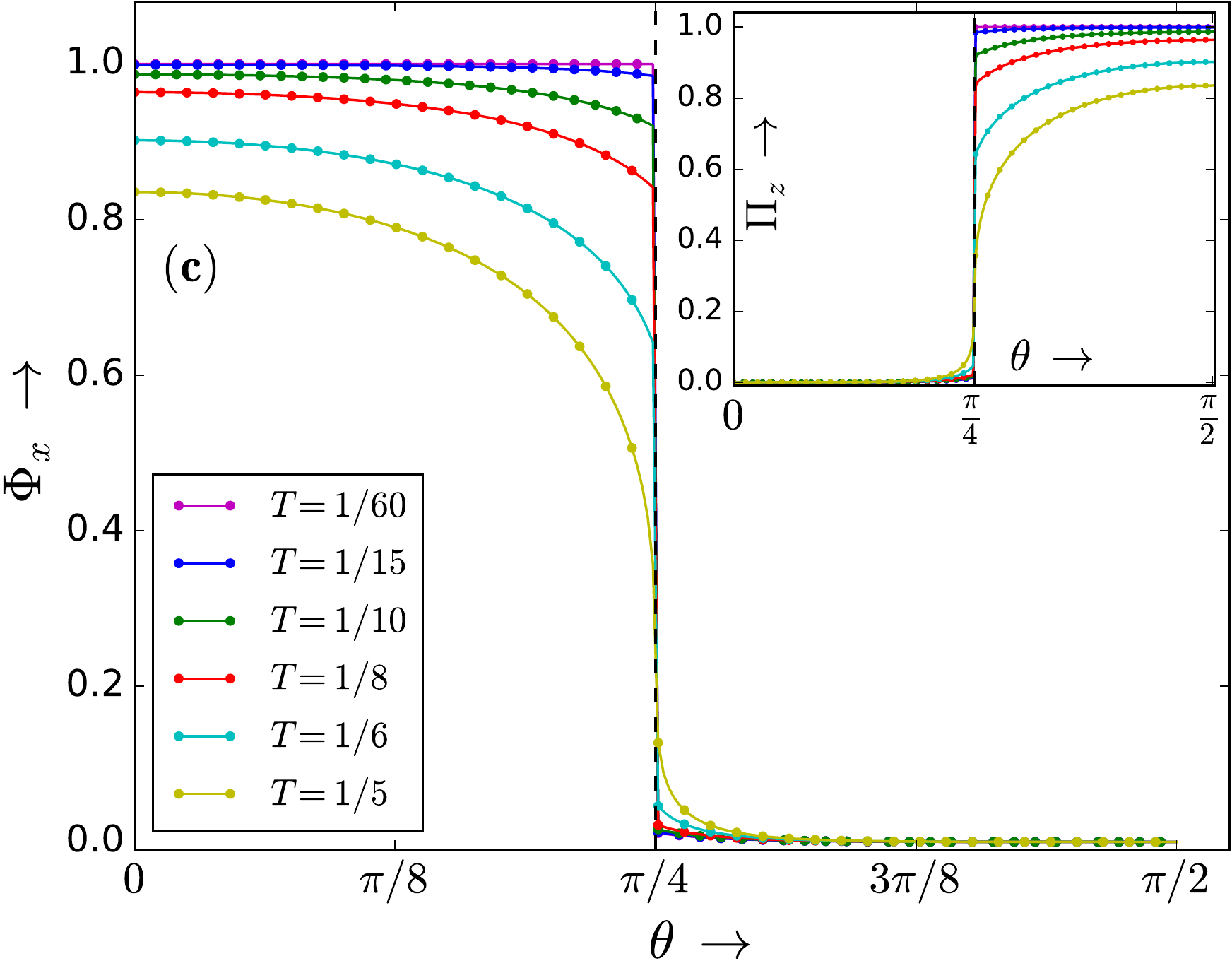}}
\caption{Various MF order parameters of eight-leg compass ladder as function of anisotropy, $\theta=\tan^{-1}(|J_{z}|/|J_{x}|)$ . The dual observables are shown in the inset figures.}
\label{figS3}
\end{figure}

\indent Next we plot various MF order parameters as function of anisotropy ($\theta$) for different values of T (see Fig.~\ref{figS3}). We observe that $\Delta_{x},\Delta_{z}$ show properties similar to four-leg case. The main thing to notice is that the discontinuity of these order parameters melts slowly and it survives up to $T\approx 1/5$, which is higher than what is observed in four-leg case ($T\approx 1/8$). 

\section{Fluctuations around mean field theory}\label{App F}
Finally we discuss qualitatively about various fluctuation effects around the four-leg ladder MFT. There are mainly two different sources of fluctuations which we have neglected in MFT~: (1) the interaction between $P$ and $W$  spins (or between `$a$' and `$b$' JW fermions), this effect is present at both $T=0$, and $T>0$. (2) thermal ($T>0$) fluctuations of static $\mathbb{Z}_{2}$ fields, $V^{z}_{j}$ . These fields are present in (a) the interaction terms (between $P$ and $W$) and (b) the quadratic MF Hamiltonian of $W$ spins ($H_{2}$). The timescale associated with collective fluctuations of JW fermions is of the order of $1/J_{x}$ or $1/J_{z}$, so that these are switched on gradually. On the other hand, the local quenches (flipping of $V^{z}_{j}$ at $T>0$) are suddenly switched on. Thus, there is a clear separation of timescales associated with different fluctuation processes. Starting with the ``fast" process (flipping of $V^{z}_{j}$), these are non-singular except around $J_{x}=J_{z}$ and $T=T_{c}$, as we will argue below in detail. Over the timescale on which the ``slow" processes (fluctuations of JW fermions) operate, there will be many such incidents of ``sudden" flipping processes. Thus the ``disorder" generated by local flips of the $V^{z}_{j}$ is annealed out at times of order of $1/J_{x}$, restoring the translational invariance. So it is reasonable to ``decouple" the two effects (1) and (2) in the spirit of an adiabatic approximation: treat the fast process first, and then consider the slow process in a medium already renormalized by the fast fluctuations. 
\\

\noindent(1):\ {\it Fluctuations arising from the interaction between JW fermions}\ :
\\

We have neglected the following interaction Hamiltonian in MFT description,
\begin{align}
V_{f}=&\frac{J_{x}}{4}\sum_{j=1}^{N}\bigg[(p^{x}_{j,1}p^{x}_{j+1,1}-\Delta^{x}_{j,1})(w^{x}_{j}w^{x}_{j+1}-\Theta^{x}_{j,1})\nonumber\\
&\hspace{0.6cm}+(p^{x}_{j,2}p^{x}_{j+1,2}-\Delta^{x}_{j,2})(v^{z}_{j}v^{z}_{j+1}w^{x}_{j}w^{x}_{j+1}-\Theta^{x}_{j,2})\bigg]\nonumber\\
&+\frac{J_{z}}{4}\sum_{j=1}^{N}\bigg[(p^{z}_{j,1}-\Delta^{z}_{j,1})(w^{z}_{j}-\Omega^{z}_{j})\nonumber\\
&\hspace{2.5cm}+(p^{z}_{j,2}-\Delta^{z}_{j,2})(w^{z}_{j}-\Omega^{z}_{j})\bigg]\label{F.1}
\end{align}
Here $p^{\mu}_{jl}=2P^{\mu}_{jl}$, $w^{\mu}_{j}=2W^{\mu}_{j}$, and $v^{\mu}_{j}=2V^{\mu}_{j}$. We have argued in the main text that if we take $v^{z}_{j}=\pm 1\ \forall j$  (which is rigorously true at $T=0$), then for a periodic four-leg ladder, we have $\Delta^{x}_{jl}\equiv \Delta_{x}$, $\Delta^{z}_{jl}\equiv \Delta_{z}$, $\Theta^{x}_{jl}\equiv\Theta_{x}$, and $\Omega_{j}^{z}\equiv \Omega_{z}$. We continue to assume this spatially uniform configuration of $v^{z}_{j}$ for $T>0$, according to the reasons explained above. The MF Hamiltonians corresponding to $p^{\mu}_{jl}$ are same for both $l=1,2$ and the interaction is only between $p$ and $w$. So we consider only one of them for further discussions and use common notation $p_{j}^{\mu}$ for both $l=1,2$. So the above Hamiltonian now reads the following,
\begin{align}
V_{f}=&\frac{J_{x}}{4}\sum_{j=1}^{N}\bigg[p^{x}_{j}p^{x}_{j+1}w^{x}_{j}w^{x}_{j+1}-2\Delta_{x}w^{x}_{j}w^{x}_{j+1}-\Theta_{x}p^{x}_{j}p^{x}_{j+1}\bigg]\nonumber\\
&+\frac{J_{z}}{4}\sum_{j=1}^{N}\big[p^{z}_{j}w^{z}_{j}-2\Delta_{z}w^{z}_{j}-\Omega_{z}p^{z}_{j}\big]\label{F.2}
\end{align}
Apart from the four-spin Ising type interaction (proportional to $J_{x}$), the rest is {\it precisely} a two-leg QCM ladder, but now with ``chain-dependent" exchanges and magnetic fields.
\begin{align}
 -\sum_{j=1}^{N}\big[J_{p}p^{x}_{j}p^{x}_{j+1}+&J_{w}w^{x}_{j}w^{x}_{j+1}\big]+\frac{J_{z}}{4}\sum_{j=1}^{N}p^{z}_{j}w^{z}_{j}\nonumber\\
 &-\sum_{j=1}^{N}\big[\mu_{p}p^{z}_{j}+\mu_{w}w^{z}_{j}\big]
\end{align}
Here $J_{p}=J_{x}\Theta_{x}/4$, $J_{w}=J_{x}\Delta_{x}/2$, $\mu_{p}=J_{z}\Omega_{z}/4$, and $\mu_{w}=J_{z}\Delta_{z}/2$ . Now it's easy to see that the fluctuations in the $zz$-sector are suppressed for all $\mu_{\sigma}\neq 0,\ \sigma=p,w$. While the remaining $xx$-part can order at $T=0$ (i.e. $\chi_{xx}(q,\omega)$ can diverge at $\omega=0$, $T=0$ for $q=0$ or $\pi$), it cannot order at $T>0$. Moreover, even with inclusion of the four-spin term, having $J_{p}\neq J_{w}$ and $\mu_{p}\neq\mu_{w}$ ensures the non-closure of the spin excitation gap. Thus neither the $xx$ nor the $zz$ fluctuations get singular. This qualitatively implies that MF results are stable against fluctuations.\\

\indent While the argument above is a symmetry-based one, actual computation of the renormalization caused by $V_{f}$ is rather involved. We now present a perturbative argument that can be made self-consistent, and which reinforces the above conclusions.\\
\indent We apply JW fermionization \eqref{A.1} on Eq.~\eqref{F.2},
\begin{widetext}
\begin{align}
V_{f}=&\bigg[\frac{J_{x}}{4}\sum_{j=1}^{N}(a_{j}-a_{j}^{\dag})(a_{j+1}+a_{j+1}^{\dag})(b_{j}-b_{j}^{\dag})(b_{j+1}+b_{j+1}^{\dag})+\frac{J_{z}}{4}\sum_{j=1}^{N}\big(2a^{\dag}_{j}a_{j}-1\big)\big(2b^{\dag}_{j}b_{j}-1\big)\bigg]\nonumber\\
&-\bigg[\frac{J_{x}}{4}\sum_{j=1}^{N}\big[2\Delta_{x}(b_{j}-b_{j}^{\dag})(b_{j+1}+b_{j+1}^{\dag})+\Theta_{x}(a_{j}-a_{j}^{\dag})(a_{j+1}+a_{j+1}^{\dag})\big]+\frac{J_{z}}{4}\sum_{j=1}^{N}\big[2\Delta_{z}(2b^{\dag}_{j}b_{j}-1)+\Omega_{z}(a^{\dag}_{j}a_{j}-1)\big]\bigg]
\end{align}
\end{widetext}
The terms inside first square bracket represents the quartic interactions between $a$ and $b$ fermions. It contains several terms, some of the interaction vertices don't even preserve the fermion numbers (like $a^{\dag}a^{\dag}b^{\dag}b$). The terms inside second square bracket denotes quadratic ``external fields". We write the above interaction in momentum space with spinor notations for fermionic fields. we define $\psi^{\dag}_{k} \equiv (a^{\dag}_{k},\ a_{-k})$, $\phi^{\dag}_{k}\equiv (b^{\dag}_{k},\ b_{-k})$, 
\begin{align}
V_{f}=&\frac{1}{2N}\sum_{kk'q}\sum_{\alpha\beta\gamma\delta}\mathcal{V}^{\alpha\beta\gamma\delta}_{kk'q}\psi^{\dag}_{k+q,\alpha}\phi^{\dag}_{k'-q,\gamma}\phi_{k',\delta}\psi_{k,\beta}\nonumber\\
&-\sum_{k}\sum_{\alpha\beta}\big[h_{a,k}^{\alpha\beta}\psi^{\dag}_{k\alpha}\psi_{k\beta}+h_{b,k}^{\alpha\beta}\phi^{\dag}_{k\alpha}\phi_{k\beta}\big]\label{F.5}
\end{align} 
Our target is to find how the single particle Green's functions (SPGF) of `$a$' and `$b$' get modified by \eqref{F.5}, the modified SPGF will cause corrections in different MF averages. We define the matrix Green's functions for `a' fermions as $\mathcal{G}^{a}_{\alpha\beta}(k,\tau)=-\langle T_{\tau}\lbrace \psi_{k,\alpha}(\tau)\psi^{\dag}_{k,\beta}(0)\rbrace\rangle$ and similar for `b' fermions, $\mathcal{G}^{b}_{\alpha\beta}(k,\tau)=-\langle T_{\tau}\lbrace \phi_{k,\alpha}(\tau)\phi^{\dag}_{k,\beta}(0)\rbrace\rangle$. The bare/ non-interacting SPGF are following,
\begin{align}
&\mathcal{G}^{0,\sigma}_{11}(k,i\omega)=\frac{(u_{k}^{\sigma})^{2}}{i\omega-E^{\sigma}_{k}}+\frac{(v_{k}^{\sigma})^{2}}{i\omega+E^{\sigma}_{k}}\nonumber\\
&\mathcal{G}^{0,\sigma}_{21}(k,i\omega)=\frac{\Delta_{k}^{\sigma}}{2E_{k}^{\sigma}}\bigg[\frac{1}{i\omega-E^{\sigma}_{k}}-\frac{1}{i\omega+E^{\sigma}_{k}}\bigg]
\end{align}
The remaining two are $\mathcal{G}^{0,\sigma}_{22}(k,i\omega)$ $=-\big[\mathcal{G}^{0,\sigma}_{11}(k,i\omega)\big]^{*}$, $\mathcal{G}^{0,\sigma}_{12}(k,i\omega)$ $=\big[\mathcal{G}^{0,\sigma}_{21}(k,i\omega)\big]^{*}$.
We find that the Feynman diagrams arising from ``external fields" in $V_{f}$ will always cancel with the zero-momentum transfer Hartree diagrams arising from the quartic interactions in the perturbative expansion of SPGF. In this way, the leading order diagrams are those of second order in $V_{f}$. It is straightforward to check that second order perturbative corrections (without Hartree diagrams) have the following general structure,
\begin{align}
\sim\frac{1}{\beta N}\sum_{q,iq}\sum_{\alpha'\beta'}&\sum_{\alpha''\beta''}\bigg[\chi^{(b),p}_{\alpha'\beta'\alpha''\beta''}(q,i\omega)\nonumber\\
&\hspace{-0.8cm}\times\mathcal{G}^{a}_{\alpha\alpha'}(p,ip)\mathcal{G}^{a}_{\alpha''\beta''}(p-q,ip-iq)\mathcal{G}^{a}_{\beta'\beta}(p,ip)\bigg]
\end{align}
the generalized susceptibility or so called bubble diagram is expressed following,
\begin{align}
&\chi^{(b),p}_{\alpha'\beta'\alpha''\beta''}(q,i\omega)=\frac{1}{\beta N}\int_{0}^{\beta}d\tau e^{i\omega\tau}\sum_{k'k''}\sum_{\gamma\delta}\sum_{\gamma'\delta'}\bigg[\mathcal{V}^{\alpha'\beta'\gamma\delta}_{p-q,k',q}\nonumber\\
&\times\mathcal{V}^{\alpha''\beta''\gamma'\delta'}_{p-q,k'',-q}\bigg\langle T_{\tau} \big\lbrace \phi^{\dag}_{k'-q,\gamma}\phi_{k',\delta}(\tau)\phi^{\dag}_{k''+q,\gamma'}\phi_{k'',\delta'}(0)\big\rbrace\bigg\rangle\bigg]
\end{align}
Explicit computation of these diagrams are straightforward but lengthy. We use here a simple physical argument. The excitation spectrum of `$a$' and `$b$' JW fermions are always gapped for all $J_{z}/J_{x}$ and $0\leq T< T_{c}$ , which means that single particle density of states (DOS), $D(\omega)$ is zero at low energies (for $\omega\leq \Delta$, the energy gap). This will put a infrared cut-off scale for all energy/ frequency integrations in the above Feynman diagrams. So as a result, the diagrams will not diverge and the pole structure of the SPGF remains preserved with some quantitative renormalization of the energy gap (like in Ref. \cite{Chen2}). As a whole, the MFT phase diagram qualitatively remains same in the presence of these interactions. 
\\

\noindent(2):\ {\it Effect of thermal fluctuations of $V^{z}_{j}$}\ :
\\

\begin{figure}
\centering
\includegraphics[height=55mm,width=75mm]{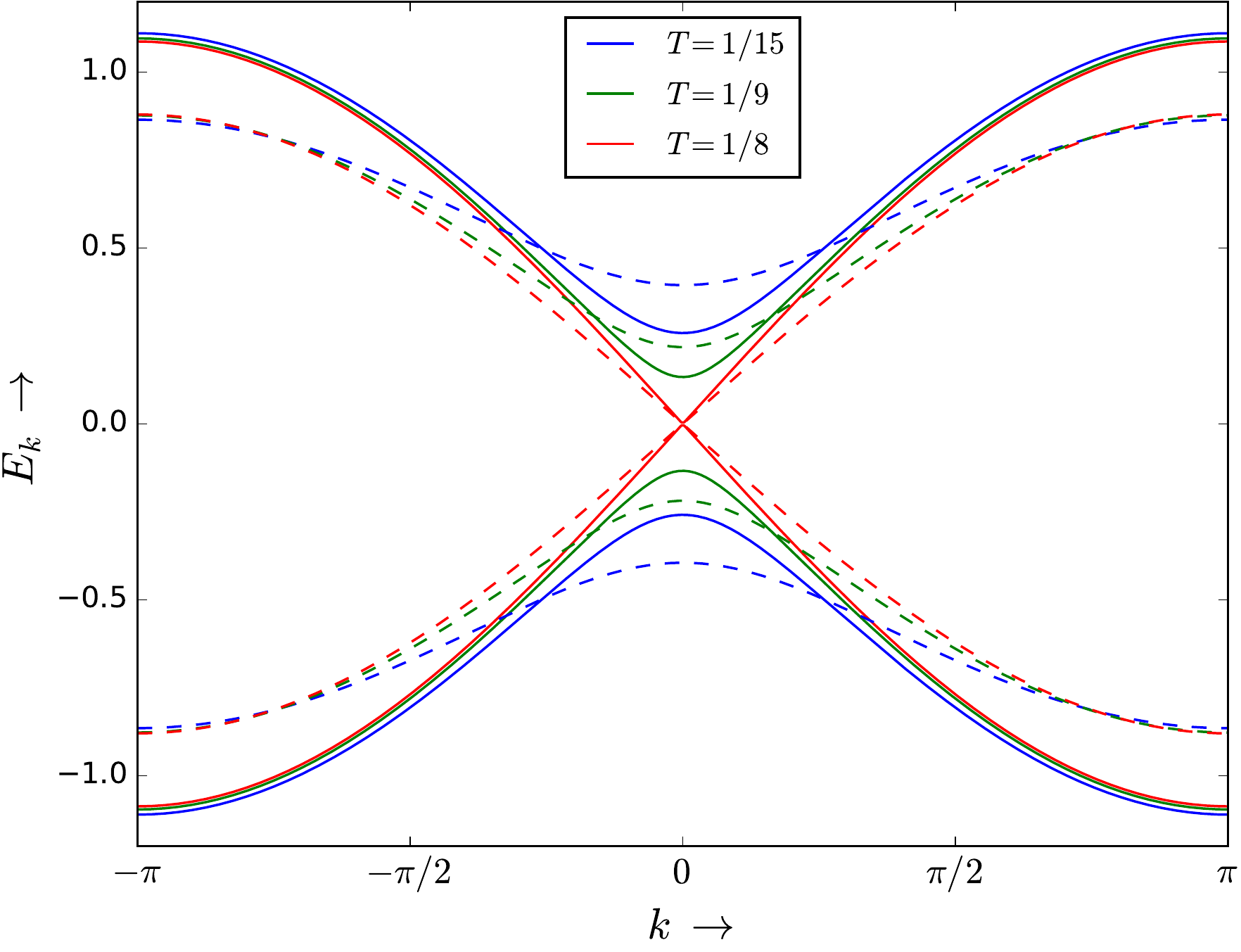}
\caption{Excitation energy ($E_{k}$) vs. momentum ($k$) plot for `$a$' (continuous lines) and `$b$' (dashed lines) JW fermions (or $P$ and $W$ Ising chains) at $J_{x}=J_{z}$ and three different $T$ values.  Close to $T_{c}\approx 0.125$, elementary Bogoliubov excitations are gapless, the spectrum goes like $E_{k}\sim v|k|$, near $k=0$.}
\label{figS4}
\end{figure}
A glance at the Hamiltonian Eq. \eqref{eq7} shows that while spatial uniformity of the bond-product of $\mathbb{Z}_{2}$ variables $v^{z}_{j}$ enables solubility of the MFT equations that lead to our results, fluctuations in these variables at finite $T$ are akin to sudden local quenches. Indeed, we can see from Eq.~\eqref{eq7} that a flip of a single $v^{z}_{j}$, say from $+1$ to $-1$ (or vice-versa), acts as a ``suddenly switched on" potential for the JW fermions. However, since these are described by a fully gapped $p$-wave superconductor, the orthogonality catastrophe (OC) associated with sudden switching of such a local potential in a Fermi sea is suppressed by the $p$-wave gap. This is true everywhere, except at a single point $J_{x}=J_{z}$ and $T=T_{c}$, where the gapless $1d$ like fermionic spectrum (see Fig.~\ref{figS4}) causes an OC to occur. Had this occurred at $T=0$, it would have generated an infrared singularity and invalidated any picture based on well-defined JW fermions, or of $p$-wave pairs formed from them \cite{AlessandroSilva}. However, at finite $T$, it is well known that this singularity is smeared by a scale $k_{B}T$, $k_{B}=$ Boltzmann constant. The resulting Doniach-Sunjic lineshape \cite{Doniach} carries the memory of the infrared singularity, and would correspond to a branch cut, instead of a renormalized pole structure, in the spin-fluctuation propagators. 
\\
\indent To make the above statements more rigorous, we consider the following dynamical spin correlation function,
\begin{align}
G_{ij}(t)&=\langle T^{z}_{i,2}(t)T^{z}_{j,2}(0)\rangle=\langle V^{x}_{i}(t)V^{x}_{j}(0)\rangle \label{F9}
\end{align}
Here the average is computed using the density matrix, $\rho=e^{-\beta H_{MF}}/Z$ with $Z=\text{Tr}(e^{-\beta H_{MF}})$. Here $H_{MF}=\sum_{l=1,2}H^{(l)}_{p}+H_{w}[\lbrace v^{z}_{j}\rbrace]$, and
\begin{align}
& H_{p}^{(l)}=J_{x}\sum_{j=1}^{N}(1+\Theta^{x}_{jl})P^{x}_{jl}P^{x}_{j+1,l}+\frac{J_{z}}{2}\sum_{j=1}^{N}(1+\Omega^{z}_{jl})P^{z}_{jl}\nonumber\\
& H_{w}[\lbrace v^{z}_{j}\rbrace]=J_{x}\sum_{j=1}^{N}\big[\Delta^{x}_{j,1}+(v^{z}_{j}v^{z}_{j+1})\Delta^{x}_{j,2}\big]W^{x}_{j}W^{x}_{j+1}\nonumber\\
&\ \ \ \ \ \ \ \ \ \  \  \ +\frac{J_{z}}{2}\sum_{j=1}^{N}(\Delta^{z}_{j,1}+\Delta^{z}_{j,2})W^{z}_{j}\label{F10}
 \end{align}
 We see $[H_{MF}, v^{z}_{j}]=0$ for all $j$. So, while computing the trace in the common eigenbasis of $H_{MF}$ and $\lbrace v^{z}_{j}\rbrace$, we need to flip $v^{z}_{j}$ at the same location twice to remain in a particular $\lbrace v^{z}_{j}\rbrace$. This makes the correlator completely local. Thus, $G_{ij}(t)=\delta_{ij}\langle V^{x}_{j}(t)V^{x}_{j}(0)\rangle$. Now, 
\begin{align*}
\langle |V^{x}_{i}(t)V^{x}_{j}(0)|\rangle =\langle |\ \cdot \big(e^{iH_{MF}t}V^{x}_{i}e^{-iH_{MF}t}V^{x}_{i}|\rangle)
\end{align*}
The physical interpretation is clear from the expression. We flip $V^{z}_{i}$ of the ``initial" state at $t'=0$, then time evolve the resulting state from $t'=0$ to $t'=t$, then further flip $V^{z}_{i}$ at $t'=t$, and time evolve backwards from $t$ to $0$ to get the ``final" state. This correlation function denotes the overlap between the ``final" and ``initial" states, the so called ``fidelity".
We write the average in Eq.~\eqref{F9} as following,
\begin{align}
\langle V^{x}_{j}(t)&V^{x}_{j}(0)\rangle=\frac{1}{Z}\bigg[\text{Tr}'_{p,w}\big(e^{-\beta H_{MF}[\lbrace v^{z}_{j}\rbrace=\pm 1]}V^{x}_{j}(t)V^{x}_{j}(0)\big)\nonumber\\
&+\sum'_{\lbrace v^{z}_{j}\rbrace}\text{Tr}_{p,w}\big(e^{-\beta H_{MF}[\lbrace v^{z}_{j}\rbrace]}V^{x}_{j}(t)V^{x}_{j}(0)\big)\bigg]\label{F11}
\end{align}
and
\begin{align}
Z=\text{Tr}'_{p,w}\big(e^{-\beta H_{MF}[\lbrace v^{z}_{j}\rbrace=\pm 1]}\big)+\sum'_{\lbrace v^{z}_{j}\rbrace}\text{Tr}_{p,w}\big(e^{-\beta H_{MF}[\lbrace v^{z}_{j}\rbrace]}\big)\label{F12}
\end{align}
The first two terms of Eqs.~\eqref{F11}, \eqref{F12} denote sum of all states of $P^{\alpha},\ W^{\beta}$ spins in the subspace where all $v^{z}_{j}=\pm 1$ (uniform configurations of $v^{z}_{j}$). The second terms of Eqs.~\eqref{F11}, \eqref{F12} denote sum over all remaining non-uniform configurations of $v^{z}_{j}$, where one or more $v^{z}_{j}$ spins are flipped. These configurations naturally increase the disorder in the system, thus play less dominant role (with respect to the first terms of Eqs.~\eqref{F11}, \eqref{F12}) where system has some kind of average order ($\Delta_{x}, \Delta_{z}$, $\Theta_{x}$, $\Omega_{z}$). These high energy configurations (random $\lbrace v^{z}_{j}\rbrace$ configurations) will play an important role only near and above $T_{c}$, where the system transits to a disordered phase. \\
\indent Below $T_{c}$, we assume that the effect of these non-uniform configurations of $\lbrace v^{z}_{j}\rbrace$ have been captured by renormalizing various mean-fields (also the $T_{c}$), i.e. $\Delta_{jl}^{x}\rightarrow \tilde{\Delta}^{x}_{jl}$, $\Delta_{jl}^{z}\rightarrow \tilde{\Delta}^{z}_{jl}$, $\Theta_{jl}^{x}\rightarrow \tilde{\Theta}^{x}_{jl}$, $\Omega_{jl}^{z}\rightarrow \tilde{\Omega}^{z}_{jl}$, and $T_{c}\rightarrow \tilde{T}_{c}(<T_{c}^{MF})$. These renormalized mean-fields and $\tilde{T}_{c}$ are unknown, unless one could treat the effect of $v^{z}_{j}$ fluctuations exactly. This renormalization of various parameters (below $T_{c}$) is justified because the JW fermion excitations are gapped, so infrared singularities are suppressed as we will see below. So we approximate Eqs.~\eqref{F11} and \eqref{F12} as
\begin{align}
\langle V^{x}_{j}(t)V^{x}_{j}(0)\rangle\approx \frac{\text{Tr}'_{p,w}\big(e^{-\beta \tilde{H}_{MF}[\lbrace v^{z}_{j}\rbrace=\pm 1]}V^{x}_{j}(t)V^{x}_{j}(0)\big)}{\text{Tr}'_{p,w}\big(e^{-\beta \tilde{H}_{MF}[\lbrace v^{z}_{j}\rbrace=\pm 1]}\big)}
\end{align}
Here the tilde sign on $H_{MF}$ tells that we have replaced all the mean-fields by renormalized parameters and put all $v^{z}_{j}=\pm 1$. In addition, the above approximation is fully sensible close to $0\leq T<<T_{c}$.\\
\indent Now we consider the effect of flipping a single $v^{z}_{j}$ at $j=j'$ on the Hamiltonian $H_{MF}$ and on various order parameters. After it has been flipped at $j'$,
\begin{align*}
&\tilde{\Theta}^{x}_{j',1}=4\langle W^{x}_{j'}W^{x}_{j'+1}\rangle\ \ \tilde{\Theta}_{j',2}^{x}=4\langle v^{z}_{j'}v^{z}_{j'+1}W^{x}_{j'}W^{x}_{j'+1}\rangle
\end{align*}
Here the trace operations are performed over all $\lbrace v^{z}_{j}\rbrace$, i.e. using Eqs.~\eqref{F11} and \eqref{F12}, thus giving all the renormalized parameters. We see that $\tilde{\Theta}_{j,1}^{x}$ and $\tilde{\Theta}^{x}_{j,2}$ are different in general, once all possible configurations of $v^{z}_{j}$ are considered in the trace. 
$\tilde{\Omega}^{z}_{j'l}=4\langle Q^{x}_{j'l}Q^{x}_{j',l+1}\rangle=4\langle Q^{x}_{j',1}Q^{x}_{j',2}\rangle=2\langle W^{z}_{j'}\rangle\equiv \Omega^{z}_{j'}$ 
This quantity is $l$ independent, because of just four legs and periodic boundary conditions along z direction. Due to differences between $\Theta^{x}_{j,1}$ and $\Theta^{x}_{j,2}$, the Hamiltonians $H^{(l=1)}_{p}$ and $H^{(l=2)}_{p}$ are now different [see Eq.~\eqref{F10}]. So, $\tilde{\Delta}^{x}_{j,1}\neq \tilde{\Delta}^{x}_{j,2}$ and $\tilde{\Delta}^{z}_{j,1}\neq \tilde{\Delta}^{z}_{j,2}$ ($\tilde{\Delta}^{x}_{jl}=4\langle P^{x}_{jl}P^{x}_{j+1,l}\rangle$, $\tilde{\Delta}^{z}_{jl}=2\langle P^{z}_{jl}\rangle$). Before considering the effect of flipping $v^{z}_{j}$ on $H_{w}$, we use the Kramer-Wannier duality for $W$ spins, 
\begin{align*}
W^{z}_{j}\rightarrow 2\tilde{W}^{x}_{j}\tilde{W}^{x}_{j+1}, \ \ \ \ W^{x}_{j}W^{x}_{j+1}\rightarrow \frac{1}{2}\tilde{W}^{z}_{j}
\end{align*} 
This duality transformation only places ``disorder" ($v^{z}_{j}$) on the local sites rather than on the links, otherwise does not effect any local dynamics. The Hamiltonian $H_{\tilde{w}}$ (in the dual spin language) after the flip of $v^{z}_{j}$ reads,
\begin{align}
&H_{\tilde{w}}[\lbrace v^{z}\rbrace, -v^{z}_{j'}]=\bigg[J_{z}\sum_{j=1}^{N}(\tilde{\Delta}^{z}_{j,1}+\tilde{\Delta}^{z}_{j,2})\tilde{W}^{x}_{j}\tilde{W}^{x}_{j+1}\nonumber\\
&+\frac{J_{x}}{2}\sum_{j=1}^{N}(\tilde{\Delta}^{x}_{j,1}+\tilde{\Delta}^{x}_{j,2})\tilde{W}^{z}_{j}\bigg]+\frac{J_{x}}{2}(\delta V_{j'})\tilde{W}^{z}_{j'}
\end{align}
Here $\delta V_{j'}=(\tilde{\Delta}^{x}_{j',1}-\tilde{\Delta}^{x}_{j',2})-(\tilde{\Delta}^{x}_{j',1}+\tilde{\Delta}^{x}_{j',2})=-2\tilde{\Delta}^{x}_{j',2}$ denotes the change in local potential due to flipping of $v^{z}_{j'}$. To simplify the picture further and to understand the leading order effect, we discard the $j$ dependence of various mean-fields (but keep the $l$ dependence). So,
\begin{align}
&H_{\tilde{w}}[\lbrace v^{z}_{j}\rbrace, -v^{z}_{j'}]\approx \bigg[J_{z}(\tilde{\Delta}^{z}_{1}+\tilde{\Delta}^{z}_{2})\sum_{j=1}^{N}\tilde{W}^{x}_{j}\tilde{W}^{x}_{j+1}\nonumber\\
&+\frac{J_{x}}{2}(\tilde{\Delta}^{x}_{1}+\tilde{\Delta}^{x}_{2})\sum_{j=1}^{N}\tilde{W}^{z}_{j}\bigg]+\frac{J_{x}}{2}(\delta V)\tilde{W}^{z}_{j'}
\end{align}
\begin{align}
H^{(l)}_{p}\approx J_{x}(1+\tilde{\Theta}^{x}_{l})\sum_{j=1}^{N}P^{x}_{jl}P^{x}_{j+1,l}+\frac{J_{z}}{2}(1+\tilde{\Omega}^{z})\sum_{j=1}^{N}P^{z}_{jl}
\end{align}
Here $\tilde{\Delta}^{\alpha}_{l}=\frac{1}{N}\sum_{j=1}^{N}\tilde{\Delta}^{\alpha}_{jl}$, with $\alpha=x,z$ and $l=1,2$. Similarly, we define $\tilde{\Theta}^{x}_{l}$ and $\tilde{\Omega}^{z}$ also. These replacements are reasonable because all these order parameters are protected by LDS. So at low $T<T_{c}$, their spatial fluctuations (along $x$) are relatively less (except some extreme points like $J_{x}\rightarrow 0$, or $J_{z}\rightarrow 0$) compared to very high $T$ where everything becomes disordered.\\
\indent Because $P$ and $W$ spins do not interact with each other in $H_{MF}$, the trace operations for $P$ and $W$ are decoupled both in the numerator and denominator. So,
\begin{align}
\langle V^{x}_{j}(t)V^{x}_{j}(0)\rangle= \frac{\text{Tr}'_{\tilde{w}}\big(e^{-\beta \tilde{H}_{\tilde{w}}[\lbrace v^{z}_{j}\rbrace=\pm 1]}V^{x}_{j}(t)V^{x}_{j}(0)\big)}{\text{Tr}'_{\tilde{w}}\big(e^{-\beta \tilde{H}_{\tilde{w}}[\lbrace v^{z}_{j}\rbrace=\pm 1]}\big)}
\end{align}
Here
\begin{align}
\tilde{H}_{\tilde{w}}[\lbrace v^{z}_{j}\rbrace=\pm 1]&=J_{z}(\tilde{\Delta}^{z}_{1}+\tilde{\Delta}^{z}_{2})\sum_{j=1}^{N}\tilde{W}^{x}_{j}\tilde{W}^{x}_{j+1}\nonumber\\
&+\frac{J_{x}}{2}(\tilde{\Delta}^{x}_{1}+\tilde{\Delta}^{x}_{2})\sum_{j=1}^{N}\tilde{W}^{z}_{j}
\end{align}
So,
$H_{\tilde{w}}[\lbrace v^{z}_{j}\rbrace, -v^{z}_{j'}]=\tilde{H}_{\tilde{w}}[\lbrace v^{z}_{j}\rbrace=\pm 1]+\frac{J_{x}}{2}(\delta V)\tilde{W}^{z}_{j'}$. Now rescale the coupling parameters with respect to $J_{z}(\tilde{\Delta}^{z}_{1}+\tilde{\Delta}^{z}_{2})/4$ , define $g=\frac{J_{x}(\tilde{\Delta}^{x}_{1}+\tilde{\Delta}^{x}_{2})}{J_{z}(\tilde{\Delta}^{z}_{1}+\tilde{\Delta}^{z}_{2})}$, $\delta g=\frac{J_{x}\delta V}{J_{z}(\tilde{\Delta}^{z}_{1}+\tilde{\Delta}^{z}_{2})}$. We further denote $\tilde{H}_{\tilde{w}}[\lbrace v^{z}_{j}\rbrace=\pm 1]=H_{0}$ and $2(\delta g)\tilde{W}^{z}_{j'}=V$. Then using the cumulant expansion for Green's function \cite{Mahan}, we see
\begin{align}
G_{jj}(t)=\langle e^{iH_{0}t}e^{-i(H_{0}+V)t}\rangle&=\langle T_{t}\big\lbrace e^{-i\int_{0}^{t}dt' V(t')}\big\rbrace\rangle\nonumber\\
&\equiv e^{\sum_{l=1}^{\infty}F_{l}(t)}.
\end{align}
Here 
\begin{align}
F_{l}(t)=\frac{(-1)^{l}}{l}\int_{0}^{t}dt_{1}....dt_{n}\langle T_{t} V(t_{1})...V(t_{n})\rangle_{connected}
\end{align}
we can choose the location of the sudden quench at $j'=0$, then
\begin{align*}
V=2(\delta g)\tilde{W}^{z}_{0}=\frac{\delta g}{L}\sum_{k,k'}(b^{\dag}_{k}b_{k'}-b_{k}b^{\dag}_{k'})
\end{align*}
Now, we expand up to second order cumulant,
\begin{align}
F_{1}(t)=\frac{it}{L}(\delta g)\sum_{k}\big((u^{b}_{k})^{2}-(v^{b}_{k})^{2}\big)\tanh{(\beta E^{b}_{k}/2)}
\end{align}
\begin{align}
&F^{(1)}_{2}(t)=\frac{it}{L^{2}}(\delta g)^{2}\sum_{k,k}\bigg[\frac{1}{2}\big(u^{b}_{k}u^{b}_{k'}-v^{b}_{k}v^{b}_{k'}\big)^{2}\frac{(n^{b}_{k}-n^{b}_{k'})}{E^{b}_{k'}-E^{b}_{k}}\nonumber\\
&\ \ \ \ \ \ \ +\frac{1}{2}\big(u^{b}_{k}v^{b}_{k'}+u^{b}_{k'}v^{b}_{k}\big)^{2}\frac{(1-n^{b}_{k}-n^{b}_{k'})}{E^{b}_{k}+E^{b}_{k'}}\bigg]
\end{align}
These terms are linear in $t$, enters into the coherent part of the spin correlation function, $G_{jj}(t)$. The second order cumulants (non-linear in $t$) are given following,
\begin{align}
&F^{'}_{2}(t)=-\frac{(\delta g)^{2}}{L^{2}}\times\nonumber\\
&\sum_{k,k'}\bigg[(u^{b}_{k}u^{b}_{k'}-v^{b}_{k}v^{b}_{k'})^{2}\frac{n^{b}_{k}(1-n^{b}_{k'})}{(E^{b}_{k'}-E^{b}_{k})^{2}}(1-e^{-it(E^{b}_{k'}-E^b_{k})})\nonumber\\
&+\frac{1}{2}(u^{b}_{k}v^{b}_{k'}+u^{b}_{k'}v^{b}_{k})^{2}\bigg\lbrace\frac{(1-n^{b}_{k})(1-n^{b}_{k'})}{(E^{b}_{k}+E^{b}_{k'})^{2}}(1-e^{-it(E^{b}_{k}+E^b_{k'})})\nonumber\\
&+\frac{n^{b}_{k}n^{b}_{k'}}{(E^{b}_{k}+E^{b}_{k'})^{2}}(1-e^{it(E^{b}_{k}+E^b_{k'})})\bigg\rbrace\bigg]\label{F23}
\end{align}
Here $u_{k}^{b}, v^{b}_{k}$, and $E^{b}_{k}$ have same expression as Eq.~\eqref{A.3}, except, now $t_{b}=\frac{J_{z}}{2}(\tilde{\Delta}^{z}_{1}+\tilde{\Delta}^{z}_{2})$ and $\mu_{b}=J_{x}(\tilde{\Delta}^{x}_{1}+\tilde{\Delta}^{x}_{2})$.\\
\indent Now to simplify Eq.~\eqref{F23}, we replace the Fermi distribution, $n^{b}_{k}=(e^{\beta E^{b}_{k}}+1)^{-1}$ simply by zero. This approximation is completely valid for $\Delta>k_{B}T$ ($\Delta$= energy gap of the JW fermion spectrum). It does not mean although that the $T$ dependence is completely neglected as various MF order parameters carry the $T$ dependence implicitly ($\tilde{\Delta}^{\alpha}_{l}\equiv \tilde{\Delta}^{\alpha}_{l}(T)$, $\alpha=x,z$). After this simplification, 
\begin{align}
F^{'}_{2}(t)=-\int_{0}^{\infty}\frac{d\omega}{\omega^{2}}R(\omega)(1-e^{-i\omega t}) \label{F24}
\end{align}
with
\begin{align}
R(\omega)=\frac{(\delta g)^{2}}{2L^{2}}\sum_{k,k'}(u^{b}_{k}v^{b}_{k'}+u^{b}_{k'}v^{b}_{k})^{2}\delta(\omega-E^{b}_{k}-E^{b}_{k'})
\end{align}
Due to gapped nature of the energy spectrum $E^{b}_{k}$, the Dirac delta function gives non-zero contribution for $|1-g|/2\leq \omega\leq |1+g|/2$. So, when the JW fermionic excitations are gapped (for $g\neq 1$, i.e. (a) $J_{x}=J_{z},\ T<\tilde{T}_{c}$ or (b) $J_{x}\neq J_{z}$, any $T$), the above integral \eqref{F24} has finite upper and lower cut-offs, thus avoiding infrared singularity due to the ``sudden quench". The pole structure of $G_{jj}$ remains preserved, with only quantitative renormalization. When $J_{x}=J_{z}$, $T\sim\tilde{T}_{c}$, we have $g=\frac{\tilde{\Delta}^{x}_{1}+\tilde{\Delta}^{x}_{2}}{\tilde{\Delta}^{x}_{1}+\tilde{\Delta}^{x}_{2}}\bigg|_{T\sim\tilde{T}_{c}}\sim 1$ (We have seen earlier that at $T=T_{c}^{MF}$, $\Delta_{x}=\Delta_{z}$ or the nematic order ($\sim |\Delta_{x}-\Delta_{z}|$) vanishes. Similarly, here we take $\tilde{\Delta}_{l}^{x}=\tilde{\Delta}^{z}_{l}$ at $T=\tilde{T}_{c}$). At sufficiently long times, $t\rightarrow\infty$ \cite{AlessandroSilva, Mahan},
\begin{align}
F'_{2}(\infty)\approx -\bigg(\frac{\delta g}{2\pi}\bigg)^{2}\ln{|g-1|}
\end{align}
Thus, $G_{jj}(t\rightarrow\infty)\rightarrow 0$ as $g\rightarrow 1$ (or $J_{x}=J_{z}$, $T\rightarrow\tilde{T}_{c}$). Exactly, at $g=1$ ($J_{x}=J_{z}$,\ $T=\tilde{T}_{c}$), $G_{jj}(t)$ develops a power law behaviour in time (leading order) \cite{AlessandroSilva, Mahan},
\begin{align}
G_{jj}(t)\sim e^{-i\epsilon t} (it)^{-\big(\frac{\delta g}{2\pi}\big)^{2}}
\end{align}
So, the leading order effect of this ``sudden quench" are (1) vanishing ``quasi-particle residue" $\sim e^{F^{'}_{2}(\infty)}$ and (2) power law time correlations. 
The effect of the Fermi factors (neglected in the above calculations) will have the ``smearing" effect (on a scale $\sim k_{B}T$) as we have argued at the beginning of this section. \\
\indent So, the above discussion shows that the sudden local quenches associated with fluctuations of the $\mathbb{Z}_{2}$ variables ($V^{z}_{j}$) do not invalidate MFT results qualitatively either, except close to $J_{x}=J_{z}$ and $T\sim\tilde{T}_{c}$.


\begin{thebibliography}{34}%




\makeatletter
\providecommand \@ifxundefined [1]{%
 \@ifx{#1\undefined}
}%
\providecommand \@ifnum [1]{%
 \ifnum #1\expandafter \@firstoftwo
 \else \expandafter \@secondoftwo
 \fi
}%
\providecommand \@ifx [1]{%
 \ifx #1\expandafter \@firstoftwo
 \else \expandafter \@secondoftwo
 \fi
}%
\providecommand \natexlab [1]{#1}%
\providecommand \enquote  [1]{``#1''}%
\providecommand \bibnamefont  [1]{#1}%
\providecommand \bibfnamefont [1]{#1}%
\providecommand \citenamefont [1]{#1}%
\providecommand \href@noop [0]{\@secondoftwo}%
\providecommand \href [0]{\begingroup \@sanitize@url \@href}%
\providecommand \@href[1]{\@@startlink{#1}\@@href}%
\providecommand \@@href[1]{\endgroup#1\@@endlink}%
\providecommand \@sanitize@url [0]{\catcode `\\12\catcode `\$12\catcode
  `\&12\catcode `\#12\catcode `\^12\catcode `\_12\catcode `\%12\relax}%
\providecommand \@@startlink[1]{}%
\providecommand \@@endlink[0]{}%
\providecommand \url  [0]{\begingroup\@sanitize@url \@url }%
\providecommand \@url [1]{\endgroup\@href {#1}{\urlprefix }}%
\providecommand \urlprefix  [0]{URL }%
\providecommand \Eprint [0]{\href }%
\providecommand \doibase [0]{http://dx.doi.org/}%
\providecommand \selectlanguage [0]{\@gobble}%
\providecommand \bibinfo  [0]{\@secondoftwo}%
\providecommand \bibfield  [0]{\@secondoftwo}%
\providecommand \translation [1]{[#1]}%
\providecommand \BibitemOpen [0]{}%
\providecommand \bibitemStop [0]{}%
\providecommand \bibitemNoStop [0]{.\EOS\space}%
\providecommand \EOS [0]{\spacefactor3000\relax}%
\providecommand \BibitemShut  [1]{\csname bibitem#1\endcsname}%
\let\auto@bib@innerbib\@empty
\bibitem [{\citenamefont {Imada}\ \emph
  {et~al.}(1998{\natexlab{b}})}]{Imada-Fujimori}%
  \BibitemOpen
 \bibfield  {author} {\bibinfo {author} {\bibfnamefont {M.}~\bibnamefont
  {Imada}}, \bibinfo {author} {\bibfnamefont {A.}~\bibnamefont {Fujimori}}, and \bibinfo {author} {\bibfnamefont {Y.}~\bibnamefont {Tokura}},\ }
  \href {\doibase 10.1103/RevModPhys.70.1039} {\bibfield
  {journal} {\bibinfo  {journal} {Rev. Mod. Phys.}\ }\textbf {\bibinfo
  {volume} {70}},\ \bibinfo {pages} {1039--1263} (\bibinfo {year} {1998})}\BibitemShut
  {NoStop}%
\bibitem [{\citenamefont {R.~Moessner}\ and\ \citenamefont
  {S.~L.~Sondhi}(2001{\natexlab{b}})}]{Moessner-Sondhi}%
  \BibitemOpen
  \bibfield  {author} {\bibinfo {author} {\bibfnamefont {R.}~\bibnamefont
  {Moessner}}\ and\ \bibinfo {author} {\bibfnamefont {S.~L.}~\bibnamefont
  {Sondhi}},~}
  \href {\doibase 10.1103/PhysRevLett.86.1881} {\bibfield  {journal}
  {\bibinfo  {journal} {Phys. Rev. Lett.}\ }\textbf {\bibinfo {volume} {86}},\
  \bibinfo {pages} {1881--1884} (\bibinfo {year} {2001}{\natexlab{b}})}\BibitemShut
  {NoStop}%
\bibitem [{\citenamefont {Lacroix}\ \emph{et.~al.}(2011)}]{Lacorix}%
  \BibitemOpen
  \href {\doibase 10.1007/978-3-642-10589-0}
  {\emph
  {\bibinfo {title} {Introduction to Frustrated Magnetism. Materials, Experiments, Theory}}},\ eds.
  \bibfield  {editor} {\bibinfo {editor} {\bibfnamefont {C.}~\bibnamefont
  {Lacorix}},\ {\bibfnamefont {P.}~\bibnamefont
  {Mendels}},\ and\ {\bibfnamefont {F.}~\bibnamefont
  {Mila}}},\ Springer Series in Solid State Sciences,\ vol. 164,\ (\bibinfo  {publisher} {Springer-Verlag Berlin Heidelberg},\ 
  \bibinfo {year} {2011})\BibitemShut {NoStop}%
\bibitem [{\citenamefont {A.~Kitaev} (2006{\natexlab{b}})}]{Kitaev}%
  \BibitemOpen
  \bibfield  {author} {\bibinfo {author} {\bibfnamefont {A.}~\bibnamefont
  {Kitaev}},~}
  \href {\doibase 10.1016/j.aop.2005.10.005} {\bibfield  {journal}
  {\bibinfo  {journal} {Annals of Physics}\ }\textbf {\bibinfo {volume} {321}},\
  \bibinfo {pages} {2--111} (\bibinfo {year} {2006}{\natexlab{b}})}\BibitemShut
  {NoStop}%
\bibitem [{\citenamefont {K.~I.~Kugel}\ and\ \citenamefont
  {D.~I.~Khomskii}(1982{\natexlab{b}})}]{Kugel-Khomskii}%
  \BibitemOpen
  \bibfield  {author} {\bibinfo {author} {\bibfnamefont {K.~I.}~\bibnamefont
  {Kugel}}\ and\ \bibinfo {author} {\bibfnamefont {D.~I.~}\ \bibnamefont
  {Khomskii}},~}
  \href {\doibase 10.1070/pu1982v025n04abeh004537} {\bibfield  {journal}
  {\bibinfo  {journal} {Soviet Physics Uspekhi}\ }\textbf {\bibinfo {volume} {25}},\
  \bibinfo {pages} {231--256} (\bibinfo {year} {1982}{\natexlab{b}})}\BibitemShut
  {NoStop}%
\bibitem [{\citenamefont {Zohar~Nussinov}\ and\ \citenamefont
  {Jeroen van den Brink}(2015{\natexlab{b}})}]{Nussinov-review}%
  \BibitemOpen
  \bibfield  {author} {\bibinfo {author} {\bibfnamefont {Z.}~\bibnamefont
  {Nussinov}}\ and\ \bibinfo {author} {\bibfnamefont {J.~van den }\ \bibnamefont
  {Brink}},~}
  \href {\doibase 10.1103/RevModPhys.87.1} {\bibfield  {journal}
  {\bibinfo  {journal} {Rev. Mod. Phys.}\ }\textbf {\bibinfo {volume} {87}},\
  \bibinfo {pages} {1--59} (\bibinfo {year} {2015}{\natexlab{b}})}\BibitemShut
  {NoStop}%
\bibitem [{\citenamefont {Dou\ifmmode \mbox{\c{c}}\else \c{c}\fi{}ot}\ \emph
  {et~al.}(2005{\natexlab{b}})}\citenamefont {Dou\ifmmode \mbox{\c{c}}\else \c{c}\fi{}ot}, \citenamefont {Feigel'man},\
  and\ \citenamefont {Ioffe}]{Doucot}%
  \BibitemOpen
  \bibfield  {author} {\bibinfo {author} {\bibfnamefont {B.}~\bibnamefont
  {Dou\ifmmode \mbox{\c{c}}\else \c{c}\fi{}ot}}, \bibinfo {author} {\bibfnamefont {M. V.}~\bibnamefont {Feigel'man}}, \bibinfo {author} {\bibfnamefont {L. B.}~\bibnamefont {Ioffe}}, and \bibinfo {author} {\bibfnamefont {A. S.}~\bibnamefont {Ioselevich}},\ }
  \href {\doibase 10.1103/PhysRevB.71.024505} {\bibfield
  {journal} {\bibinfo  {journal} {Phys. Rev. B}\ }\textbf {\bibinfo
  {volume} {71}},\ \bibinfo {pages} {024505} (\bibinfo {year} {2005})}\BibitemShut
  {NoStop}%
\bibitem [{\citenamefont {Nussinov}\ and\ \citenamefont
  {Fradkin}(2005{\natexlab{b}})}]{Nussinov}%
  \BibitemOpen
  \bibfield  {author} {\bibinfo {author} {\bibfnamefont {Z.}~\bibnamefont
  {Nussinov}}\ and\ \bibinfo {author} {\bibfnamefont {E.}\ \bibnamefont
  {Fradkin}},~}
  \href {\doibase 10.1103/PhysRevB.71.195120} {\bibfield  {journal}
  {\bibinfo  {journal} {Phys. Rev. B}\ }\textbf {\bibinfo {volume} {71}},\
  \bibinfo {pages} {195120} (\bibinfo {year} {2005}{\natexlab{b}})}\BibitemShut
  {NoStop}%
\bibitem [{\citenamefont {Vidal}\ \emph
  {et~al.}(2005{\natexlab{b}})}]{Vidal}%
  \BibitemOpen
 \bibfield  {author} {\bibinfo {author} {\bibfnamefont {J.}~\bibnamefont
  {Vidal}}, \bibinfo {author} {\bibfnamefont {R.}~\bibnamefont {Thomale}}, \bibinfo {author} {\bibfnamefont {K. P.}~\bibnamefont {Schmidt}}, and \bibinfo {author} {\bibfnamefont {S.}~\bibnamefont {Dusuel}},\ }
  \href {\doibase 10.1103/PhysRevB.80.081104} {\bibfield
  {journal} {\bibinfo  {journal} {Phys. Rev. B}\ }\textbf {\bibinfo
  {volume} {80}},\ \bibinfo {pages} {081104(R)} (\bibinfo {year} {2009})}\BibitemShut
  {NoStop}%
\bibitem [{\citenamefont {Chen}\ and\ \citenamefont
  {Hu}(2007{\natexlab{b}})}]{Chen1}%
  \BibitemOpen
  \bibfield  {author} {\bibinfo {author} {\bibfnamefont {H.~D.}~\bibnamefont
  {Chen}}\ and\ \bibinfo {author} {\bibfnamefont {J.}\ \bibnamefont
  {Hu}},\ }
  \href {\doibase 10.1103/PhysRevB.76.193101} {\bibfield  {journal}
  {\bibinfo  {journal} {Phys. Rev. B}\ }\textbf {\bibinfo {volume} {76}},\
  \bibinfo {pages} {193101} (\bibinfo {year} {2007}{\natexlab{b}})}\BibitemShut
  {NoStop}%
\bibitem [{\citenamefont {Brzezicki}\ \emph
  {et~al.}(2007{\natexlab{b}})}]{Andrzej1}%
  \BibitemOpen
 \bibfield  {author} {\bibinfo {author} {\bibfnamefont {W.}~\bibnamefont
  {Brzezicki}}, \bibinfo {author} {\bibfnamefont {J.}~\bibnamefont {Dziarmaga}}, and \bibinfo {author} {\bibfnamefont {A. M.}~\bibnamefont {Ole\'s}},\ }
  \href {\doibase 10.1103/PhysRevB.75.134415} {\bibfield
  {journal} {\bibinfo  {journal} {Phys. Rev. B}\ }\textbf {\bibinfo
  {volume} {75}},\ \bibinfo {pages} {134415} (\bibinfo {year} {2007})}\BibitemShut
  {NoStop}%
\bibitem [{\citenamefont {Brzezicki}\ and\ \citenamefont
  {Andrzej M.}(2009{\natexlab{b}})}]{Andrzej2}%
  \BibitemOpen
  \bibfield  {author} {\bibinfo {author} {\bibfnamefont {W.}~\bibnamefont
  {Brzezicki}}\ and\ \bibinfo {author} {\bibfnamefont {A. M.}\ \bibnamefont
  {Ole\'s}},\ }
  \href {\doibase 10.1103/PhysRevB.76.193101} {\bibfield  {journal}
  {\bibinfo  {journal} {Phys. Rev. B}\ }\textbf {\bibinfo {volume} {80}},\
  \bibinfo {pages} {014405} (\bibinfo {year} {2009}{\natexlab{b}})}\BibitemShut
  {NoStop}%
\bibitem [{\citenamefont {Mattis}\ and\ \citenamefont
  {Nam}(1972{\natexlab{b}})}]{Mattis}%
  \BibitemOpen
  \bibfield  {author} {\bibinfo {author} {\bibfnamefont {D. C.}~\bibnamefont
  {Mattis}}\ and\ \bibinfo {author} {\bibfnamefont {S. B.}\ \bibnamefont
  {Nam}},\ }
  \href {\doibase 10.1063/1.1666120} {\bibfield  {journal}
  {\bibinfo  {journal} {Journal of Mathematical Physics}\ }\textbf {\bibinfo {volume} {13}},\
  \bibinfo {pages} {1185-1189} (\bibinfo {year} {1972}{\natexlab{b}})}\BibitemShut
  {NoStop}%
\bibitem [{\citenamefont {O. Hart}\ \emph
  {et~al.}(2020{\natexlab{b}})}]{Hart}%
  \BibitemOpen
 \bibfield  {author} {\bibinfo {author} {\bibfnamefont {O.}~\bibnamefont
  {Hart}}, \bibinfo {author} {\bibfnamefont {S.}~\bibnamefont {Gopalakrishnan}}, and \bibinfo {author} {\bibfnamefont {C.}~\bibnamefont {Castelnovo}},\ }
  \href {https://arxiv.org/abs/2009.00618} {\bibfield
  {journal} {\bibinfo  {journal} {arXiv:2009.00618 [cond-mat.str-el]}\ }(\bibinfo {year} {2020})}\BibitemShut
  {NoStop}%
\bibitem [{\citenamefont {Chen}\ \emph
  {et~al.}(2007{\natexlab{b}})}]{Chen2}%
  \BibitemOpen
 \bibfield  {author} {\bibinfo {author} {\bibfnamefont {H.~D.}~\bibnamefont
  {Chen}}, \bibinfo {author} {\bibfnamefont {C.}~\bibnamefont {Fang}}, \bibinfo {author} {\bibfnamefont {J.}~\bibnamefont {Hu}},\ and\ \bibinfo {author} {\bibfnamefont {H.}~\bibnamefont {Yao}},\ }
  \href {\doibase 10.1103/PhysRevB.75.144401} {\bibfield  {journal}
  {\bibinfo  {journal} {Phys. Rev. B}\ }\textbf {\bibinfo {volume} {75}},\
  \bibinfo {pages} {144401} (\bibinfo {year} {2007}{\natexlab{b}})}\BibitemShut
  {NoStop}%
\bibitem [{\citenamefont {Dorier}\ \emph
  {et~al.}(2005{\natexlab{b}})}]{Dorier}%
  \BibitemOpen
 \bibfield  {author} {\bibinfo {author} {\bibfnamefont {J.}~\bibnamefont
  {Dorier}}, \bibinfo {author} {\bibfnamefont {F.}~\bibnamefont {Becca}}, and \bibinfo {author} {\bibfnamefont {F.}~\bibnamefont {Mila}},\ }
  \href {\doibase 10.1103/PhysRevB.72.024448} {\bibfield  {journal}
  {\bibinfo  {journal} {Phys. Rev. B}\ }\textbf {\bibinfo {volume} {72}},\
  \bibinfo {pages} {024448} (\bibinfo {year} {2005}{\natexlab{b}})}\BibitemShut
  {NoStop}%
\bibitem [{\citenamefont {Batista}\ and\ \citenamefont
  {Nussinov}(2005{\natexlab{b}})}]{Batista}%
  \BibitemOpen
  \bibfield  {author} {\bibinfo {author} {\bibfnamefont {C. D.}~\bibnamefont
  {Batista}}\ and\ \bibinfo {author} {\bibfnamefont {Z.}\ \bibnamefont
  {Nussinov}},\ }
  \href {\doibase 10.1103/PhysRevB.72.045137} {\bibfield  {journal}
  {\bibinfo  {journal} {Phys. Rev. B}\ }\textbf {\bibinfo {volume} {72}},\
  \bibinfo {pages} {045137} (\bibinfo {year} {2005}{\natexlab{b}})}\BibitemShut
  {NoStop}%
\bibitem{footnote1}Here we are considering ferromagnetic compass interactions ($J_{x},J_{z}<0$). The antiferromagnetic ($J_{x},J_{z}>0$) case is related to previous one by a trivial rotation of spin basis on one sublattice. 
\bibitem [{\citenamefont {Rynbach}\ \emph
  {et~al.}(2010{\natexlab{b}})}]{Rynbach}%
  \BibitemOpen
 \bibfield  {author} {\bibinfo {author} {\bibfnamefont {A. van}~\bibnamefont
  {Rynbach}}, \bibinfo {author} {\bibfnamefont {S.}~\bibnamefont {Todo}}, and \bibinfo {author} {\bibfnamefont {S.}~\bibnamefont {Trebst}},\ }
  \href {\doibase 10.1103/PhysRevLett.105.146402} {\bibfield  {journal}
  {\bibinfo  {journal} {Phys. Rev. Lett.}\ }\textbf {\bibinfo {volume} {105}},\
  \bibinfo {pages} {146402} (\bibinfo {year} {2010}{\natexlab{b}})}\BibitemShut
  {NoStop}%
\bibitem [{\citenamefont {Or\'us}\ \emph
  {et~al.}(2009{\natexlab{b}})}]{Orus}%
  \BibitemOpen
 \bibfield  {author} {\bibinfo {author} {\bibfnamefont {R.}~\bibnamefont
  {Or\'us}}, \bibinfo {author} {\bibfnamefont {Andrew C.}~\bibnamefont {Doherty}}, and \bibinfo {author} {\bibfnamefont {G.}~\bibnamefont {Vidal}},\ }
  \href {\doibase 10.1103/PhysRevLett.102.077203} {\bibfield  {journal}
  {\bibinfo  {journal} {Phys. Rev. Lett.}\ }\textbf {\bibinfo {volume} {102}},\
  \bibinfo {pages} {077203} (\bibinfo {year} {2009}{\natexlab{b}})}\BibitemShut
  {NoStop}%
\bibitem [{\citenamefont {Brzezicki}\ and\ \citenamefont
  {Andrzej}(2005{\natexlab{b}})}]{Brzezicki}%
  \BibitemOpen
  \bibfield  {author} {\bibinfo {author} {\bibfnamefont {W.}~\bibnamefont
  {Brzezicki}}\ and\ \bibinfo {author} {\bibfnamefont {A. M.}\ \bibnamefont
  {Ole\'s}},\ }
  \href {\doibase 10.1103/PhysRevB.82.060401} {\bibfield  {journal}
  {\bibinfo  {journal} {Phys. Rev. B}\ }\textbf {\bibinfo {volume} {82}},\
  \bibinfo {pages} {060401(R)} (\bibinfo {year} {2010}{\natexlab{b}})}\BibitemShut
  {NoStop}%
\bibitem [{\citenamefont {L. D. Faddeev}\ and\ \citenamefont
  {L. A. Takhtajan}(1981{\natexlab{b}})}]{Faddeev}%
  \BibitemOpen
  \bibfield  {author} {\bibinfo {author} {\bibfnamefont {L.~D.}~\bibnamefont
  {Faddeev}}\ and\ \bibinfo {author} {\bibfnamefont {L.~A.}\ \bibnamefont
  {Takhtajan}},\ }
  \href {\doibase 10.1016/0375-9601(81)90335-2} {\bibfield  {journal}
  {\bibinfo  {journal} {Physics Letters A}\ }\textbf {\bibinfo {volume} {85}},\
  \bibinfo {pages} {375 - 377} (\bibinfo {year} {1981}{\natexlab{b}})}\BibitemShut
  {NoStop}%
\bibitem [{\citenamefont {F. Trousselet}\ \emph
  {et~al.}(2012{\natexlab{b}})}]{Trousselet}%
  \BibitemOpen
 \bibfield  {author} {\bibinfo {author} {\bibfnamefont {F.}~\bibnamefont
  {Trousselet}}, \bibinfo {author} {\bibfnamefont {A.~M.}~\bibnamefont {Ole\'s}}, \bibinfo {author} {\bibfnamefont {P.~}\bibnamefont {Horsch}},\ }
  \href {\doibase 10.1103/PhysRevB.86.134412} {\bibfield  {journal}
  {\bibinfo  {journal} {Phys. Rev. B}\ }\textbf {\bibinfo {volume} {86}},\
  \bibinfo {pages} {134412} (\bibinfo {year} {2012}{\natexlab{b}})}\BibitemShut
  {NoStop}%
\bibitem [{\citenamefont {S. Wenzel}\ and\ \citenamefont
  {W. Janke}(2008{\natexlab{b}})}]{Wenzel}%
  \BibitemOpen
  \bibfield  {author} {\bibinfo {author} {\bibfnamefont {S.}~\bibnamefont
  {Wenzel}}\ and\ \bibinfo {author} {\bibfnamefont {W.}\ \bibnamefont
  {Janke}},\ }
  \href {\doibase 10.1103/PhysRevB.78.064402} {\bibfield  {journal}
  {\bibinfo  {journal} {Phy. Rev. B}\ }\textbf {\bibinfo {volume} {78}},\
  \bibinfo {pages} {064402} (\bibinfo {year} {2008}{\natexlab{b}})}\BibitemShut
  {NoStop}%
\bibitem [{\citenamefont {J. Oitmaa}\ and\ \citenamefont
  {C. J. Hamer}(2011{\natexlab{b}})}]{Oitmaa}%
  \BibitemOpen
  \bibfield  {author} {\bibinfo {author} {\bibfnamefont {J.}~\bibnamefont
  {Oitmaa}}\ and\ \bibinfo {author} {\bibfnamefont {C.~J.}\ \bibnamefont
  {Hamer}},\ }
  \href {\doibase 10.1103/PhysRevB.83.094437} {\bibfield  {journal}
  {\bibinfo  {journal} {Phy. Rev. B}\ }\textbf {\bibinfo {volume} {83}},\
  \bibinfo {pages} {094437} (\bibinfo {year} {2011}{\natexlab{b}})}\BibitemShut
  {NoStop}%
\bibitem [{\citenamefont {P.~Czarnik}\ \emph
  {et~al.}(2016{\natexlab{b}})}]{Czarnik-Supplementary}%
  \BibitemOpen
 \bibfield  {author} {\bibinfo {author} {\bibfnamefont {P.}~\bibnamefont
  {Czarnik}}, \bibinfo {author} {\bibfnamefont {J.}~\bibnamefont {Dziarmaga}}, \bibinfo {author} {\bibfnamefont {A.~M.~}\bibnamefont {Ole\'s}},\ }
  \href {\doibase 10.1103/PhysRevB.93.184410} {\bibfield  {journal}
  {\bibinfo  {journal} {Phys. Rev. B}\ }\textbf {\bibinfo {volume} {93}},\
  \bibinfo {pages} {184410} (\bibinfo {year} {2016}{\natexlab{b}})}\BibitemShut
  {NoStop}%
\bibitem [{\citenamefont {Anderson P. W.}\ (2006{\natexlab{b}})}]{Anderson}%
  \BibitemOpen
 \bibfield  {author} {\bibinfo {author} {\bibfnamefont {P.~W.}~\bibnamefont
  {Anderson}}, }
  \href {\doibase 10.1038/nphys388} {\bibfield  {journal}
  {\bibinfo  {journal} {{\it Nature Phys}}\ }\textbf {\bibinfo {volume} {2}},\
  \bibinfo {pages} {626–630} (\bibinfo {year} {2006}{\natexlab{b}})}\BibitemShut
  {NoStop}%
\bibitem [{\citenamefont {Chen}\ and\ \citenamefont
  {Nussinov}(2008{\natexlab{b}})}]{Chen-Nussinov}%
  \BibitemOpen
  \bibfield  {author} {\bibinfo {author} {\bibfnamefont {H.~D.}~\bibnamefont
  {Chen}}\ and\ \bibinfo {author} {\bibfnamefont {Z.}\ \bibnamefont
  {Nussinov}},\ }
  \href {\doibase 10.1088/1751-8113/41/7/075001} {\bibfield  {journal}
  {\bibinfo  {journal} {Journal of Physics A: Mathematical and Theoretical}\ }\textbf {\bibinfo {volume} {41}},\
  \bibinfo {pages} {075001} (\bibinfo {year} {2008}{\natexlab{b}})}\BibitemShut
  {NoStop}%
\bibitem [{\citenamefont {Nevidomskyy}\ (2011{\natexlab{b}})}]{Nevidomskyy}%
  \BibitemOpen
 \bibfield  {author} {\bibinfo {author} {\bibfnamefont {Andriy H.}~\bibnamefont
  {Nevidomskyy}},\ }
  \href {https://arxiv.org/abs/1104.1747} {\bibfield  {journal}
  {\bibinfo  {journal} {arXiv:1104.1747 [cond-mat.str-el]}\ }(\bibinfo {year} {2011}{\natexlab{b}})}\BibitemShut
  {NoStop}%
\bibitem [{\citenamefont {C. Chen}\ \emph
  {et~al.}(2009{\natexlab{b}})}]{C.Chen}%
  \BibitemOpen
 \bibfield  {author} {\bibinfo {author} {\bibfnamefont {C.-C.}~\bibnamefont
  {Chen}}, \bibinfo {author} {\bibfnamefont {B.}~\bibnamefont {Moritz}}, \bibinfo {author} {\bibfnamefont {J.~}\bibnamefont {van den Brink}},\ \bibinfo {author} {\bibfnamefont {T.~P.~}\bibnamefont {Devereaux}},\ and \bibinfo {author} {\bibfnamefont {R.~R.~P.~}\bibnamefont {Singh}},\ }
  \href {\doibase 10.1103/PhysRevB.80.180418} {\bibfield  {journal}
  {\bibinfo  {journal} {Phys. Rev. B}\ }\textbf {\bibinfo {volume} {80}},\
  \bibinfo {pages} {180418(R)} (\bibinfo {year} {2009}{\natexlab{b}})}\BibitemShut
  {NoStop}%
\bibitem [{\citenamefont {J. Nasu}\ and\ \citenamefont
  {S. Ishihara}(2012{\natexlab{b}})}]{Nasu}%
  \BibitemOpen
  \bibfield  {author} {\bibinfo {author} {\bibfnamefont {J.}~\bibnamefont
  {Nasu}}\ and\ \bibinfo {author} {\bibfnamefont {S.}\ \bibnamefont
  {Ishihara}},\ }
  \href {\doibase 10.1209/0295-5075/97/27002} {\bibfield  {journal}
  {\bibinfo  {journal} {{EPL} (Europhysics Letters)}}\ \textbf {\bibinfo {volume} {97}},\
  \bibinfo {pages} {27002} (\bibinfo {year} {2012}{\natexlab{b}})}\BibitemShut
  {NoStop}%
\bibitem [{\citenamefont {Rajiv R.P. Singh}\ (2009{\natexlab{b}})}]{Rajiv R. Singh}%
  \BibitemOpen
 \bibfield  {author} {\bibinfo {author} {\bibfnamefont {Rajiv R.~P.}~\bibnamefont
  {Singh}},}
  \href {https://arxiv.org/abs/0903.4408} {\bibfield  {journal}
  {\bibinfo  {journal} {arXiv:0903.4408 [cond-mat.str-el]}\ }(\bibinfo {year} {2009}{\natexlab{b}})}\BibitemShut
  {NoStop}%
\bibitem [{\citenamefont {R. Applegate}\ \emph
  {et~al.}(2010{\natexlab{b}})}]{Applegate}%
  \BibitemOpen
 \bibfield  {author} {\bibinfo {author} {\bibfnamefont {R.}~\bibnamefont
  {Applegate}}, \bibinfo {author} {\bibfnamefont {J.}~\bibnamefont {Oitmaa}}, and \bibinfo {author} {\bibfnamefont {R. R. P.}~\bibnamefont {Singh}},\ }
  \href {\doibase 10.1103/PhysRevB.81.024505} {\bibfield  {journal}
  {\bibinfo  {journal} {Phys. Rev. B}\ }\textbf {\bibinfo {volume} {81}},\
  \bibinfo {pages} {024505} (\bibinfo {year} {2010}{\natexlab{b}})}\BibitemShut
  {NoStop}%
\bibitem [{\citenamefont {M. Takahashi} (1989{\natexlab{b}})}]{Takahashi}%
  \BibitemOpen
 \bibfield  {author} {\bibinfo {author} {\bibfnamefont {M.}~\bibnamefont
  {Takahashi}},\ }
  \href {\doibase 10.1103/PhysRevB.40.2494} {\bibfield  {journal}
  {\bibinfo  {journal} {Phys. Rev. B}\ }\textbf {\bibinfo {volume} {40}},\
  \bibinfo {pages} {2494--2501} (\bibinfo {year} {1989}{\natexlab{b}})}\BibitemShut
  {NoStop}%
\bibitem [{\citenamefont {Schlappa, J.}\ \emph
  {et~al.}(2012{\natexlab{b}})}]{Schlappa}%
  \BibitemOpen
 \bibfield  {author} {\bibinfo {author} {\bibfnamefont {J.}~\bibnamefont
  {Schlappa}}, \bibinfo {author} {\bibfnamefont {K.}~\bibnamefont {Wohlfeld}}, \bibinfo {author} {\bibfnamefont {K.}~\bibnamefont {Zhou}}\ \emph
  {et~al.},\ }
  \href {\doibase 10.1038/nature10974} {\bibfield  {journal}
  {\bibinfo  {journal} {{\it Nature}}\ }\textbf {\bibinfo {volume} {485}},\
  \bibinfo {pages} {82-85} (\bibinfo {year} {2012}{\natexlab{b}})}\BibitemShut
  {NoStop}%
\bibitem [{\citenamefont {Nussinov}\ and\ \citenamefont
  {Oritz}(2009{\natexlab{b}})}]{Nussinov2009}%
  \BibitemOpen
  \bibfield  {author} {\bibinfo {author} {\bibfnamefont {Z.}~\bibnamefont
  {Nussinov}}\ and\ \bibinfo {author} {\bibfnamefont {G.}\ \bibnamefont
  {Oritz}},\ }
  \href {\doibase 10.1073/pnas.0803726105} {\bibfield  {journal}
  {\bibinfo  {journal} {Proceedings of the National Academy of Sciences\ }}\textbf {\bibinfo {volume} {106}},\
  \bibinfo {pages} {16944--16949} (\bibinfo {year} {2009}{\natexlab{b}})}\BibitemShut
  {NoStop}%
\bibitem [{\citenamefont {Subir Sachdev}(2011)}]{Subir Sachdev-Supplementary}%
  \BibitemOpen
  \bibfield  {author} {\bibinfo {author} {\bibfnamefont {S.}\ \bibnamefont
  {Sachdev}},\ }\bibfield  {title} \href {https://doi.org/10.1017/CBO9780511973765} {\emph {\bibinfo {title} {Quantum Phase Transitions},}} {2nd edition,} {\bibinfo {publisher} {Cambridge University Press, Cambridge}} {(\bibinfo {year} {2011})}\BibitemShut {NoStop}%
\bibitem [{\citenamefont {Cenke Xu}\ and\ \citenamefont
  {J. E. Moore}(2005{\natexlab{b}})}]{Xu-Moore-Supplementary}%
  \BibitemOpen
  \bibfield  {author} {\bibinfo {author} {\bibfnamefont {C.}~\bibnamefont
  {Xu}}\ and\ \bibinfo {author} {\bibfnamefont {J.~E.~}\ \bibnamefont
  {Moore}},\ }
  \href {\doibase https://doi.org/10.1016/j.nuclphysb.2005.04.003} {\bibfield  {journal}
  {\bibinfo  {journal} {Nuclear Physics B}\ }\textbf {\bibinfo {volume} {716}},\
  \bibinfo {pages} {487 - 508} (\bibinfo {year} {2005}{\natexlab{b}})}\BibitemShut
  {NoStop}%
\bibitem [{\citenamefont {Eytan Barouch}\ \citenamefont {Barry M. McCoy}(1971{\natexlab{b}})}]{Barouch-Supplementary}%
  \BibitemOpen
 \bibfield  {author} {\bibinfo {author} {\bibfnamefont {E.}~\bibnamefont
  {Barouch}} and \bibinfo {author} {\bibfnamefont {B.~M.}~\bibnamefont {McCoy}},\ }
  \href {\doibase 10.1103/PhysRevA.3.786} {\bibfield  {journal}
  {\bibinfo  {journal} {Phys. Rev. A}\ }\textbf {\bibinfo {volume} {3}},\
  \bibinfo {pages} {786--804} (\bibinfo {year} {1971}{\natexlab{b}})}\BibitemShut
  {NoStop}%
\bibitem{footnote1-Supplementary} M2, D2 denotes respectively the second set of Mattis's transformations and mean-field decoupling procedure.
\bibitem [{\citenamefont {Alessandro Silva}(2008{\natexlab{b}})}]{AlessandroSilva}%
  \BibitemOpen
  \bibfield  {author} {\bibinfo {author} {\bibfnamefont {A.}~\bibnamefont
  {Silva}},\ }
  \href {\doibase 10.1103/PhysRevLett.101.120603} {\bibfield  {journal}
  {\bibinfo  {journal} {Phys. Rev. Lett.}\ }\textbf {\bibinfo {volume} {101}},\
  \bibinfo {pages} {120603} (\bibinfo {year} {2008}{\natexlab{b}})}\BibitemShut
  {NoStop}%
\bibitem [{\citenamefont {Sebastian Doniach}(2016{\natexlab{b}})}]{Doniach}%
  \BibitemOpen
  \bibfield  {author} {\bibinfo {author} {\bibfnamefont {S.}~\bibnamefont
  {Doniach}},\ }
  \href {\doibase } {\bibfield  {journal}
  {\bibinfo  {journal} {Journal of Physics: Condensed Matter}\ }\textbf {\bibinfo {volume} {28}},\
  \bibinfo {pages} {421005} (\bibinfo {year} {2016}{\natexlab{b}})}\BibitemShut
  {NoStop}%
\bibitem [{\citenamefont {Gerald D. Mahan}(2000)}]{Mahan}%
  \BibitemOpen
  \bibfield  {author} {\bibinfo {author} {\bibfnamefont {G. D.}\ \bibnamefont
  {Mahan}},\ }\bibfield  {title} \href {https://www.springer.com/gp/book/9780306463389} {\emph {\bibinfo {title} {Many-Particle Physics},}} {3rd edition,} {\bibinfo {publisher} {Springer, US}} {(\bibinfo {year} {2000})}\BibitemShut {NoStop}%
\end{thebibliography}
\end{document}